\documentclass[11pt,reqno]{article}
\usepackage{amsmath}
\usepackage{amsfonts}
\usepackage{amssymb}
\usepackage{latexsym}
\usepackage[dvips]{graphicx}
\usepackage{epsf}

\allowdisplaybreaks \numberwithin{equation}{section}

\newcommand{\be}{\begin{equation}}
\newcommand{\ee}{\end{equation}}
\newcommand{\bea}{\begin{eqnarray}}
\newcommand{\eea}{\end{eqnarray}}
\newcommand{\nn}{\nonumber}
\newcommand{\Zom}{\mathbb{Z}}

\newcommand{\Nom}{\mathbb{N}}

\newcommand{\tr}{{\rm tr}}
\newcommand{\Tr}{{\rm Tr}}

\newcommand{\cC}{\mathcal{C}}
\newcommand{\cD}{\mathcal{D}}
\newcommand{\cF}{\mathcal{F}}

\newcommand{\cJ}{\mathcal{J}}
\newcommand{\cL}{\mathcal{L}}
\newcommand{\cM}{\mathcal{M}}
\newcommand{\cN}{\mathcal{N}}
\newcommand{\cO}{\mathcal{O}}
\newcommand{\cR}{\mathcal{R}}
\newcommand{\cS}{\mathcal{S}}

\newcommand{\cZ}{\mathcal{Z}}

\newcommand{\fp}{f^{\prime}}
\newcommand{\fpp}{f^{\prime \prime}}
\newcommand{\Pt}{\tilde{\Phi}}

\newcommand{\unit}{{\bf{1}}}

\newcommand{\half}{\tfrac{1}{2}}

\newcommand{\bb}{\bar{b}}
\newcommand{\cb}{\bar{c}}
\newcommand{\gb}{\bar{g}}
\newcommand{\hb}{\bar{h}}
\newcommand{\vb}{\bar{v}}

\newcommand{\I}{\mathrm{i}}
\newcommand{\e}{\mathrm{e}}
\newcommand{\p}{\partial}

\newcommand{\Cb}{\bar{C}}
\newcommand{\Db}{\bar{D}}
\newcommand{\Ib}{\bar{I}}

\newcommand{\Kb}{\bar{K}}
\newcommand{\Pb}{\bar{P}}
\newcommand{\Rb}{\bar{R}}

\newcommand{\Gammab}{\bar{\Gamma}}
\newcommand{\eb}{\bar{\eta}}

\newcommand{\gt}{\tilde{g}}

\newcommand{\Wt}{\widetilde{W}}

\newcommand{\ub}{\bar{\upsilon}}
\newcommand{\mb}{\bar{\mu}}

\newcommand{\bZ}{{\bf{Z}}}

\begin{document}
\begin{flushright} \small
ITP--UU--07/63 \\ SPIN--07/48 \\SPHT--T07/154
\end{flushright}
\bigskip

\begin{center}
 {\LARGE\bfseries On the renormalization group flow of $f(R)$-gravity }
\\[10mm]
Pedro F. Machado$^1$ and Frank Saueressig$^{1,2}$ \\[3mm]
$^1${\small\slshape
Institute for Theoretical Physics \emph{and} Spinoza Institute \\
Utrecht University, 3508 TD Utrecht, The Netherlands \\
{\upshape\ttfamily P.Machado@phys.uu.nl} }
\\[3mm]
$^2${\small\slshape
Service de Physique Th\'eorique, CEA Saclay \\
91191 Gif-Sur-Yvette Cedex, France \\
{\upshape\ttfamily F.S.Saueressig@phys.uu.nl} }\\

\end{center}
\vspace{5mm}

\hrule\bigskip

\centerline{\bfseries Abstract} \medskip
\noindent
We use the functional renormalization group equation for quantum gravity
to construct a non-perturbative flow equation for
modified gravity theories of the form $S = \int \! d^dx \sqrt{g} f(R)$.
Based on this equation we show that certain gravitational interactions
monomials can be consistently decoupled from the renormalization group (RG) flow and 
reproduce recent results on the asymptotic safety conjecture. 
The non-perturbative RG flow of non-local extensions of the Einstein-Hilbert 
truncation including $\int \! d^dx \sqrt{g}\ln(R)$
 and $\int \! d^dx \sqrt{g} R^{-n}$ interactions is investigated in detail.
The inclusion of such interactions resolves the infrared singularities plaguing the RG
trajectories with positive cosmological constant in previous truncations. 
In particular, in some $R^{-n}$-truncations all physical trajectories emanate from a Non-Gaussian (UV) fixed point and are well-defined on all RG scales.  
The RG flow of the $\ln(R)$-truncation
contains an infrared attractor which  drives a positive
cosmological constant to zero dynamically.
\bigskip
\hrule\bigskip
\newpage
\tableofcontents
\newpage
\section{Introduction}
Theories of $f(R)$-gravity, in which the gravitational Lagrangian is based
 on an arbitrary function of the curvature scalar,
 have recently attracted considerable attention within the cosmology literature.
One of the prime reasons for this originates from the observation \cite{cosmoacc}
that non-local terms which become dominant as $R$ 
decreases can provide a natural explanation for the observed
late-time acceleration of our universe, without the need of introducing dark matter 
(see, e.g., \cite{ON} for a review and more references).
By combining these non-local interactions with higher derivative
curvature terms, $f(R)$-gravity can furthermore generate an inflationary phase in the
early universe while at the same time satisfying the experimental bounds of gravity tests at
 solar system scales \cite{via}.

These phenomenological successes make it interesting to investigate $f(R)$-gravity
also from a quantum gravity perspective. A framework which naturally
 lends itself to such an analysis is the functional renormalization group
 equation (FRGE) for quantum gravity \cite{MR}.\footnote{For a recent pedagogical introduction see \cite{RSrev}.} This equation allows a detailed
 investigation of the non-perturbative renormalization group (RG) flow of gravitational theories
\cite{percadou,oliver1,frank1,oliver2,oliver3,oliver4,souma,frank2,prop,oliverbook,perper1,codello,essential,lit1,lit2,hier,CPR}, and has also been used to study the consequences of scale-dependent (running) gravitational coupling constants to black hole physics \cite{BH1,BH2}, cosmology \cite{cosmo1,cosmo2,cosmo3,cosmo4,cosmo5}, galaxy rotation curves \cite{GRC,GRC2}, and 
collider signatures of TeV-scale gravity models at the LHC \cite{LHC1,LHC2,LHC3}.

The main ingredient of the FRGE is the effective 
average action $\Gamma_k$ \cite{EAArev1,EAArev2}, a Wilson-type, coarse grained effective action which is constructed along similar lines to the ordinary effective action.
Its construction starts from the standard generating functional for connected Greens functions $W$, obtained via the path integral. The usual action is then augmented by the addition of an infrared (IR) cutoff $\Delta_k S$, which provides a momentum dependent mass-squared term $\cR_k(p^2)$ for the quantum field mode with momentum $p$: for $p^2 \gg k^2$, the cutoff function $\cR_k(p^2)$ vanishes, so that the high-momentum modes are integrated out in the usual way. For $p^2 \ll k^2$, we have $\cR_k(p^2) \propto k^2$ so that the integration over the low-momentum modes is effectively suppressed by a 
mass-squared term proportional to $k^2$. This modification induces a scale-dependence in the generating functional $W_k$, which is then related to $\Gamma_k$ by a (modified) Legendre transform.
 
One of the main virtues of $\Gamma_k$ is that its scale-dependence is governed by an exact FRGE. Schematically, this equation takes the form 
\be\label{FRGE}
\p_t \Gamma_k = \frac{1}{2} \, {\Tr} \frac{\p_t \cR_k}{\Gamma^{(2)}_k + \cR_k} \, ,
\ee
with $\Gamma^{(2)}_k$ denoting the second variation of $\Gamma_k$ and $t := \ln(k)$. 
This equation defines a
 one-parameter family of effective field theories $\{ \Gamma_k, \check{k} \le k \le \hat{k}\}$,
 with $\hat{k}$ a UV cutoff scale and $\check{k}$, the effective or coarse graining scale.
 This family interpolates between the classical action $S$,
 imposed as an initial condition $\Gamma_{\hat{k}} = S$ at the scale $\hat{k}$, and the
coarse grained effective action $\Gamma_{\check{k}}$ which provides an effective description
of physical processes with typical momentum $\check{k}$. Presupposing that the limit $\check{k} \rightarrow 0$ exists (which is a priori not clear for a theory involving massless degrees of freedom), the standard effective action is obtained as  
 $\Gamma = \Gamma_0 = \lim_{k \rightarrow 0} \Gamma_k$. 

The framework outlined above provides a very powerful tool for quantizing a fundamental theory. In this case, one imposes the initial condition $\Gamma_{\hat{k}} = S$ for the classical action $S$, uses the FRGE to compute $\Gamma_k$ for $k < \hat{k}$ and then takes the limits $k \rightarrow 0$ and $\hat{k} \rightarrow \infty$. The defining property of a fundamental theory, loosely speaking, is that the limit $\hat{k} \rightarrow \infty$ exists.
Typically, taking this limit requires the existence of a UV fixed point of the RG flow. In the context of the FRGE for quantum gravity, there is by now good evidence that such a Non-Gaussian fixed point (NGFP) indeed exists,
defining a predictive fundamental theory of gravity along the lines of the asymptotic safety scenario proposed by Weinberg \cite{wein1} (see \cite{Maxrev,livrev,Perrev} for recent reviews).\footnote{Some features associated with the NGFP have also been observed in the apparently orthogonal Causal Dynamical Triangulations (CDT) approach to quantum gravity \cite{AJL1,AJL2,AJL3}, thus possibly hinting at a hidden relation between the FRGE and CDT. Moreover, the existence of the NGFP is also supported by studies of the symmetry reduced Euclidean path integral \cite{max}.}. 

Moreover, the FRGE can also be used to investigate the structure of the gravitational effective action in the IR irrespectively of the existence of a fundamental theory. Here, one starts from an effective theory at the scale $\hat{k}$ and uses the FRGE to evolve the theory down to scales $k < \hat{k}$, successively integrating out the quantum fluctuations ``shell by shell''. In particular, it allows one to address the proposals made in \cite{non-loc1,non-loc2,non-loc3} that quantum gravity gives rise to strong RG effects in the IR which are responsible for the tiny value of $\Lambda$. At the level of the gravitational RG flow, this would most likely be reflected by
the existence of an IR fixed point, which attracts the RG flow for $k \rightarrow 0$. 
Such an IR fixed point could provide a dynamical solution to the cosmological constant problem,
if $\Lambda$ is driven to zero irrespectively of its initial value at the scale $\hat{k}$.

The main limitation of the FRGE, however, is that it cannot be solved exactly.
 In order to extract some non-perturbative information from it,
 one therefore resorts to the approximation technique
 of truncating $\Gamma_k$. That is, one makes an ansatz
 for $\Gamma_k$ which only retains a finite (or infinite) subset of all possible interaction
monomials compatible with the symmetries of the theory.
Introducing $k$-dependent coupling constants multiplying the interaction monomials,
 this ansatz is substituted into the exact FRGE. By projecting
the resulting equation onto the subspace spanned by the truncation, one can then construct
the $\beta$-functions for the scale-dependent couplings without 
 evoking perturbation theory in a small parameter. Depending on the number
of couplings included in the ansatz, this procedure leads either to 
a coupled system of ordinary differential equations (finite number of couplings) 
or to a partial differential equation (infinite number of couplings) governing the scale-dependence of the 
truncated $\Gamma_k$. Consistency of the truncation ansatz thereby requires the extraction of all terms proportional to the interaction monomials spanning the truncation ansatz from the traces appearing on the RHS of the FRGE.

\begin{figure}[t]
\leavevmode
\hskip 14mm
\epsfxsize=13cm
\epsfysize=8.9cm
\epsfbox{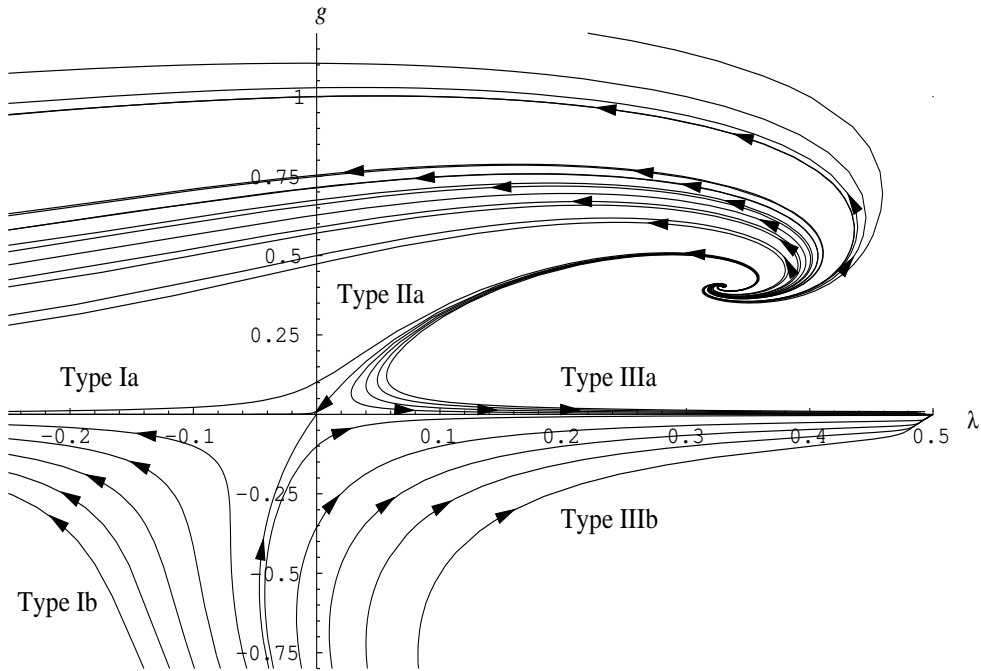}
\vskip 2mm
\caption{
The RG flow of 4-dimensional quantum gravity in the Einstein-Hilbert truncation. The arrows point in the direction of increasing coarse graining, i.e., towards decreasing $k$.}
\label{fig0}
\end{figure}

Finding good truncation subspaces generally poses a delicate problem, requiring some physics intuition about the interaction terms which are relevant for describing the physics at momentum scale $k^2$, and usually involves some educated guesswork. Contrary to perturbation theory, where the quality of the approximation is controlled by an expansion parameter, approximating the full RG flow of the theory by a truncation does not allow one to judge the quality of the truncation a priori.\footnote{A posteriori, however, there are indirect tests, such as, e.g., analyzing the (unphysical) cutoff function dependence of observable quantities \cite{RSrev}.}  
So far, most work on gravitational RG flows has been carried out in the Einstein-Hilbert truncation, in which the gravitational part of $\Gamma_k[g]$ is of the Einstein-Hilbert form,
\be\label{EHT}
\Gammab_k^{\rm EH}[g] = \frac{1}{16 \pi G_k} \int d^dx \sqrt{g} \{ -R + 2 \Lambda_k \} \, ,
\ee 
with $G_k$ and $\Lambda_k$ denoting the running Newton's constant and cosmological constant, respectively.
The associated $\beta$-functions governing the RG flow of the corresponding dimensionless couplings $g_k = G_k k^{d-2}, \lambda_k = \Lambda_k/k^2$, have been first derived in \cite{MR}, and its solutions, classified in \cite{frank1}, are shown in Figure \ref{fig0}. There, the most prominent features are the UV attractive NGFP at 
$g^* >0, \lambda^* >0$ and the boundary of the coupling constant space at $\lambda = 1/2$, which causes the
termination of the type IIIa trajectories (positive cosmological constant in the IR) at a finite value $k_{\rm term} > 0$.
Now, due to the truncation approximation, there is no a priori reason
that the NGFP and the $\lambda=1/2$-line carry over into the full theory
space. Indeed, while there exists substantial evidence that the NGFP is
not a truncation artefact and is present in the full theory, 
 we will show that the $\lambda =
1/2$-line can be generically resolved with the inclusion of non-local
curvature terms in the truncation ansatz.

One of the main results of this paper is the extension of the $\beta$-functions obtained in the Einstein-Hilbert truncation to a partial differential equation governing the non-perturbative RG flow of $f(R)$-gravity theories of the form
\be\label{gravaction}
\Gammab_k[g] = (16 \pi G_k)^{-1} \! \int \! d^dx \sqrt{g} \, f_k(R) \, ,
\ee
where $f_k(R)$ is an arbitrary, scale-dependent function of the scalar curvature.\footnote{At the classical level (or for fixed values $k$), certain gravitational actions of this form can be mapped to the Einstein Hilbert action coupled to one additional scalar field \cite{Maeda:1988ab}. While the later formulation is very useful in analyzing the solutions of the theory, the conformal rescaling which relates the two dual formulations introduces a $k$-dependence of the metric and scalar field in the dual frame, making the derivation of the RG equation in the dual frame rather cumbersome. Thus, we will derive the RG equation based on $f_k$ in the following.} This is achieved by performing the transverse-traceless (TT) decomposition of the metric fluctuations and combining the construction of the FRGE with ``type B cutoff'' \cite{oliver1} (which we will denote as TT-cutoff in the sequel) with the ideas outlined in \cite{CPR}. Subsequently, 
we use this flow equation to study the RG flow of specific non-local extensions of the Einstein-Hilbert action
\be\label{logRmodel}
\Gammab_k[g] = \frac{1}{16 \pi G_k} \int d^dx \sqrt{g} \{ -R + 2 \Lambda_k + 16 \pi G_k \, \ub_k \, \ln(R) \} \, ,
\ee 
and
\be\label{Rinvmodel}
\Gammab_k[g] = \frac{1}{16 \pi G_k} \int d^dx \sqrt{g} \{ -R + 2 \Lambda_k + 16 \pi G_k \, \mb_k \, R^{-n} \} 
\; , \qquad n \ge 1 \in \Zom \, .
\ee 
The non-local terms included in the truncations \eqref{logRmodel} and \eqref{Rinvmodel} serve as prototypes for interactions which become dominant for small values of the curvature scalar. In this respect, the primary goal of this paper is the qualitative understanding of the RG flow for the new non-local couplings $\ub_k, \mb_k$. The truncations \eqref{logRmodel} and \eqref{Rinvmodel} thereby constitute simple models which allow one to investigate if and how non-local interactions of this type affect the gravitational RG flow. Demanding that the truncated $\bar{\Gamma}_k$ give rise to stable cosmological solutions, which (most likely) do not exist for classical actions of the form \eqref{logRmodel} and \eqref{Rinvmodel}, requires the extension of the truncation ansatz by including higher (positive) powers of the scalar curvature. Based on the work on $R^2$-truncations we expect that these higher derivative couplings can appear rather naturally along the RG trajectory, so that passing on to phenomenological acceptable models is straightforward once the dynamical origin of the non-local interactions is established. 

Studing the RG flow based on the truncations \eqref{logRmodel} and \eqref{Rinvmodel} also constitutes a natural extension of the results  \cite{frank2}, where theories containing non-local volume terms were considered. The study of the RG flow of the couplings $\ub_k, \mb_k$ hence provides a natural next step in this program, laying the groundwork for analyzing truncations including the (probably more physical) non-local interactions $\int d^dx \sqrt{g} R \ln(-D^2) R$ and $\int d^dx \sqrt{g} R (-D^2)^{-1} R$ proposed in \cite{non-loc1} and \cite{Wetterich:1997bz}.

The rest of the paper is organized as follows. Section \ref{sect:2} constructs the FRGE for quantum gravity utilizing an
improved geometric gauge-fixing scheme and transverse traceless decomposition. This FRGE is restricted to $f(R)$-gravity in Section \ref{sect:3}, and the tools for evaluating the corresponding operator traces are there introduced. Section \ref{Sect:4} discusses the possibilities for decoupling certain gravitational interaction monomials from the RG flow and the resolution of the IR singularities appearing in the Einstein-Hilbert truncation. Section \ref{sect:5} and \ref{sect:6} contain the detailed analysis of the RG flow arising in the truncations \eqref{logRmodel} and \eqref{Rinvmodel}, respectively. Lastly, using our results from Section \ref{sect:3}, we evaluate the operator traces for a special choice of cutoff function in $d=4$, deriving a partial differential equation governing the RG flow of $f(R)$-gravity in Section \ref{Sect:7}, before concluding with a short summary and discussion of our results in Section \ref{sect:8}. The technical details on the heat-kernel techniques and threshold functions employed in the construction of the RG equations are collected in Appendix \ref{App:B} and \ref{App:C}, respectively.

\section{The improved FRGE for quantum gravity}
\label{sect:2}
%
We start by constructing the flow equation for $f(R)$-gravity, rederiving
the FRGE for quantum gravity based on the TT-cutoff \cite{oliver1} and including
the improvements \cite{CPR}. The novel features of our construction
are the implementation of a distinguished, geometrically motivated gauge-fixing condition
 and the sidestepping of the momentum dependent field redefinitions appearing in the original construction.
 As pointed out in \cite{CPR}, this leads to a significant simplification of the TT-cutoff. 
%
\subsection{The geometric gauge-fixing condition}
\label{sect:2.1}
In order to assure that our FRGE respects diffeomorphism invariance, we follow \cite{MR} and define a scale-dependent  Euclidean functional integral $\bZ_k$ using the background field method. Thereby, the quantum metric $\gamma_{\mu\nu}$ is decomposed into an arbitrary but fixed background metric $\gb_{\mu \nu}$ and a (not necessarily small) fluctuation field $h_{\mu \nu}$ 
\be\label{Qmetric}
\gamma_{\mu\nu} = \gb_{\mu \nu} + h_{\mu \nu} \, .
\ee
The formal expression for the generating functional $\bZ_k$ then includes integrations over the quantum fluctuations $h_{\mu\nu}$ and the Faddeev-Popov ghosts $C^\mu$, $\Cb_\mu$,
\be\label{fctint}
\begin{split}
\bZ_k[{\rm sources}] = & \int \cD h_{\mu \nu} \cD C^\mu \cD \Cb_\mu \\
& \times \exp\left[ -S[h+\gb] - S_{\rm gf}[h;\gb] - S_{\rm gh}[h,C,\Cb;\gb] - \Delta_k S[h,C,\Cb;\gb] - S_{\rm source} \right] \, .
\end{split}
\ee
Here, $S[\gamma] = S[h+\gb]$ is a diffeomorphism invariant action functional which, for the time being, is assumed to be positive definite. Furthermore, $S_{\rm gf}[h;\gb]$ denotes the gauge-fixing term,
\be\label{Sgf}
S_{\rm gf}[h; \gb] = \frac{1}{2 \alpha} \int d^dx \sqrt{\gb} \, \gb^{\mu \nu} \, F_\mu[h; \gb] \,  F_{\nu}[h; \gb] \, ,
\ee
which depends on the gauge-fixing parameter $\alpha$ and implements the gauge-fixing condition $F_\mu[h, \gb] = 0$. 
For the construction of the FRGE, it is particularly convenient to choose a $F_\mu[h; \gb]$ which is linear in $h_{\mu \nu}$. Explicitly, we take
\be\label{geogauge}
F_\mu[h;\gb] = \sqrt{2} \, \kappa \, \big( \Db^\nu \, h_{\mu \nu} - \frac{1 + \rho}{d} \, \Db_\mu \, h^\nu{}_{\nu} \big) \, ,
\ee
with $\kappa = (32 \pi G)^{-1/2}$. Here and in the following, barred and unbarred quantities are constructed with respect to the background metric and the full quantum metric, respectively. In particular, $\Db_\mu$ denotes the covariant derivative based on $\gb_{\mu \nu}$, while $D_\mu$ is covariant with respect to $\gamma_{\mu \nu}$. The constant $\rho$  parameterizes the freedom of implementing different gauge choices, with the harmonic or de Donder gauge employed in \cite{oliver1} being obtained by setting $\rho = \tfrac{d}{2} - 1$. In the following, it will however turn out that it is more convenient to set $\rho = 0$, as this results in drastic simplifications. The ghost action arising from the 
gauge fixing \eqref{geogauge} is constructed in the standard way and reads
\be\label{Sgh}
S_{\rm gh}[h, C, \Cb; \gb] = - \sqrt{2} \, \int \, d^dx \sqrt{\gb} \, \Cb_\mu  \, \cM^{\mu}{}_\nu \, C^\nu \, , 
\ee
with the Faddeev-Popov operator
\be
\cM^{\mu}{}_{\nu} = \Db^\rho \, \gamma^\mu{}_\nu \, D_\rho + \Db^\rho \, \gamma_{\rho \nu} D^\mu - 2 \, \frac{1 + \rho}{d} \, \Db_\mu \, \gb^{\rho \sigma} \, \gamma_{\rho \nu} D_\sigma \, .
\ee
The remaining terms in \eqref{fctint}, $\Delta_k S[h,C,\Cb;\gb]$ and $S_{\rm Source}$, contain the scale-dependent IR cutoff and the source terms for the metric fluctuations and ghosts, respectively, and will be discussed in Subsection \ref{sect:2.3}.
%
\subsection{The transverse traceless decomposition}
\label{sect:2.2}
%
One of the main technical difficulties in practical applications of the FRGE for gauge theories and
gravity is the construction of the IR cutoff $\Delta_k S[h,C,\Cb;\gb]$ and the inversion of 
 $\Gamma^{(2)}_k + \cR_k$, which requires the (partial) diagonalization of these 
operators. With the TT-cutoff, this is facilitated by performing the transverse-traceless (TT) decomposition  \cite{York} of the metric fluctuations and the ghost fields.\footnote{This decomposition is optional
when considering the Einstein-Hilbert truncation with $\alpha = 1$, but mandatory when higher powers of the curvature scalar are included in the truncation ansatz.} Here, the metric fluctuations $h_{\mu \nu}$ are decomposed into a trace part $h^{\rm tr}_{\mu \nu} = \tfrac{1}{d} \gb_{\mu \nu} \phi$, encoded in the scalar $\phi$, a transverse vector $\xi_\mu$, the longitudinal part of which is given by $\sigma$, and the transverse traceless tensor $h_{\mu \nu}^{\rm T}$ according to 
\be\label{TTmetric}
h_{\mu \nu} = h_{\mu \nu}^{\rm T} + \Db_\mu \xi_\nu + \Db_\nu \xi_\mu + \Db_\mu \Db_\nu \sigma - \frac{1}{d} \gb_{\mu \nu} \Db^2 \sigma
+ \frac{1}{d} \gb_{\mu \nu} \phi \, , 
\ee
with the component fields subject to the constraints
\be
\gb^{\mu \nu} h_{\mu \nu}^{\rm T} = 0 \, , \quad \Db^\mu h_{\mu \nu}^{\rm T} = 0 \, , \quad \Db^\mu \xi_\mu = 0 \, , \quad \phi = \gb_{\mu \nu} h^{\mu \nu} \, . 
\ee  
Similarly, the ghost fields $C^\mu, \Cb_\mu$ are decomposed into their transverse and longitudinal parts $C^{{\rm T} \mu}, \Cb^{\rm T}_{\mu}$ and $\eta, \bar{\eta}$ as
\be\label{TTghost}
\Cb_{\mu} = \Cb^{\rm T}_{\mu} + \Db_{\mu} \eb \, , \quad C_{\mu} = C^{\rm T}_{\mu} + \Db_{\mu} \eta \, , \qquad \Db^\mu \, \Cb_\mu^{\rm T} = 0 \, , \quad \Db^\mu \, C_{\mu}^{\rm T} = 0 \, .
\ee

From eqs.\ \eqref{TTmetric} and \eqref{TTghost}, it is obvious that not all modes of the component fields contribute to the metric and ghost fields. In the metric decomposition, the constant mode of the $\sigma$-vectors, $\cC_\mu = \Db_\mu \sigma$ satisfying the conformal Killing equation
\be
\Db_\mu \cC_\nu + \Db_\nu \cC_\mu - \frac{2}{d} \gb_{\mu \nu} \Db_\mu \cC^\mu = 0 \, ,
\ee
and transversal vectors solving the Killing equation
\be
\Db_\mu \xi_\nu + \Db_\nu \xi_\mu  = 0 \, 
\ee
do not contribute to $h_{\mu \nu}$. Analogously, the ghost fields $\Cb_{\mu}, C^\mu$ are independent of the constant modes of $\eta, \eb$. These modes are unphysical and must be excluded from the spectrum. In fact, this is also a necessary requirement for the invertibility of the Jacobians arising from the coordinate transformations \eqref{TTmetric} and \eqref{TTghost}, as we will see below. 

A virtue of the TT-decomposition is that the component fields (almost) provide an orthogonal basis for the quantum fluctuations. Evaluating the scalar product in the ghost sector gives
\be\label{ipghost}
\langle \Cb , C \rangle := \int d^dx \sqrt{\gb} \, \Cb_\mu \, C^\mu 
= \int d^dx \sqrt{\gb} \, \big\{ \Cb_{\mu}^{\rm T} \, C^{\mu {\rm T}} - \eb \, \Db^2 \, \eta \} \, , 
\ee
establishing that this decomposition is orthogonal. In the metric sector, on the other hand, the inner product becomes
\be\label{ipmet}
\begin{split}
\langle h^{(1)} , h^{(2)} \rangle := & \int d^dx \sqrt{\gb} \, h_{\mu \nu}^{(1)} \, \gb^{\mu \rho} \, \gb^{\nu \sigma} \, h_{\rho \sigma}^{(2)} \\
= & \int d^dx \sqrt{\gb} \Big[ 
h_{\mu \nu}^{(1){\rm T}} \, h^{(2) {\rm T} \mu \nu } 
- 2 \, \xi_\mu^{(1)}  \, \big[ \gb^{\mu \nu} \Db^2 + \Rb^{\mu \nu}  \big] \, \xi_\nu^{(2)}
-2 \xi_\mu^{(1)} \Rb^{\mu\nu} \Db_\nu \sigma^{(2)} \\
& -2 \xi_\mu^{(2)} \Rb^{\mu\nu} \Db_\nu \sigma^{(1)}
+ \sigma^{(1)}  \big[ \frac{d-1}{d} (\Db^2)^2 + \Db_\mu \Rb^{\mu \nu} \Db_\nu \big] \sigma^{(2)}
+ \frac{1}{d} \phi^{(1)} \phi^{(2)} \Big]\,,
\end{split}
\ee
which is orthogonal up to the $\xi_\mu$-$\sigma$-mixed terms. We note that these crossterms vanish, however, for the Einstein backgrounds (satisfying $\Rb_{\mu \nu} = C \gb_{\mu \nu}$) considered in the next section, 
$\int d^dx \sqrt{\gb} \, \xi_\mu \Rb^{\mu\nu} \Db_\nu \sigma = - C \int d^dx \sqrt{\gb} \, \sigma \Db^\mu \xi_\mu = 0$.
Thus, in that case, the component fields do provide an orthogonal basis for the quantum fluctuations. 

The inner products \eqref{ipmet} and \eqref{ipghost} can be used to compute the Jacobians of the coordinate transformations \eqref{TTmetric} and \eqref{TTghost}. For this purpose, we follow \cite{motto1} and consider the Gaussian integral over $h_{\mu \nu}$ and $\Cb_\mu, C^\mu$. In the metric sector, this yields
\be\label{Jac1}
\begin{split}
& \int \! \cD h_{\mu\nu} \,  \exp\left[ - \half \langle h,h\rangle \right] = \\ & \;\; J_{\rm grav} \! \int \! \cD h^{\rm T}_{\mu \nu} \cD \xi_\mu \cD \sigma \cD \phi \,  
\exp\Big[ - \half \int d^dx \sqrt{\gb}\left\{ h^{\rm T}_{\mu\nu} h^{{\rm T} \mu \nu} + \tfrac{1}{d} \phi^2 + [\xi_\mu , \sigma] \, M^{(\mu, \nu)} \, [\xi_\nu , \sigma]^{\rm T} \right\}
\Big] \, ,
\end{split}
\ee
with $M^{(\mu, \nu)}$ a $(d+1) \times (d+1)$-matrix differential operator whose first $d$ columns act on transverse spin one fields $\xi_\mu$ and whose last column acts on the spin zero fields $\sigma$. The corresponding matrix can be read off from \eqref{ipmet},
\be
M^{(\mu, \nu)} = \left[ 
\begin{array}{cc}
- 2 \, \big[ \gb^{\mu \nu} \Db^2 + \Rb^{\mu \nu}  \big] & -2 \Rb^{\mu\lambda} \Db_\lambda \\
2 \Db_\lambda \Rb^{\lambda \nu} & \frac{d-1}{d} (\Db^2)^2 + \Db_\mu \Rb^{\mu \nu} \Db_\nu \\
\end{array}
\right] \, .
\ee
The Jacobian $J_{\rm grav}$ is found by performing the Gaussian integrals in \eqref{Jac1} and, up to an infinite normalization constant, which can be absorbed into the normalization of the measure, reads
\be\label{Jacgrav}
J_{\rm grav} = \Big( \det{}_{\rm (1T,0)}^\prime\left[ M^{(\mu, \nu)} \right] \Big)^{1/2} \,.
\ee
Here, the subscript ${\rm (1T,0)}$ refers to the matrix structure explained above and the prime indicates that the unphysical modes are left out. The Jacobian introduced by the change of coordinates in the ghost sector \eqref{TTghost} can be determined analogously. Setting
\be
\begin{split}
\int \! \cD C^\mu \, \cD \Cb_\mu & \exp[- \langle \Cb, C \rangle ] = \\
& J_{\rm gh} \int \! \cD C^{{\rm T}\mu} \, \cD \Cb^{\rm T}_{\mu} \, \cD \eta \, \cD \eb 
\exp\left[- \int d^dx \sqrt{\gb} \left\{ \Cb_{\mu}^{\rm T} \, C^{\mu {\rm T}} - \eb \, \Db^2 \, \eta \right\} \right] \,  
\end{split}
\ee
and performing the Grassmann functional integrals results in
\be\label{Jacghost}
\begin{split}
J_{\rm gh} = & \, \left( 
{\det}^\prime_0[-\Db^2]
\right)^{-1}\,.
\end{split}
\ee

In order to include the contribution from the Jacobians in the partition function \eqref{fctint}, we essentially copy
the Fadeev-Popov trick and introduce a set of auxiliary fields to exponentiate the determinants.
This leads to an additional term $S_{\rm aux}[\mbox{aux.\ fields}; \gb]$ in $\bZ_k$.
Let us first consider $J_{\rm grav}$ arising from the metric sector. Introducing the transverse ghost $\cb^{\rm T}_\mu, c^{{\rm T}\mu}$, a ``longitudinal'' Grassmann scalar $\bb, b$, a transverse vector field $\zeta^{\rm T}_\mu$ and a real auxiliary scalar $\omega$, we can use standard results on Gaussian integration to write \eqref{Jacgrav} as
\be\label{Jacmet}
\begin{split}
J_{\rm grav} & =  
\! \int \! \cD c_\mu^{\rm T} \cD \cb^{{\rm T} \mu} \, \cD b \, \cD \bar{b} \, 
\cD \zeta^{\rm T}_\mu \, \cD \omega \, \\ & \hspace{1cm} \times
\exp\left[ - \int d^dx \sqrt{\gb} 
\left\{ [\zeta^{{\rm T}}_\mu, \omega] \big[ M^{(\mu ,\nu)} \big]^\prime  [\zeta^{{\rm T}}_\nu, \omega]^{\rm T} +
 [\cb^{{\rm T} }_\mu, \bar{b}] \big[ M^{(\mu ,\nu)} \big]^\prime  [c^{\rm T}_\nu, b]^{\rm T} 
\right\} \right] \, .
\end{split}
\ee
The Jacobian from the ghost sector \eqref{Jacghost} is exponentiated by introducing a new complex scalar field $s, \bar{s}$,
\be\label{Jacgh}
J_{\rm gh} = \int \! \cD  \bar{s} \, \cD s \, \exp\left[ -  \int \, d^dx \, \sqrt{\gb} \,
 \bar{s} \, [-\Db^2]^\prime \, s \right] \, .
\ee
By combining the results \eqref{Jacmet} and \eqref{Jacgh}, we then find
\be\label{Saux}
\begin{split}
S_{\rm aux}[\zeta^{\rm T}, \, & \,\bar{s}, s,  \omega, \cb^{{\rm T}}, c^{\rm T}, \bar{b}, b; \gb] = \\  
& \int \, d^dx \, \sqrt{\gb} \, \Big\{
[\zeta^{{\rm T}}_\mu, \omega] \big[ M^{(\mu ,\nu)} \big]^\prime  [\zeta^{{\rm T}}_\nu, \omega]^{\rm T} +
 [\cb^{{\rm T} }_\mu, \bar{b}] \big[ M^{(\mu ,\nu)} \big]^\prime  [c^{\rm T}_\nu, b]^{\rm T} 
 + \bar{s} \, [-\Db^2]^\prime \, s   \Big\} \,  .
\end{split}
\ee
Here, the primes are placed as a reminder that the unphysical modes have been excluded from the auxiliary fields.
%
\subsection{The FRGE in terms of component fields}
\label{sect:2.3}
%
After discussing the new ingredients for the improved FRGE with TT-cutoff, we are now
in the position to construct the corresponding flow equation.
To lighten our notation, let us first introduce the index sets
\be\label{qind}
I_1 = \{ h^{\rm T}, \xi, \sigma, \phi \} \; , \quad I_2 = \{ \Cb^{\rm T}, C, \eb, \eta \} \; , \quad 
I_3 = \{ \zeta^{\rm T}, \omega, \bar{s}, s, \cb^{{\rm T}}, c^{\rm T}, \bar{b}, b \} \, , 
\ee
and a shorthand notation for the quantum fields,
\be\label{Qfields}
\chi = \Big\{h^{\rm T}, \xi, \sigma, \phi, \zeta^{\rm T}, \omega, \bar{s}, s, C, \eta, c^{\rm T}, b, \Cb^{\rm T}, \eb, \cb^{{\rm T}}, \bar{b} \Big\} \, .
\ee

The two remaining ingredients in the scale-dependent partition function \eqref{fctint} which have not yet been specified are the IR cutoff $\Delta_k S$ and the source terms. The IR cutoff is constructed in such a way that, for all fields (including the auxiliaries), the integration over the $p^2 = - \Db^2$-eigenmodes with large eigenvalues $p^2 \gg k^2$ are unaffected while the contributions of the modes with small eigenvalues $p^2 \ll k^2$ are suppressed. To this end, $\Delta_kS$ provides a momentum-dependent mass term which is quadratic in the fields \eqref{Qfields}
\be\label{IRcutoff}
\Delta_kS = 
\half \!\! \sum_{\zeta_1,\zeta_2 \in I_1} \langle \zeta_1, (\cR_k^{\rm grav})_{\zeta_1 \zeta_2} \zeta_2 \rangle
+ \half \!\! \sum_{\psi_1,\psi_2 \in I_2} \langle \psi_1, (\cR_k^{\rm gh})_{\psi_1 \psi_2} \psi_2 \rangle
+ \half \!\! \sum_{\varsigma_1,\varsigma_2 \in I_3} \langle \varsigma_1, (\cR_k^{\rm aux})_{\varsigma_1 \varsigma_2} \varsigma_2 \rangle \, .
\ee
The operators $\cR_k^{\rm grav}, \cR_k^{\rm gh}$, and $\cR_k^{\rm aux}$ depend on the
background metric only. In order to implement the desired suppression behavior, they must vanish for
 $p^2/k^2 \rightarrow \infty$ (and in particular for $k^2 \rightarrow 0$) while, for
 $p^2/k^2 \rightarrow 0$, they must behave as $\cR_k \rightarrow \cZ_k k^2$, with $\cZ_k$
 a possibly matrix valued k-dependent function. In addition, hermiticity requires that
 $(\cR_k)_{\zeta_2\zeta_1} = (\cR_k)^\dagger_{\zeta_1\zeta_2}$ and
 $(\cR_k)_{\psi_2\psi_1} = - (\cR_k)^\dagger_{\psi_1\psi_2}$ for bosonic and Grassmann
valued fields, respectively. Furthermore, $(\cR_k)_{\psi_1\psi_2} = 0$ if both $\psi_1,\psi_2 \in \{ C^{\rm T}, \eta, c^{\rm T}, b\}$ or $\{\Cb^{\rm T}, \bar{\eta}, \bar{c}^{\rm T}, \bar{b} \}$. In the sequel we will denote the three terms in \eqref{IRcutoff} by $\Delta_kS_{\rm grav}, \Delta_kS_{\rm gh}$ and $\Delta_kS_{\rm aux}$, respectively.

Finally, we specify $S_{\rm source}$, which introduces source terms for all component fields, again including the auxiliaries. Denoting the sources for the bosonic field and ghost fields by $J$ and $K$, we set
\be\label{eq:2.27}
\begin{split}
S_{\rm source} = & \, 
- \sum_{\zeta \in I_1} \langle J_\zeta, \zeta \rangle
 - \langle J_{\zeta^{\rm T}}, \zeta^{\rm T} \rangle
- \langle J_{\omega}, \omega \rangle 
- \langle \bar{J}_{\bar{s}}, \bar{s} \rangle
- \langle J_{s}, s \rangle 
\\ & \,
- \sum_{\psi \in \{C^{\rm T}, \eta, c^{\rm T}, b \}} \!\!\! \langle \bar{K}_\psi, \psi \rangle
- \sum_{\psi \in \{\Cb^{\rm T}, \eb, \cb^{\rm T}, \bb\}}  \!\!\!\langle {K}_\psi, \psi \rangle \, .
\end{split}
\ee
We hereafter denote the sources collectively by 
\be\label{Jsource}
\cJ = \Big\{ J_{h^{\rm T}}, J_\xi, J_\sigma, J_\phi,J_{\zeta^{\rm T}},J_{\omega},\bar{J}_{\bar{s}}, J_{s}, \Kb_{C^{\rm T}}, \Kb_{\eta},\Kb_{c^{\rm T}}, \Kb_{b}, K_{\Cb^{\rm T}}, K_{\eb}, K_{\cb^{\rm T}}, K_{\bb} \Big\} \, ,
\ee
which allows us to write \eqref{eq:2.27} schematically as 
\be\label{Ssource}
S_{\rm source}[\cJ;\gb] = - \langle \cJ, \chi \rangle \, .
\ee
The relations between the component sources and the standard sources for the metric fluctuations $h_{\mu \nu}, \Cb_\mu, C^\mu$ have been worked out in detail in \cite{oliver1}, where it was shown that the standard sources can be reconstructed from \eqref{Jsource}. 

With these results at hand, we can now write the scale-dependent partition function \eqref{fctint} in 
terms of the component fields,
\be\label{fctint2}
\begin{split}
\bZ_k[\cJ; \gb] = & \int 
\cD h^{\rm T} \, \cD\xi \, \cD \sigma \, \cD \phi \, \cD \zeta^{\rm T} \, \cD \omega \, \cD s \cD \bar{s} \, 
\cD C^{{\rm T}\mu}  \cD \Cb^{\rm T}_\mu \, \cD \eta \cD \eb \, 
\cD c^{{\rm T} \mu} \cD  \cb^{{\rm T}}_\mu \, \cD b \cD \bar{b} \\
& \qquad \quad \times \exp\left[ -S[h+\gb] - S_{\rm gf}[h;\gb] - S_{\rm gh} - \Delta_k S -S_{\rm aux} - S_{\rm source} \right] \, .
\end{split}
\ee
Note that the $k$-dependence on the RHS of this equation arises solely through the IR cutoff $\Delta_k S$.
The derivation of the improved FRGE then proceeds completely analogously to \cite{oliver1}. 
Based on \eqref{fctint2}, the scale-dependent generating functional for the connected Green's functions is given by
\be\label{genfct}
W_k[\cJ;\gb] = \ln \bZ_{k}[\cJ;\gb] \, .
\ee
From this, the classical component fields are obtained as 
\be
\varphi_i := \langle \chi_i \rangle = \frac{1}{\sqrt{\gb}} \, \frac{\delta W_k}{\delta \cJ^i} \, ,
\ee
where we denote the classical counterparts of the quantum fields \eqref{Qfields} collectively by
\be
\varphi = \Big\{ 
\hb^{\rm T}, \bar{\xi}, \bar{\sigma}, \bar{\phi}, \bar{\zeta}^{\rm T}, \bar{\omega}, \bar{t}, t, v^{\rm T}, \varrho, u^{\rm T}, \tau, \bar{v}^{\rm T}, \bar{\varrho}, \bar{u}^{{\rm T}}, \bar{\tau}
\Big\} \,.
\ee
The classical counterparts of the fundamental fields $h_{\mu \nu}$, $\Cb^\mu$ and $C_\mu$ can be reconstructed by substituting the corresponding component fields into the formulas for the TT-decomposition. For later purposes, we also define the classical analogue of the quantum metric \eqref{Qmetric}
\be\label{metclass}
g_{\mu \nu} := \langle \gamma_{\mu \nu} \rangle = \gb_{\mu \nu} + \hb_{\mu \nu} \, .
\ee
Subsequently, the scale-dependent effective action is constructed as the Legendre-transform on $W_k$ with respect to the component sources $\cJ$, taking the latter as functions of the classical fields\footnote{As was shown in \cite{oliver1}, taking the Legendre transform with respect to the sources for the (physical) component fields is equivalent to taking the Legendre transform with respect to the sources for the fundamental fields.}
\be\label{legtrav}
\tilde{\Gamma}_k\left[ \varphi ; \gb \right] = \langle \cJ , \varphi \rangle - W_k \left[\cJ;\gb\right] \, .
\ee
The effective average action $\Gamma_k$ is then defined as the difference between $\tilde{\Gamma}_k$ and the cutoff action with the classical fields inserted,
\be\label{effavact}
\Gamma_k[\varphi;\gb] :=  \tilde{\Gamma}_k[\varphi;\gb] - \Delta_kS[\varphi;\gb] \, .
\ee

The derivation of the FRGE governing the $k$-dependence of $\Gamma_k[\varphi; \gb]$ proceeds in several steps. First, we differentiate \eqref{genfct} with respect to $t = \ln(k)$
\be\label{Wkder}
\begin{split}
-\p_t W_k = & \,
\half {\rm Tr}^\prime \left[ \sum_{\zeta_1,\zeta_2 \in I_1} \langle \zeta_1 \otimes \zeta_2 \rangle \p_t(\cR_k)_{\zeta_1 \zeta_2}\right]
+ \half {\rm Tr}^\prime \left[ \sum_{\psi_1,\psi_2 \in I_2} \langle \psi_1 \otimes \psi_2 \rangle \p_t(\cR_k)_{\psi_1 \psi_2}\right] \\
& \, + \half {\rm Tr}^\prime \left[ \sum_{\chi_1,\chi_2 \in I_3} \langle \chi_1 \otimes \chi_2 \rangle \p_t(\cR_k)_{\chi_1 \chi_2}\right] \, ,
\end{split}
\ee
where we expressed the RHS in a matrix notation. We then introduce the Hessian of the effective action,
\be
\left( \tilde{\Gamma}^{(2)}_k\right)^{ij}(x,y) := (-1)^{[j]} \frac{1}{\sqrt{\gb(x) \gb(y)}} \, \frac{\delta^2 \tilde{\Gamma}_k}{\delta \varphi_i(x) \delta \varphi_j(y)} \, ,
\ee
with $[j] = 0,1$ for commuting fields $\varphi_j$ and Grassmann fields $\varphi_j$, respectively. This Hessian is the inverse of the connected two-point function
\be
\left( G_k\right)_{ij}(x,y) := \langle \chi_i \chi_j \rangle - \varphi_i(x) \varphi_j(y) = \frac{1}{\sqrt{\gb(x) \gb(y)}} \frac{\delta^2 W_k}{\delta \cJ^i(x) \delta \cJ^j(y)}\,,
\ee
in the sense that 
\be
\int \! d^dy \sqrt{\gb(y)} \left( G_k\right)_{ij}(x,y) \left( \tilde{\Gamma}^{(2)}_k\right)^{jl}(y,z)
= \delta_i^l \frac{\delta(x-z)}{\sqrt{\gb(z)}} \, .
\ee
Using these relations, we can express the expectation values $\langle \chi_i(x) \chi_j(y) \rangle$ in eq.\ \eqref{Wkder} through $\left( \tilde{\Gamma}^{(2)}_k\right)_{ij}^{-1}(y,z) + \phi_i(x) \phi_j(y)$. Finally, performing the Legendre transform \eqref{legtrav} and subtracting $\Delta_kS[\varphi;\gb]$ yields the desired FRGE for $\Gamma_k[\varphi;\gb]$
\be\label{Floweq}
\begin{split}
\p_t \Gamma_k[\varphi;\gb] = & \, 
\half {\rm Tr}^\prime \Big[ \sum_{\zeta_1,\zeta_2 \in \Ib_1} \left( \Gamma^{(2)}_k + \cR_k \right)^{-1}_{\zeta_1 \zeta_2} \p_t(\cR_k)_{\zeta_1 \zeta_2}\Big] \\ &
+ \half {\rm Tr}^\prime \Big[ \sum_{\psi_1,\psi_2 \in \Ib_2} \left( \Gamma^{(2)}_k + \cR_k \right)^{-1}_{\psi_1 \psi_2} \p_t(\cR_k)_{\psi_1 \psi_2}\Big] \\
& \, + \half {\rm Tr}^\prime \Big[ \sum_{\chi_1,\chi_2 \in \Ib_3} \left( \Gamma^{(2)}_k + \cR_k \right)^{-1}_{\chi_1 \chi_2} \p_t(\cR_k)_{\chi_1 \chi_2}\Big] \, .
\end{split}
\ee
Here, we again used a matrix notation on the RHS, and the barred index sets run over the classical fields,
\be\label{clind}
\Ib_1 = \{ \bar{h}^{\rm T}, \bar{\xi}, \bar{\sigma}, \bar{\phi} \} \; , \quad 
\Ib_2 = \{ \vb^{\rm T}, v^{\rm T}, \bar{\varrho}, \varrho \} \; , \quad 
\Ib_3 = \{ \bar{\zeta}^{\rm T}, \bar{\omega}, \bar{t}, t, \bar{u}^{{\rm T}}, u^{\rm T}, \bar{\tau}, \tau \} \, . 
\ee
The first, second and third term in \eqref{Floweq} thus encode the contributions from the gravitational, ghost and auxiliary fields, respectively. 

Eq.\ \eqref{Floweq} is the desired exact FRGE for the effective average action employing the improved TT-cutoff. Compared to \cite{oliver1}, the new feature is the appearance of the third trace term, capturing the contribution of the auxiliary fields, which replaces the momentum dependent field redefinitions in the metric and ghost sector. As it will turn out in the next section, this form of
the FRGE is very convenient when studying truncations involving a general function $f_k(R)$.
%
\section{Constructing the flow equation of $f(R)$-gravity}
\label{sect:3}
%
After deriving the exact FRGE \eqref{Floweq} in the last section, we now proceed by projecting the resulting RG flow on
truncation subspaces spanned by arbitrary functions of the curvature scalar. In view of recent applications 
of RG methods to gravity theories including extra dimensions, we do not fix the space-time dimension $d$ and keep the shape of the IR cutoff generic.
%
\subsection{The truncation ansatz}
%
Let us start by specifying our truncation ansatz for $\Gamma_k[\varphi; \gb]$, which
we take to be of the form
\be\label{trunc}
\Gamma_k[\varphi; \gb] = \bar{\Gamma}_k[g] + S_{\rm gf}[g-\gb;\gb] + S_{\rm gh}[g-\gb, \Cb^{\rm T},C^{\rm T}, \bar{\eta}, \eta;\gb] + S_{\rm aux} \, ,
\ee
with
\be\label{Ansatz}
\bar{\Gamma}_k[g] = \frac{1}{16 \pi G_k} \! \int \! d^dx \sqrt{g} \, f_k(R) \, .
\ee
This ansatz captures the RG flow of $f(R)$-gravity by promoting the classical function $f(R)$ to be RG scale-dependent.
The RG effects in the gauge-fixing, ghost and auxiliary sector 
are neglected by treating $S_{\rm gf}, S_{\rm gh}$ and $S_{\rm aux}$ as classical,
 i.e., not containing any scale-dependent couplings.
A similar form of this truncation ansatz (without resorting to the auxiliary field formulation) has already
been considered in the pioneering paper \cite{MR}, where it was shown that neglecting the RG running in
$S_{\rm gf}$ and $S_{\rm gh}$ is, to leading order, compatible with the modified Ward identities of the theory.
Moreover, even when treating $S_{\rm gf}$ as classical, the RG effects in this sector can be included
by taking $\alpha = 0$, which corresponds to a fixed point of the RG flow \cite{Lit2}.

When substituting the truncation ansatz \eqref{trunc} into the FRGE \eqref{Floweq} we find that
 the gravitational trace only receives contributions from 
\be\label{Ggrav}
\Gamma^{\rm grav}_k[g;\gb] := \bar{\Gamma}_k[g] + S_{\rm gf}[g-\gb;\gb] \, , 
\ee
while the ghost and auxiliary traces are evaluated based on the classical actions $S_{\rm gh}$ and $S_{\rm aux}$,
\be\label{Floweq2}
\begin{split}
\p_t \Gamma_k[\varphi;\gb] = & \, 
\half {\rm Tr}^\prime \Big[ \sum_{\zeta_1,\zeta_2 \in \Ib_1} \left( \Gamma^{{\rm grav}(2)}_k + \cR_k \right)^{-1}_{\zeta_1 \zeta_2} \p_t(\cR_k)_{\zeta_1 \zeta_2}\Big] \\ &
+ \half {\rm Tr}^\prime \Big[ \sum_{\psi_1,\psi_2 \in \Ib_2} \left( S_{\rm gh}^{(2)} + \cR_k \right)^{-1}_{\psi_1 \psi_2} \p_t(\cR_k)_{\psi_1 \psi_2}\Big] \\
& \, + \half {\rm Tr}^\prime \Big[ \sum_{\chi_1,\chi_2 \in \Ib_3} \left( S_{\rm aux}^{(2)} + \cR_k \right)^{-1}_{\chi_1 \chi_2} \p_t(\cR_k)_{\chi_1 \chi_2}\Big] \, .
\end{split}
\ee
The index sets here are given by \eqref{clind}.

Our main task now is to evaluate the traces appearing on the RHS of \eqref{Floweq2}. 
This is done as follows. We first compute the Hessians of 
the various terms appearing in \eqref{trunc} in Subsection \ref{sect:3.2a}. 
This computation can be simplified by choosing the background metric $\gb_{\mu \nu}$
as the one-parameter family of metrics on the $d$-sphere $S^d$ (parameterized by the radius
$r$ or, equivalently, the curvature scalar of the sphere), as this suffices to distinguish
different functions of the curvature scalar. 
For these backgrounds, the unphysical modes discussed in Subsection \ref{sect:2.2} correspond to
the lowest $-\Db^2$-eigenmode ($l=1$) of the transverse vector $\xi_\mu$, the two lowest eigenmodes
($l = 0,1$) for $\sigma$, and, similarly, the lowest mode ($l=1$) of the transverse ghost field $C^{{\rm T} \mu}, \Cb^{\rm T}_{\mu}$ as well as the two lowest modes ($l=0,1$) for the scalar ghost $\eta,\bar{\eta}$.
Using Table \ref{t.2} in Appendix \ref{App:B}, one can explicitly check that these modes are annihilated by the 
operators appearing in the Jacobians \eqref{Jacgrav} and \eqref{Jacghost} restricted to $S^d$. 
In Subsection \ref{sect3.2},
these results are used to construct the IR cutoff operators $\Delta S_k$ explicitly.
 Substituting those results back into \eqref{Floweq2}, we find cancellations between several
 terms appearing on the RHS, so that the final result \eqref{traces} takes a surprisingly simple form.
%
\subsection{Computing the Hessian $\Gamma_k^{(2)}$}
\label{sect:3.2a}
%
We start by deriving the term quadratic in the metric fluctuations $h$ arising from $\Gammab_k[g=\gb+h;\gb]$, \eqref{Ansatz}. 
 For this purpose, we expand $\Gammab_k[\gb+h;\gb]$ in a Taylor series in $h$,
\be
\Gammab_k[\gb+h;\gb] = \Gammab_k[\gb;\gb] + \cO(h) + \Gammab_k^{\rm quad}[\gb+h;\gb] + \cO(h^3)\,.
\ee 
Here, $\Gammab_k^{\rm quad}[\gb+h;\gb]$ is found by taking the second variation of $\Gammab_k[\gb+h;\gb]$ with respect to $h$ and setting $g = \gb$ afterwards. Applying the chain rule, we obtain
\be
\begin{split}
\delta^2 \Gammab_k =  \frac{1}{16 \pi G_k} \int \!  d^dx 
\Big[ f_k(R)\delta^2 \sqrt{g}  + f_k^\prime(R)\left( 2 (\delta \sqrt{g}) (\delta R) + \sqrt{g}  (\delta^2 R) \right)  + \sqrt{g} f_k^{\prime \prime}(R) (\delta R)^2 \Big]  ,
\end{split}
\ee
where, here and in the following, the prime denotes a derivative with respect to $R$, i.e., $f_k^\prime(R) = \frac{\p f_k(R)}{\p R}$, etc. We then substitute in the above expression the variations of the metric and Ricci scalar, which can, e.g., be found in
\cite{Oliverdis}. Note that the result drastically simplifies once we set $g = \gb$ with a spherically symmetric background metric. Performing the TT decomposition of  
$h_{\mu \nu}$ according to \eqref{TTmetric}, 
 a lengthy but straightforward computation yields
\be\label{Gexp}
\begin{split}
\Gammab_k^{\rm quad}[h;\gb] = \frac{1}{32 \pi G_k} \, \int d^dx \sqrt{\gb} \, \cL_k \, ,
\end{split}
\ee
with
\bea\nn\label{Gquad}
\cL_k & = & \half \, h_{\mu \nu}^{\rm T} \, \Big[ \fp_k \, \Db^2 - f_k + \tfrac{2 (d-2)}{d (d-1)} \, \Rb \, \fp_k \Big] h^{\mu \nu {\rm T}} \\ \nn
&& + \tfrac{1}{d} \, \left( d \, f_k - 2 \, \Rb \, \fp_k \right) \, 
\xi_\mu \, \Big[ \Db^2 + \tfrac{1}{d} \, \Rb \Big] \, \xi^\mu \\
&& + \tfrac{1}{4 d^2} \, \phi \, \Big[ 
4 (d-1)^2 \fpp_k \Db^4 
- 2 (d-1) \left( (d-2) \fp_k - 4 \, \Rb \, \fpp_k \right) \Db^2 \\ \nn && \qquad \qquad \quad 
+ (d-2)(d\,f_k - 4 \, \Rb \, \fp_k) + 4 \, \Rb^2 \, \fpp_k \Big] \, \phi \\ \nn
&& + \tfrac{1}{2 d^2} \, \sigma \, \Big[
2 (d-1)^2 \fpp_k \Db^8 - (d-1) ((d-2) \fp_k - 4 \Rb \fpp_k) \Db^6 \\ \nn && \qquad \qquad \quad
- \left( d(d-1)f_k - \Rb(d\fp_k + 2 \Rb \fpp_k) \right) \Db^4 - \Rb (df_k -2\Rb\fp_k) \Db^2
\Big] \sigma \\ \nn
&& - \tfrac{1}{d^2} \, \phi \Big[ 
2 (d-1)^2 \fpp_k \Db^6 - (d-1) ((d-2) \fp_k - 4 \Rb \fpp_k) \Db^4 - \Rb( (d-2) \fp_k - 2 \Rb \fpp_k) \Db^2
\Big] \sigma \, .
\eea
Here the barred quantities are constructed from the background metric $\gb$ and we have suppressed the argument of $f_k(\Rb)$.

Following the procedure outlined above, we also extract the quadratic terms arising from $S_{\rm gf}, S_{\rm gh}$ and $S_{\rm aux}$. Applying the TT decomposition in the first two cases yields 
\bea\label{SgfTT}
S_{\rm gf}^{\rm quad} & = & \frac{\kappa^2}{\alpha} \int d^dx \sqrt{\gb} 
  \, \Big\{  
\xi_{\mu} \, \Big[ \Db^2 + \frac{1}{d} \, \Rb \, \Big]^2 \, \xi^\mu 
 - \frac{1}{d^2} \, \sigma \, \Big[ \left( (d-1) \Db^2 + \Rb \right)^2 \, \Db^2 \Big] \, \sigma 
\\ \nn && \qquad \qquad  \qquad \quad
- \frac{\rho^2}{d^2} \, \phi \, \Db^2 \, \phi 
+ \frac{\rho}{d^2} \, \phi \, \Big[ (d-1) \Db^4 + \Rb \, \Db^2 \, \Big] \, \sigma \Big\} \, , \\ \label{SghTT}
S_{\rm gh}^{\rm quad} & = & \, - \sqrt{2} \int \, d^dx \, \sqrt{\gb} \, \Big\{ \, \Cb^{\rm T}_\mu \, \big[ \Db^2 + \tfrac{1}{d} \, \Rb \big] \, C^{{\rm T} \mu}
- \frac{2}{d} \, \bar{\eta} \, \big[ (d-1-\rho)\,  \Db^4 + \Rb \, \Db^2 \big] \,  \eta
 \Big\} \, ,
\eea
while substituting the spherical symmetric background metric simplifies $S_{\rm aux}$ to
\be\label{Saux2}
\begin{split}
S_{\rm aux}^{\rm quad} = &   
\int \, d^dx \, \sqrt{\gb} \, \Big\{
\zeta^{{\rm T}\mu} \, \big[  - \Db^2 - \tfrac{1}{d} \Rb \big]^\prime \, \zeta_\mu^{\rm T} 
+ \cb^{{\rm T} \mu} \, \big[ - \Db^2 - \tfrac{1}{d} \Rb \big]^\prime \, c^{\rm T}_\mu \\
& \qquad \qquad \qquad + \bar{s} \, [-\Db^2]^\prime \, s 
+ \bar{b} \, \big[  \Db^4 + \tfrac{1}{d-1} \Rb \Db^2 \big]^{\prime \prime} \, b 
+ \omega \, \big[  \Db^4 + \tfrac{1}{d-1} \Rb \Db^2 \big]^{\prime \prime} \, \omega  \Big\} \,  .
\end{split}
\ee
Here, the primes on the operators indicate that the corresponding number of lowest eigenmodes are excluded from
the physical fields. Note that, for the special case $d=4$, eqs.\ \eqref{Gexp}, \eqref{SgfTT} and \eqref{SghTT} are in precise agreement with earlier findings \cite{Cognola:2005de}.  
We also remark that, while in the harmonic gauge used in \cite{oliver1} performing the TT-decomposition in the ghost sector is optional, working in the gauge \eqref{geogauge} requires this decomposition in order to diagonalize the quadratic fluctuations.
\subsection{Adapting the cutoff operators}
\label{sect3.2}
%
Our next task is to explicitly construct the IR cutoff operators \eqref{IRcutoff}.
For this, it is useful to split $\cR_k(p^2)$ into its matrix part and a scalar
function $R_k$, encoding the momentum-dependent mass term,
\be\label{eq:3.12}
\big[ \cR_k(p^2) \big]_{\varphi_1 \varphi_2} 
= \big[ \cZ_k \big]_{\varphi_1 \varphi_2} \, R_k(p^2) 
= \big[ \cZ_k \big]_{\varphi_1 \varphi_2} \, k^2 R^{(0)}(-\Db^2/k^2) \, .
\ee
The dimensionless profile functions
$R^{(0)}(-\Db^2/k^2)$ interpolate between $R^{(0)}(0) = 1$ and $R^{(0)}(\infty) = 0$
and otherwise allow one to adjust the ``shape'' of the momentum-dependent mass-squared term -
see Appendix \ref{C.2} for an explicit choice of a profile function specifying a cutoff-scheme. 
The aim of the IR cutoff is the regularization of Hessians
\be
\big[\Gamma^{(2)}_k \big]_{\varphi_1 \varphi_2} = \big[ f( - \Db^2, k , \ldots) \big]_{\varphi_1 \varphi_2} 
\ee
in such a way that all covariant derivatives appear in the form
\be\label{Pdef}
\Pb_k = - \Db^2 + R_k(p^2) \, .
\ee
Thus, as a result of the IR regularization, 
\be\label{rule}
\big[ \Gamma^{(2)}_k + \cR_k \big]_{\varphi_1 \varphi_2} = 
\big[ f( - \Db^2 + R_k, k , \ldots) \big]_{\varphi_1 \varphi_2}
\ee
depends on the covariant Laplacian $-\Db^2$ through the combination \eqref{Pdef} only.

With these prerequisites, we can now explicitly construct $\Delta_kS$ for the truncation \eqref{trunc}. In order to simplify the resulting expressions, we use the gauge-freedom retained in \eqref{Sgf} to set\footnote{We note that the limit $\alpha \rightarrow 0$ has to be implemented with care and, strictly speaking, should be taken only after constructing the flow equation to avoid singular expressions. In any case, a more careful treatment of this matter confirms the validity of the final result \eqref{traces}.}
\be\label{gaugefix}
\rho = 0, \quad \alpha = 0 \, .
\ee
As eq.\ \eqref{Cgrav} below will illustrate, this gauge choice leads to a factorization of the physical and gauge degrees of freedom contained in $h_{\mu \nu}$ in such a way that the physical degrees of freedom are parameterized by $h_{\mu \nu}^{\rm T}$ and $\phi$, while the gauge degrees of freedom are confined to $\xi^{\rm T}_\mu$ and its longitudinal component, $\sigma$. Thus, from a geometrical point of view, \eqref{gaugefix} aligns the gauge orbits with the directions $\xi^{\rm T}_\mu, \sigma$. This feature will later lead to a considerable simplification of the flow equation. In particular, we will find that the matrix structure of the operator $\Gamma^{(2)}_k + \cR_k$ diagonalizes, so that the latter can be easily inverted.

This said, let us start with constructing the IR cutoff for the auxiliary fields. Considering \eqref{Saux2}, it is straightforward to establish that the prescription \eqref{rule} is implemented by
\be\label{IRaux}
\begin{split}
\Delta_k S_{\rm aux} =  \int d^dx \sqrt{\gb} & \, \Big\{
\zeta^{{\rm T}\mu} R_k \zeta_\mu^{\rm T}
+ \cb^{{\rm T}\mu} R_k c_\mu^{\rm T}
+ \bar{s} R_k s
+ \bb \left[ \Pb_k^2 - \tfrac{1}{d-1} \Rb R_k - \Db^4 \right] b \\ & \quad
+ \omega \left[ \Pb_k^2 - \tfrac{1}{d-1} \Rb R_k - \Db^4 \right] \omega
\Big\} \, .
\end{split}
\ee
Analogously, starting from \eqref{SghTT}, we find the IR cutoff in the ghost sector,
\be\label{IRgh}
\begin{split}
\Delta_k S_{\rm gh} = \sqrt{2} \int d^dx \sqrt{\gb} & \, \Big\{
\Cb^{\rm T}_\mu R_k C^{{\rm T} \mu} + \tfrac{2(d-1)}{d} \eb 
\left[ P_k^2 - \tfrac{1}{d-1} \Rb R_k - (-\Db^2)^2 \right] \eta
\Big\} \, .
\end{split}
\ee
The construction of $\Delta_k S_{\rm grav}$ is slightly more involved. Here, we first observe that, for $\rho = 0$, eq.\ \eqref{SgfTT} diagonalizes and contains the fields $\xi_\mu$ and $\sigma$ only. Combining \eqref{Gexp} and \eqref{SgfTT} then gives 
\bea\label{Cgrav}
\Gamma^{\rm grav;quad}_k \!\!\! & = & \!\!\!  \frac{1}{32 \pi G_k} \int d^dx \sqrt{\gb} 
\, \Big\{
\half \, h_{\mu \nu}^{\rm T} \, \Big[ \fp_k \, \Db^2 - f_k + \tfrac{2 (d-2)}{d (d-1)} \, \Rb \, \fp_k \Big] h^{\mu \nu {\rm T}} \\ \nn
&& \!\!\! \qquad \qquad \qquad+ \tfrac{1}{4 d^2} \, \phi \, \Big[ 
4 (d-1)^2 \fpp_k \Db^4 
- 2 (d-1) \left( (d-2) \fp_k - 4 \, \Rb \, \fpp_k \right) \Db^2 \\ \nn
&& \!\!\! \qquad \qquad \qquad \qquad \qquad
+ (d-2)(d\,f_k - 4 \, \Rb \, \fp_k) + 4 \, \Rb^2 \, \fpp_k \Big] \, \phi 
\Big\} \\ \nn
&& \!\!\! + \frac{\kappa^2}{\alpha} \int d^dx \sqrt{\gb} 
  \, \Big\{ \, 
\xi_{\mu} \, \Big[ \Db^2 + \frac{1}{d} \, \Rb \, \Big]^2 \, \xi^\mu 
 - \frac{1}{d^2} \, \sigma \, \Big[ \left( (d-1) \Db^2 + \Rb \right)^2 \, \Db^2 \Big] \, \sigma 
+ \cO(\alpha) \Big\} \, .
\eea
Compared to the terms appearing in $S_{\rm gf}$, the $\xi\xi, \sigma\sigma$ and $\phi\sigma$-contributions from $\Gammab^{\rm quad}_k$ are suppressed by one power of $\alpha$ and are indicated by $\cO(\alpha)$ in the formula above.  
In the limit $\alpha \rightarrow 0$, these terms do not contribute to the flow equation and will therefore be neglected
in the following discussion. Again following the rule \eqref{rule}, we find the 
 IR cutoff in the metric sector,
\bea\label{IRgrav}
\Delta_k S_{\rm grav}\!\!\!  & = &\!\!\!  \frac{1}{32 \pi G_k} \int d^dx \sqrt{\gb}
\Big\{ \half h_{\mu \nu}^{\rm T} \, \Big[ - \fp_k \, R_k \Big] h^{\mu \nu {\rm T}} 
\\ \nn
&& \!\!\! \qquad \qquad + \tfrac{1}{4 d^2} \, \phi \, \Big[ 
4 (d-1)^2 \fpp_k (\Pb_k^2 - \Db^4)  
+ 2 (d-1) \left( (d-2) \fp_k - 4  \Rb  \fpp_k \right) R_k 
\Big] \, \phi \Big\} \\ \nn
&& \!\!\! + \frac{\kappa^2}{\alpha} \int d^dx \sqrt{\gb} 
  \, \Big\{ \, 
\xi_{\mu} \, \Big[ \Pb_k^2 - \Db^4 - \frac{2}{d} \Rb R_k \, \Big] \, \xi^\mu \\ \nn
&& \!\!\! \qquad \qquad 
 + \frac{1}{d^2} \, \sigma \, \Big[ \left( (d-1) \Pb_k - \Rb \right)^2 \, \Pb_k -
\left( (d-1) (-\Db)^2 - \Rb \right)^2 \, (-\Db^2)
 \Big] \, \sigma 
+ \cO(\alpha) \Big\} \, .
\eea
This result completes the adaptation of the IR cutoff to the truncation \eqref{trunc}. From the results \eqref{IRaux}, \eqref{IRgh} and \eqref{IRgrav}, it is then straightforward to read off the entries of the cutoff operator $\cR_k$ and construct  $\Gamma^{(2)}_k + \cR_k$, which indeed assume a diagonal form in field space. For convenience, their entries are summarized in Table \ref{t.1}.
\begin{table}[t]
\begin{center}
\begin{tabular}{l|l|l}
Index & Cutoff operator $\cR_k$ & Matrix element of $\Gamma^{(2)}_k + \cR_k$ \\ \hline
$h^{\rm T} h^{\rm T}$ 
& $- \, \frac{1}{32 \pi G_k} f_k^\prime \, R_k $
& $\frac{1}{32 \pi G_k} \Big[ - f_k^\prime \, \Pb_k - f_k + \frac{2(d-2)}{d(d-1)} \, \Rb \, f_k^\prime\Big] \bigg. $ \\[1.1ex]
$\phi \phi$
& $\frac{1}{64 \pi G_k d^2} \tilde{\cR}^{\phi\phi}_k$
& $\frac{1}{64 \pi G_k d^2} \widetilde{\Gamma}^{(2) \phi \phi}_k$ \\[1.1ex]
$\xi \xi$
& $\frac{1}{16 \pi G_0 \alpha} \Big[ \Pb_k^2 - \Db^4 - \frac{2}{d} \Rb R_k \Big] $
& $\frac{1}{16 \pi G_0 \alpha} \Big[ \Pb_k - \frac{1}{d} \Rb \Big]^2 $ \\[1.1ex]
$\sigma \sigma$ 
& $\frac{(d-1)^2}{16 \pi G_0 \alpha d^2} \Big[ \big( \Pb_k - \tfrac{1}{d-1} \Rb \big)^2 \Pb_k +
\big( \Db^2 + \tfrac{1}{d-1} \Rb \big)^2 \Db^2 \Big] $
& $\frac{(d-1)^2}{16 \pi G_0 \alpha d^2} \Big[ \big( \Pb_k - \tfrac{1}{d-1} \Rb \big)^2 \Pb_k \Big] $ \\[1.1ex]
$\Cb^{\rm T} C^{\rm T}$
& $\sqrt{2} R_k$
& $ \sqrt{2} \Big[ P_k - \tfrac{1}{d} \, \Rb \Big] $ \\[1.1ex] 
$\eb \eta$
& $ 2 \sqrt{2} \, \frac{d-1}{d} \, \Big[ \Pb_k^2 - \Db^4 - \frac{1}{d-1} \Rb R_k \Big]$
& $ 2 \sqrt{2} \, \frac{d-1}{d} \, \Big[ \Pb_k - \tfrac{1}{d-1} \Rb \Big] \, \Pb_k $ \\[1.1ex]
$\zeta \zeta$
& $2 R_k$
& $2 \Big[ \Pb_k - \frac{1}{d} \Rb \Big] $ \\[1.1ex]
$\cb^{\rm T} c^{\rm T}$
& $ R_k$
& $\Big[ \Pb_k - \frac{1}{d} \Rb \Big] $ \\[1.1ex]
$\bar{s} s$
& $ R_k$
& $ \Pb_k  $ \\[1.1ex]
$\bar{b} b$
& $ \Big[ \Pb_k^2 - \Db^4 - \frac{1}{d-1} \Rb R_k \Big]$
& $ \Big[ \Pb_k - \frac{1}{d-1} \Rb \Big] \, \Pb_k   $ \\[1.1ex]
$\omega \omega$
& $ 2 \, \Big[ \Pb_k^2 - \Db^4 - \frac{1}{d-1} \Rb R_k \Big]$
& $ 2 \, \Big[ \Pb_k - \frac{1}{d-1} \Rb \Big] \, \Pb_k  $ 
\end{tabular}
\end{center}
\parbox[c]{\textwidth}{\caption{\label{t.1}{Matrix entries of the operators $\cR_k$ and $\Gamma^{(2)}_k + \cR_k$ to leading order in $\alpha$. The first column indicates the indices of the matrix element in field space, while the second and third columns contain the corresponding matrix element of $\cR_k$ and $\Gamma^{(2)}_k + \cR_k$, respectively. The elements are symmetric under the change of bosonic indices, while they acquire a minus sign when Grassmann-valued indices are swapped. $\tilde{\cR}^{\phi\phi}_k$ and $\widetilde{\Gamma}^{(2) \phi \phi}_k$ are defined in eqs.\ \eqref{Rpp} and \eqref{Gpp}.}}}
\end{table}

With these results, we can now evaluate the traces in \eqref{Floweq2} for the truncation \eqref{trunc}.
Substituting the operators $\cR_k$ and $(\Gamma^{(2)}_k + \cR_k)^{-1}$ listed in Table \ref{t.1} and properly taking into account the non-physical modes, we find\footnote{A similar equation also underlies the analysis of the UV properties of $f(R)$-gravity \cite{CPR}. The corresponding equation differs from \eqref{traces} by the contribution of the single eigenmode proportional to $D_1(d,0)$. We thank R.\ Percacci and C.\ Rahmede for discussion on this point.}
\be\label{traces}
\begin{split}
\p_t \Gammab_k = & \, - \half {\rm Tr}_0^{\prime \prime} \left[ \frac{\p_t R_k}{\Pb_k - \frac{1}{d-1} \Rb} \right] 
+ D_1(d,0) \left. \frac{1}{\Pb_k} \, \p_t R_k \right|_{-\Db^2 = \Lambda_1(d,0)} 
- \half {\rm Tr}_{\rm 1T}^{\prime} \left[ \frac{\p_t R_{k} }{ \Pb_k - \frac{1}{d} \Rb }  \right] \\
& \,  + \half  {\rm Tr}_{\rm 2T}\left[ \frac{\p_t (Z_{Nk} f^{\prime}_k R_k)}
{Z_{Nk} \left( f^\prime_k \, \Pb_k + f_k - \frac{2(d-2)}{d(d-1)} \, \Rb \, f^\prime_k \right)} \right]  
 +\half  {\rm Tr}_0 \left[ \frac{\p_t \big( Z_{Nk} \widetilde{\cR}_k^{\phi \phi}\big)}
{Z_{Nk} \widetilde{\Gamma}^{(2) \phi \phi}_k} \right] \, ,
\end{split}
\ee
with $\Gammab_k$ and $f_k(\Rb)$ introduced in eq.\ \eqref{Ansatz}. Here, $Z_{Nk}$ captures the scale-dependence of Newton's constant via $G_k := Z_{Nk}^{-1} G_{\hat{k}}$, 
with $G_{\hat{k}}$ denoting its ``bare'' value. The subscripts 0,1T,2T on the operator-traces indicate that their arguments  act on fields with spin $s=0,1,2$. Furthermore, the primes imply that the corresponding number of lowest $-\Db^2$-eigenvalues are excluded from the trace. Following the notation of Appendix \ref{App:B}, the multiplicity and $-\Db^2$-eigenvalues of a given field mode are denoted by $D_l(d,s)$ and $\Lambda_l(d,s)$. For the $d$-sphere, these are summarized in Table \ref{t.2} in Appendix \ref{App:B}. Finally, we have introduced
\be\label{Rpp}
\widetilde{\cR}_k^{\phi \phi} = 4 \, (d-1)^2 f^{\prime \prime}_k \left( \Pb_k^2 - (-\Db^2)^2 \right)
+ 2 \, (d-1) \left( (d-2) \, f^\prime_k - 4 \, \Rb \, f^{\prime \prime}_k \right) R_k \, ,
\ee
and
\be\label{Gpp}
\widetilde{\Gamma}^{(2) \phi \phi}_k = 
4 \, (d-1)^2 f^{\prime \prime}_k \, \left( \Pb_k - \tfrac{1}{d-1} \Rb \right)^2 
+ 2 \, (d-1)  (d-2) \, f^\prime_k   \left( \Pb_k - \tfrac{2}{d-1} \Rb \right)
+ d(d-2) f_k  \, ,
\ee
in order to write the last scalar trace in compact form. After carrying out the variations $\Gamma^{(2)}_k[g, \ldots;\gb]$, we can also identify $g = \gb$ in \eqref{traces}, dropping the bar on the metric in the sequel. For later reference, we  denote the five terms appearing on the RHS of \eqref{traces} by $\cS_k^i$, $i = 1,\ldots,5$, respectively.

Let us close this subsection with the following observation. An interesting feature of the RG equation \eqref{traces} is that it consists of a 'universal', $f_k(R)$-independent part, and a truncation-dependent part. The former encompasses the terms $\cS^1_k, \cS^2_k, \cS^3_k$ in the first line and captures the contribution of the scalars $\eta, \sigma, s, b, \omega$ and the auxiliary vector fields $\zeta_\mu,c_\mu$. The term $\cS_k^2$ thereby arises from combining the scalar traces with a different number of unphysical modes.
The second part, which consists of the traces $\cS^4_k$ and $\cS^5_k$, stems from the gravitational degrees of freedom $h^{\rm T}_{\mu \nu}$ and $\phi$, and encodes all information about the particular form of $f_k(R)$. Retrospectively, the appearance of this structure is not surprising: it is a direct consequence of the geometrical gauge-choice \eqref{gaugefix}, which ensures that the gravitational degrees of freedom are carried by $h^{\rm T}_{\mu \nu}$ and $\phi$ only. 

\subsection{Heat-kernel techniques for evaluating operator traces}
\label{sect3.3}
%
The final step in constructing non-perturbative $\beta$-functions from the flow equation \eqref{traces} requires the explicit evaluation of the operator traces in terms of curvature invariants of the background manifold. This can be done by the standard heat-kernel techniques which are the subject of this subsection.

In order to be able to apply heat-kernel methods to our problem, we first have to complete the traces by adding and subtracting the missing modes according to
\be\label{eq:3.24}
\Tr^{\prime \ldots \prime}_{s}[W(-\Db^2)] = \Tr_{s}[W(-\Db^2)] - \sum_{l \in \{l_1, \ldots l_m\}} D_l(d,s) W(\Lambda_l(d,s))\,,
\ee
Here, $W$ is an arbitrary, smooth, operator-valued function and $s$ denotes the spin of the fields on which the covariant Laplacian acts. The discrete sum in \eqref{eq:3.24} can be evaluated using the multiplicities $D_l(d,s)$ and eigenvalues $\Lambda_l(d,s)$ for $-\Db^2$ acting on a spin $s$ field. To evaluate the complete traces, we assume that the function $W(z)$, which arises from replacing the argument of the operator $W(-\Db^2)$ by $z$, has a Fourier transform $\Wt(s)$. The trace is then expressed by this Fourier-transform,
\be\label{Ftrans}
\Tr\left[ W(-\Db^2) \right] = \lim_{\epsilon \rightarrow 0} \int_{-\infty}^{\infty} ds \, \Wt(s) \, 
\Tr\left[ \e^{-(i s-\epsilon) \Db^2 }\right] \, ,
\ee
with $\epsilon$ an infinitesimal regulator. The operator trace on the RHS can then be evaluated via the asymptotic heat-kernel expansion,
\be\label{heatexp}
\Tr\left[ \e^{-it(\Db^2 + Q) }\right]=  \left( \frac{i}{4 \pi t} \right)^{d/2} \int d^dx \sqrt{\gb} \left[ {\tr} \, a_0 - it \, {\tr} \, a_2(x,Q) - t^2 \, {\tr} \, a_4(x,Q) + \ldots \right] \, .
\ee
Here, $Q$ denotes a ``matrix potential'' which is independent of the covariant Laplacian, the symbol ${\tr}$ denotes a trace with respect to the tensor indices of the fields on which the Laplacian acts, and the $a_{2k}(x,Q)$ are the heat-kernel coefficients of the Laplacian (see Appendix \ref{App:B} for details).

Substituting the heat-kernel expansion into \eqref{Ftrans}, we obtain
\be\label{expand2}
\Tr\left[ W(-\Db^2) \right] =
\left( 4 \pi \right)^{-d/2} \int d^dx \sqrt{\gb} \, \left\{  Q_{d/2}[W]  \, \tr \, a_0 + Q_{d/2-1}[W] \, \tr \,  a_2(x,Q) + \ldots \right\} \, ,
\ee
where the functionals $Q_n[W]$ are defined via
\be\label{FTint}
Q_n[W] := \lim_{\epsilon \rightarrow 0} \int_{-\infty}^{\infty} ds \left( -is+\epsilon \right)^{-n} \, \Wt(s) \, .
\ee
Making use of the Mellin transform, $Q_n[W]$ can be reexpressed through the original function $W(z)$,
\be\label{Qfct}
Q_n[W] = \frac{(-1)^i}{\Gamma(n+i)} \int_0^\infty dz \, z^{n+i-1} \, \frac{d^i W(z)}{dz^i} \, , \quad i > -n \, , \; i \in \Nom \, .
\ee
For $n>0$ and $n \le 0 \in \Zom$, it is convenient to choose $i=0$ and $i = -n+1$, respectively. Then, assuming that $\lim_{z \rightarrow \infty} W(z) = 0$ sufficiently fast\footnote{For the functions appearing in the traces \eqref{traces}, this holds.}, \eqref{Qfct} simplifies to
\be\label{Qfct2}
\begin{split}
Q_n[W] = & \, \frac{1}{\Gamma(n)} \int_0^\infty dz \, z^{n-1} \, W(z) \, , \quad n > 0 \, , \\
Q_{-\tilde{n}}[W] = & \, (-1)^{\tilde{n}} \left. \frac{d^{\tilde{n}} W(z)}{dz^{\tilde{n}}} \right|_{z = 0} \,  , \quad \tilde{n} = - n \in \Nom \, .
\end{split}
\ee
This includes the special case $Q_0[W] = W[0]$. The second identity will play an important role when considering the full flow equation arising from \eqref{traces} in Section \ref{Sect:7}.

For $n>0$, the functionals $Q_n[W]$ resulting from \eqref{traces} are closely related to the (generalized) threshold functions introduced in Appendix \ref{App:C}. For the traces $\cS^1_k, \cS^2_k$ and $\cS_k^4$, this relation leads to the standard threshold functions \eqref{thrfct}. Defining 
\be
\cN := \left(2 g_k \right)^{-1} \, \p_t \left( g_k \, R_k \right)\,,
\ee 
with $g_k$ an arbitrary function of the cutoff scale $k$, we have
\be\label{Qthres1}
Q_n\left[ (P_k + c)^{-p} \, \cN  \right] = k^{2(n-p+1)} \left[ \Phi^p_n(c/k^2) + \half \, \p_t \ln(g_k) \, \Pt^p_n(c/k^2) \right] \, .
\ee
Typically, $g_k$ represents a (combination of) $k$-dependent coupling constants or wavefunction renormalization factors multiplying $R_k$. These enter the RHS of eq.\ \eqref{Qthres1} through their ``anomalous dimension'' $\eta_g(k) := \p_t \ln(g_k)$. With regard to the later sections, it will be useful to introduce the anomalous dimension of Newton's constant and the non-local couplings $\ub_k $ and $\mb_k$ appearing in \eqref{logRmodel} and \eqref{Rinvmodel},
\be\label{andim}
\eta_N(k) := - \p_t \ln(Z_{Nk}) \, , \quad \eta_{\ub}(k) :=  \p_t \ln(\ub_{k}) \quad \mbox{and} \quad \eta_{\mb}(k) :=  \p_t \ln(\mb_{k}) \, . 
\ee

For the last trace in \eqref{traces}, an analogous calculation relates $Q_n$ to the generalized threshold functions \eqref{genthrfct}. Using a schematic notation which highlights the $-D^2$-dependence of the trace argument, we find
\be\label{Qgen}
\begin{split}
Q_n & \left[ \frac{\p_t \left(g_k (P_k^2 - D^4) + \gt_k R_k \right)}{(u_k P_k^2 + v_k P_k + w_k)^p}\right] \\ 
& \, \quad = k^{2(n-2p+2)} \left\{ \p_t g_k \, \tilde{\Upsilon}^p_{n,0,1}(u_k,v_k/k^2,w_k/k^4) + 4 g_k \Upsilon^p_{n,1}(u_k,v_k/k^2,w_k/k^4) \right\} 
\\ & \, \qquad
+ k^{2(n-2p+1)} \left\{ \p_t \gt_k \, \tilde{\Upsilon}^p_{n,0,0}(u_k,v_k/k^2,w_k/k^4) + 2 \gt_k \Upsilon^p_{n,0}(u_k,v_k/k^2,w_k/k^4) \right\} \, .
\end{split}
\ee
With these tools, we have now all the ingredients to derive the $\beta$-functions for particular choices of $f_k(R)$.
To conclude this section, we illustrate the techniques just introduced by expanding the ``universal terms'' in the first line of \eqref{traces} up to linear order in the curvature scalar. These results will be very useful when deriving the $\beta$-functions of the $\ln(R)$- and $R^{-n}$-truncation in sections \ref{sect:5} and \ref{sect:6}, respectively.

Let us first make a general observation on the ``finite sum terms'' in \eqref{traces}, that is, on those terms which do not form a complete operator trace. For $d>2$ and spherical backgrounds, these terms do not contribute to truncation subspaces
which contain terms linear in $R$ as the highest power of the curvature scalar. This is due to the fact that
the trace terms are always proportional to the volume of the background sphere and, therefore, inherit extra inverse powers of the curvature scalar $V \propto R^{-d/2}$. Thus, contributions coming from discrete eigenvalues are effectively suppressed by a factor $R^{d/2}$ and hence, for $d > 2$, only start to affect the running of couplings multiplying $\int d^dx \sqrt{g} R^n, n>1$. Given 
this, we can see that both $\cS^2_k$ and those terms arising from the completion
of traces in \eqref{traces} do not contribute to the truncation spaces spanned by the $\ln(R)$- and $R^{-n}$-truncation. Therefore, these can be ignored in the derivation of the corresponding $\beta$-functions.

With this observation, it is now straightforward to evaluate the contribution of the first line of
the flow equation, neglecting $\cO(R^2)$ and higher curvature terms. Explicitly,
\be\label{unitrace}
\begin{split} 
\cS^1_k = & \, -
  \frac{k^d}{(4 \pi)^{d/2}} \left\{C^{\rm S}_0\,\Phi^1_{d/2} \int \! d^dx \sqrt{g} 
+ k^{-2} \big(\tfrac{C^{\rm S}_0}{d-1} \Phi^2_{d/2} + C^{\rm S}_2 \, \Phi^1_{d/2-1} \big) \int \! d^dx \sqrt{g} R \right\} \, , \\
\cS^3_k = & \, - \frac{k^d}{(4 \pi)^{d/2}} \, \Big\{C^{\rm 1T}_0\,\Phi^1_{d/2} \, \int \! d^dx \sqrt{g} + 
k^{-2} 
\left( \tfrac{C^{\rm 1T}_0}{d} \Phi_{d/2}^2 + C^{\rm 1T}_2\, \Phi_{d/2-1}^1\right) \,
 \int \! d^dx  \sqrt{g} R \Big\} \, .
\end{split}
\ee
The $d$-dependent coefficients $C_{2k}^{\rm s}$ are defined in eq.\ \eqref{B.20}. In order to lighten our notation, we have introduced 
\be\label{Phinot}
\begin{split}
\Phi_n^p:= \Phi_n^p(0) \, , \quad
\tilde{\Phi}_n^p:= \tilde{\Phi}_n^p(0) \, , \quad
\Upsilon^p_{n;m}:= \Upsilon^p_{n;m}(1,0,0) \, , \quad
\tilde{\Upsilon}^p_{n,m,l}:= \tilde{\Upsilon}^p_{n,m,l}(1,0,0)\,.
\end{split}
\ee
Here and in the following, $\Phi^p_n$, $\Pt^p_n$, $\Upsilon^p_{n;m}$ and $\tilde{\Upsilon}^p_{n,m,l}$
without argument will denote the cutoff-scheme
 dependent threshold constants defined above, while, for the
related threshold functions, the argument will always be given explicitly.

We close this subsection with the following remark. In order to evaluate the contributions of the traces \eqref{Ftrans} to our truncation, we use the early time expansion of the heat-kernel \eqref{heatexp}, expanding for small values of $s$. Based on perturbation theory, one would, at first sight, expect that an investigation of the gravitational RG flow in the IR should be based on the late-time expansion of the heat-kernel \cite{BGVZ,Barvinsky:2004he}, expanding the traces for large values $s$. Analyzing the Fourier-transformed functions $\Wt(s)$ arising from \eqref{traces}, one however finds that, thanks to the IR cutoff, $\lim_{s \rightarrow \infty} \Wt(s) = 0$ for $k > 0$.\footnote{For the exponential cutoff function, $\Wt(s)$ even vanishes exponentially fast.} Thus, even for small values of $k$, the main contribution to the integral \eqref{FTint} arises from small values $s$, so that the early time expansion of the heat kernel can also be used to investigate the gravitational RG flow of the truncations \eqref{logRmodel} and \eqref{Rinvmodel} in the IR.   
\section{General properties of the RG flow of $f(R)$-gravity}
\label{Sect:4}
%
One of the most remarkable properties of the flow equation \eqref{traces} is the fact that it is valid for any space-time dimension $d$, cutoff function $R_k$, and arbitrary function $f_k(R)$.
This makes it worthwhile to pause and highlight some general features of the resulting RG flow,
before analyzing concrete models specifying some of the quantities above. We  
start by formulating a decoupling theorem for non-local interactions in Subsection \ref{Sect4.1}, and then move to derive the properties of $f_k(R)$ required for resolving the IR singularities as, e.g., the $\lambda = 1/2$-line observed in
the Einstein-Hilbert truncation, in Subsection \ref{Sect:4.2}. 
%
\subsection{The perturbative decoupling of non-local interactions}
\label{Sect4.1}
%
We now turn towards the discussion of the $\beta$-functions for interaction monomials which are built 
from non-local curvature terms. Here, it is useful to distinguish two classes: interactions which blow up as the
curvature scalar becomes small, as, e.g., $R^{-n}$- or $\ln(R)$-terms, and non-local interactions which remain finite as $R \rightarrow 0$, like $R \ln(R)$-terms. Based on the flow equation \eqref{traces}, 
one can argue that the $\beta$-functions for the scale-dependent couplings multiplying interactions
of the first kind are always trivial, while, in the second case, they are always proportional to a coupling constant multiplying an interaction monomial containing (some power of) $\ln(R)$, so that the corresponding interactions can be consistently decoupled from the RG flow.

Let us start by investigating the $\beta$-functions for the non-local couplings of the first kind.  
For this purpose, we will consider $\Gammab_k[g]$ as a Laurent series in $R$,
\be\label{eq:4.1}
\Gammab_k[g] = \int d^dx \sqrt{g} \, 
\sum_{n \ge - n_0} \bar{\mu}_{n;k} \, R^n  
\, , \quad n_0 > 0 \, ,
\ee
corresponding to $f_k(R) = 16 \pi G_k \sum_{n \ge - n_0} \bar{\mu}_{n;k} \, R^n $.\footnote{As illustrated in the next section, this argument also applies to $\ln(R)$-terms which are not explicitly 
included in the ansatz \eqref{eq:4.1}.} 
Substituting this ansatz into \eqref{traces} and dividing by $V$, we see that the LHS of the flow equation is a Laurent series whose highest order pole is given by $R^{-n_0}$. The coefficients in this expansion are proportional to the $t$-derivatives of the non-local couplings.

Now, in order to extract the $\beta$-functions for these couplings, we have to perform a Laurent series expansion of the RHS of the RG equation. Using the techniques introduced in Subsection \ref{sect3.3}, this is done by first series expanding the arguments of the traces in $R$ and then using the heat-kernel expansion \eqref{heatexp},\eqref{expand2} to evaluate the $-D^2$-dependent traces. The first step reveals another remarkable property of the flow equation. Due to the homogeneity property of its RHS with respect to the function $f_k(R)$, expanding the arguments of the traces \emph{does not give rise} to terms containing inverse powers of $R$.
Thus, even though $f_k(R)$ contains poles as $R \rightarrow 0$, the expansion of the trace arguments only yields terms which are regular as $R \rightarrow 0$. Taking into account that the heat-kernel expansion \eqref{heatexp} also  contains positive powers of $R$ only, we find that the RHS of the flow equation is regular in $R=0$. Comparing the pole structure on both sides of the flow equation, we then conclude that the $\beta$-functions of the non-local couplings appearing in \eqref{eq:4.1} are trivial,
\be\label{res1}
\p_t \bar{\mu}_{n;k} = 0 \, , \;  -n_0 \le n < 0 \, ,
\ee
verifying our first claim. Eq.\ \eqref{res1} implies that non-local couplings of the first kind are scale-independent constants along an arbitrary RG trajectory. In particular, once we start with $\bar{\mu}_{n;\hat{k}} = 0$ at the initial scale $\hat{k}$, the RG flow does not dynamically generate non-local couplings of the form $R^{-n}$. 

The situation becomes slightly more complicated for non-local couplings of the second kind. In this case, expanding the arguments of the traces and using the heat-kernel expansion, one finds that such interactions also appear on the RHS of the flow equation. Thus, the $\beta$-functions for this type of non-local couplings are non-trivial. However, since the heat-kernel expansion \eqref{heatexp} does not contain $(\ln(R))^n$-terms, the only source for the logarithmic factors in the non-local interactions is the expansion of the traces. This implies that the corresponding non-local terms on the RHS of the flow equation are necessarily proportional to a non-local coupling constant multiplying $\ln(R)$-terms in $f_k(R)$. While it is difficult to construct the resulting $\beta$-functions explicitly, we can still  conclude that the resulting flow equation permits us to \emph{consistently set the corresponding non-local couplings to zero}. In other words, if there are no interaction terms containing (powers of) $\ln(R)$ present at the initial scale $\hat{k}$, they will not be generated by the RG flow. Note that the same argument also applies to fractional powers of the curvature tensor, which also do not arise from the heat-kernel expansion \eqref{heatexp}, so that it is possible to consistently decouple such interaction monomials as well.

This situation is reminiscent of the observations made in \cite{frank2}, where the RG flow of non-local functions $f_k(V)$ of the space-time volume $V$ were studied. There, the $\beta$-functions for the non-local couplings $V^2$ and $V \ln(V)$ were explicitly derived, with the result that, even though the $\beta$-functions of the non-local couplings were non-trivial, they could be consistently decoupled from the RG equations by setting these couplings to zero. Noting that such interaction monomials are also excluded in the heat-kernel expansion, it is tempting to speculate that {\it all coupling constants whose interaction monomials do not appear in the heat-kernel expansion can be consistently set to zero} in the flow equation.
\subsection{Resolving the IR singularities of RG trajectories with $\Lambda_0 > 0$}
\label{Sect:4.2}
%
In this subsection, we discuss a feature of the RG trajectories found in the Einstein-Hilbert
truncation (cfg.\ Figure \ref{fig0}), namely, the termination of the type IIIa trajectories at 
$\lambda_{k_{\rm term}} = 1/2$ for a finite value $k_{\rm term} > 0$. In principle,
this class of trajectories would give rise to a positive cosmological constant in the IR,
and it was argued in \cite{RW1} that the RG trajectory realized by nature is of this type.
In this light, it is an important question whether the termination of these trajectories
is due to using an insufficient truncation to describe this part of the gravitational
RG flow, and thus whether there are extensions of the Einstein-Hilbert truncation which allow
the continuation of these trajectories down to $k = 0$. 
In the remainder of this section, we will use the RG equation for $f(R)$-gravity to 
show that such extensions indeed exist.

Let us begin by sketching the origin of the IR singularity in the Einstein-Hilbert truncation
\be\label{fEH}
f_k(R) = -R + 2 \Lambda_k \, .
\ee
From the RG equation \eqref{traces}, we observe that its universal part is 
 independent of $\Lambda_k$ and therefore cannot be responsible for the
singularity. Consequently, the singularity must arise from
 the $f_k(R)$-dependent part. For conciseness, we illustrate its origin in $\cS^4_k$, noting that the analysis for $\cS^5_k$ is completely analogous.

Substituting the ansatz \eqref{fEH} into $S^4_k$, we find
\be
\begin{split}
S^{4{\rm EH}}_k = & \half \Tr_{\rm 2T} \left[
\frac{\p_t \left( Z_{Nk} R_k \right)}{Z_{Nk} \left( P_k - 2 \Lambda_k + \tfrac{d^2-3d+4}{d(d-1)} R \right)}
\right] 
=  \half \Tr_{\rm 2T} \left[
\frac{\p_t \left( Z_{Nk} R_k \right)}{Z_{Nk} \left( P_k - 2 \Lambda_k \right) }+ \cO(R)
\right] \\
= & \frac{(d-2)(d+1)}{2(4 \pi)^{d/2}} \, k^d \, 
\left[ \Phi^1_{d/2}(-2 \Lambda_k/k^2) - \half \eta_N \Pt^1_{d/2}(-2 \Lambda_k/k^2)\right] 
\int d^dx \sqrt{g} + \cO(R) \, .
\end{split}
\ee  
Here, we carried out a Taylor expansion of the argument with respect to $R$ in the first step
and then used the heat-kernel expansion \eqref{expand2}, together with \eqref{Qthres1}, to obtain the last line. Substituting 
the threshold functions for the optimized cutoff \eqref{thresopt}, this becomes\footnote{For explicitness, we use the optimized cutoff from Appendix \ref{C.2}. Other cutoff-schemes like, e.g., the exponential cutoff or the sharp cutoff lead to qualitatively similar results.}
\be
S^{4{\rm EH}}_k = \tfrac{(d-2)(d+1)}{2(4 \pi)^{d/2}}  \, 
\left[ \tfrac{1}{\Gamma(d/2+1)}  - \tfrac{1}{2 \Gamma(d/2+2)} \eta_N 
\right] \, k^d\, \frac{1}{1-2 \lambda_k} \, 
\int d^dx \sqrt{g} + \cO(R) \, ,
\ee
where we have expressed the cosmological constant through its dimensionless counterpart $\lambda_k = \Lambda_k/k^2$.
From this expression, we see that the RHS of the flow equation, and therefore also the
$\beta$-functions for $g_k, \lambda_k$, has a pole at $\lambda = 1/2$.
 An analysis of the RG flow in the Einstein-Hilbert truncation reveals that the type
 IIIa trajectories shown in Figure \ref{fig0} run into the $\lambda = 1/2$-line, which causes their termination
 at a finite value $k_{\rm term} > 0$. Our aim now is to characterize the properties of $f_k(R)$ which remove the poles 
at $\lambda = 1/2$ from the RHS of the flow equation. Again, we will focus
 on $\cS^4_k$, noting that the same mechanism applies to $\cS^5_k$.
 
The principal idea behind the resolution of the $\lambda = 1/2$ singularity is to exploit the homogeneity of the trace-argument in $f_k$ and its derivatives in such a way that the $\Lambda_k$-terms arising in the denominators are  multiplied by some power of $R$. The $\Lambda_k$-terms are then dressed up as $\Lambda_k R^\epsilon$ and thus expanded in the Taylor series. In that case, the RHS of the flow equation has \emph{no poles} of the form $(1-2 \lambda_k)^{-p}, p > 0$, and the singularity is resolved. This can be achieved if $f^\prime_k(R)$ contains a term which diverges in the limit $R \rightarrow 0$,
\be\label{cond1}
\lim_{R \rightarrow 0} f^\prime_k(R) = \bar{\mu}_k \, R^{-\epsilon} + \ldots \; , \quad \epsilon > 0 \, ,
\ee 
with $\bar{\mu}_k$ a scale-dependent coupling constant. The relation \eqref{cond1} is easily integrated to yield the corresponding condition on $f_k(R)$,
\be\label{extent}
\begin{split}
\lim_{R \rightarrow 0} f_k(R) = & \, \frac{\bar{\mu}_k}{1-\epsilon} \, R^{1-\epsilon} + \ldots \, , \; \epsilon > 0, \, \epsilon \not = 1 \quad \mbox{and} \quad 
\lim_{R \rightarrow 0} f_k(R) =  \ub_k \, \ln(R) + \ldots \, , \; \epsilon = 1 \, .
\end{split}
\ee 

Let us illustrate the workings of this mechanism in detail by adding a term of the form \eqref{extent} to the Einstein-Hilbert truncation,\footnote{For $\epsilon = 1$, one should replace $R^{1-\epsilon}$ by $\ln(R)$, which is easily done in the formulas below.} 
\be
f_k(R) = -R + 2 \Lambda_k + 16 \pi G_k \, \bar{\mu}_k \, R^{1-\epsilon} \, .
\ee
Here, $\bar{\mu}_k$ has been rescaled for later convenience. Substituting this ansatz into $\cS_k^4$, 
we find
\be
\begin{split}
\cS_k^4  
= & \, \frac{1}{2} \Tr_{\rm 2T} \left[ \frac{Z_{Nk}^{-1} \p_t \left( Z_{Nk} \, (\tfrac{16 \pi G_k \mb_k}{1-\epsilon}  - R^{\epsilon}) \, R_k \right)}{(\tfrac{16 \pi G_k \mb_k}{1-\epsilon} - R^{\epsilon})(P_k - \tfrac{2(d-2)}{d(d-1)} R) 
- R^{1+\epsilon} + 2 \Lambda_k R^{\epsilon} + 16 \pi G_k \, \bar{\mu}_k \, R } \right] \\
= & \, \frac{1}{2} \Tr_{\rm 2T} \left[ \frac{\p_t \left( \mb_k \, R_k \right)}{\mb_k \, P_k} + \cO(R^{\epsilon}, R) \right] \\
= & \, \frac{(d-2)(d+1)}{2 (4 \pi)^{d/2}} k^d \, \left[ \Phi^1_{d/2}(0) + \half \, \eta_{\mb} \, \Pt^1_{d/2}(0) \right] \int d^dx \sqrt{g} + \cO(R^{\epsilon}, R) \, .
\end{split}
\ee
Using the threshold functions for the optimized cutoff, we observe that the poles at $\lambda = 1/2$ have been removed. Thus, the singularities at $\lambda = 1/2$ vanish with the inclusion of interaction terms of the form \eqref{extent}. 
This will be further illustrated in the following sections,
where it will be shown explicitly that the truncations \eqref{logRmodel} and \eqref{Rinvmodel}
contain RG trajectories with positive cosmological constant in the IR $\Lambda_0 > 0$.

One cautious remark is now in order. 
 Once terms of the form \eqref{extent} are included in the truncation subspace, one has to be very careful about taking the limit in which their corresponding couplings are sent to zero. The reason is that taking this limit does not commute with the Taylor expansion of the trace arguments. This can be easily seen from the following simplified example. Setting
$
g(x) = g_1 x^{-1} + g_2, h(x) = h_1 x^{-1} + h_2,
$  
with $g_1, g_2, h_1, h_2$ constant, and considering the expansion of the quotient
\be
\frac{g(x)}{h(x)} = \frac{g_1 x^{-1} + g_2}{h_1 x^{-1} + h_2} = \frac{g_1  + g_2 x}{h_1  + h_2 x} =
\frac{g_1}{h_1} + \left( \frac{g_2}{h_1} - \frac{g_1 h_2}{h_1^2} \right) x + \cO(x^2)\,,
\ee
we see that taking the ``non-local couplings'' $g_1, h_1$ to zero after the expansion leads to a quite different result than doing so before the Taylor series is constructed. From this example, it is clear that considering the RG equation on a truncation subspace which involves couplings of the form \eqref{extent} and setting the new coupling to zero does not automatically yield the $\beta$-functions obtained in the Einstein-Hilbert case. That being said, in the remainder of this paper we will adopt the viewpoint that the $\beta$-functions derived from the truncations \eqref{logRmodel} and \eqref{Rinvmodel} can be extended continuously to the point of vanishing non-local coupling.

\section{Truncation I: the $\ln(R)$-truncation}
\label{sect:5}
%
Following up on our general discussion, we now continue our investigation by
analyzing the non-perturbative $\beta$-functions arising from the $\ln(R)$-truncation \eqref{logRmodel} which,
in the notation of Section \ref{sect:3}, corresponds to
\be\label{lnRansatz}
f_k(R) = - R + 2 \Lambda_k + 16 \pi G_k \, \ub_k \, \ln(R) \, .
\ee
We start by deriving the $\beta$-functions governing the scale-dependence of the couplings $G_k, \Lambda_k$ and $\ub_k$ in Subsection \ref{sect:5.1}, while their properties are investigated analytically and numerically in Subsections \ref{sect:5.2} and \ref{sect:5.3}.
%
\subsection{The $\beta$-functions of the $\ln(R)$-truncation}
\label{sect:5.1}
%
Our first task in the construction of the $\beta$-functions of the $\ln(R)$-truncation is the evaluation of the truncation-dependent traces $\cS^4_k$ and $\cS^5_k$ in \eqref{traces} up to linear order in $R$. Substituting the
ansatz \eqref{lnRansatz} and using the heat-kernel techniques of Subsection \ref{sect3.3}, this yields
\be
\begin{split}
\cS^4_k =&\,
\frac{k^d}{(4 \pi)^{d/2}} \, C^{\rm 2T}_0 \, \left\{ \Phi_{d/2}^1 + \half \eta_{\ub} \, \tilde{\Phi}_{d/2}^1 \, \right\} \, \int \! d^dx \sqrt{g} \\ 
&+ \frac{k^{d-2}}{(4 \pi)^{d/2}} \, \Big\{ 
C^{\rm 2T}_0 \, \Big[ \, \big( \tfrac{d-2}{d(d-1)}- \tfrac{\Lambda_k}{16 \pi G_k \ub_k}\big) \, \big( \eta_{\ub}  \, \tilde{\Phi}_{d/2}^2 + 2 \, \Phi_{d/2}^2 \big) 
 + \frac{k^2 \left( \eta_{\ub} + \eta_N \right)}{32 \pi G_k \, \ub_k}  \, \tilde{\Phi}_{d/2}^1 \,\Big] \\ 
& \qquad+ C^{\rm 2T}_2 \, \left[ \Phi_{d/2-1}^1 + \half \, \eta_{\ub} \, \tilde{\Phi}_{d/2-1}^1 \right]  \Big\} \, \int \! d^dx  \sqrt{g} R \, , \\
\cS^5_k =&\, \frac{k^d}{(4 \pi)^{d/2}} \left\{ 2 \Upsilon^1_{d/2,1} + \half \eta_{\ub} \tilde{\Upsilon}^1_{d/2,0,1} \right\} \int \! d^dx \sqrt{g} \\ 
&+ \frac{k^{d-2}}{(4 \pi)^{d/2}} \Big\{ 
\frac{d+2}{2(d-1)} 
\left(
2 \Upsilon^2_{d/2,2}-  \Upsilon^1_{d/2,0,0}+ \half \eta_{\ub} \big( \tilde{\Upsilon}^2_{d/2,1,1} - \tilde{\Upsilon}^1_{d/2,0,0}  
\big) \right)
\\ 
& \qquad + C_2^{\rm S} \left( 2 \Upsilon^1_{d/2-1,1}+ \half \eta_{\ub} \tilde{\Upsilon}^1_{d/2-1,0,1} \right) \Big\} \int \! d^dx  \sqrt{g} R \, ,
\end{split}
\ee
for the 2T and scalar trace, respectively. Here, 
$\eta_N(k)$ and $\eta_{\ub}(k)$ are the anomalous dimensions \eqref{andim} and the $C_{2k}^s$ are defined in \eqref{B.20}. Inserting the truncation ansatz into the LHS of the flow equation \eqref{traces}
in turn gives 
\be
\p_t \Gammab_k[g] = \frac{1}{16 \pi G_{\hat{k}}} \, \int \! d^dx \sqrt{g} \Big[ - R \, \p_t Z_{Nk} + 2 \, \p_t(Z_{Nk} \Lambda_k) \Big] 
+ \int \! d^dx \sqrt{g} \ln(R) \, \p_t \ub_k \, . 
\ee
Taking into account the contributions from the universal traces \eqref{unitrace} and equating the coefficients of the invariants $\int \! d^dx \sqrt{g}$, $\int \! d^dx \sqrt{g}R$ and $\int \! d^dx \sqrt{g} \ln(R)$ spanning our truncation space then leads to the following system of coupled differential equations
\be\label{eq:5.8}
\begin{split}
\p_t \ub_k & =  0 \, , \\
\p_t Z_{Nk} &= \frac{16 \pi G_{\hat{k}} k^{d-2}}{(4 \pi)^{d/2} - \half \, k^d \, \ub_k^{-1} \, \tilde{\Phi}^1_{d/2} \, C_0^{\rm 2T}} \Big[ 
\big(\tfrac{(d-2)(d+1)}{16 \pi G_k} \tfrac{\Lambda_k}{ \ub_k} - \tfrac{d^3 - 4d^2 + d -3}{d(d-1)} \big) \, \Phi^2_{d/2}  
\qquad \\  
& \quad \quad
+ \big( \tfrac{d^2-6}{6d} - C_2^{\rm 2T} \big) \Phi^1_{d/2-1}
- \tfrac{d+2}{d-1} \big( \Upsilon^2_{d/2,2} - \half \Upsilon^1_{d/2,1} \big) - \tfrac{1}{3} \Upsilon^1_{d/2-1,1} 
\Big] \, , \\
\p_t(Z_{Nk} \Lambda_k) & =  \frac{G_{\hat{k}} k^d}{(4 \pi)^{d/2-1}} \big[ (d^2 -3d-2) \, \Phi^1_{d/2} + 4 \Upsilon^1_{d/2,1} \big] \, .
\end{split}
\ee
The $\beta$-functions resulting from these equations are most conveniently expressed in terms of the 
 the dimensionless couplings
\be\label{dimlesscc}
g_k = k^{d-2} \, Z_{Nk}^{-1} \, G_{\hat{k}} = k^{d-2} \, G_k \, , \quad 
\lambda_k = k^{-2} \Lambda_k \, , 
\quad \upsilon_k = k^{-d} \ub_k \, . 
\ee
In terms of these, the eqs.\ \eqref{eq:5.8} imply the following set of autonomous first order differential equations 
\be\label{DGL1}
\p_t g_k = \beta_g(g, \lambda, \upsilon) \, , \quad 
\p_t \lambda_k = \beta_\lambda(g, \lambda, \upsilon) \, , \quad
\p_t \upsilon_k  = \beta_\upsilon(g, \lambda, \upsilon)\,,
\ee
where
\be\label{betafct1}
\begin{split} 
\beta_g(g, \lambda, \upsilon) = & \, \left(d-2+\eta_N \right) \, g_k \, , \\
\beta_\lambda(g, \lambda, \upsilon) = & \, - \left( 2 - \eta_N \right) \lambda_k + (4\pi)^{1-d/2} \, g_k 
\left( (d^2-3d-2) \, \Phi^1_{d/2} + 4 \, \Upsilon^1_{d/2,1} \right) \, , \\
\beta_\upsilon(g, \lambda, \upsilon) = & \, - d \, \upsilon_k \, .
\end{split}
\ee
Here, the anomalous dimension of Newton's constant is given by
\be
\begin{split}
\eta_N = & \, - \, \frac{32 \pi g_k}{2 \, (4 \pi)^{d/2} - \upsilon_k^{-1} \Pt^1_{d/2} \, C_0^{\rm 2T}}
\Big[ 
\big(\tfrac{(d-2)(d+1)}{16 \pi g_k} \tfrac{\lambda_k}{ \upsilon_k} - \tfrac{d^3 - 4d^2 + d -3}{d(d-1)} \big) \, \Phi^2_{d/2}  
\qquad \\  
& \quad \quad
+ \big( \tfrac{d^2-6}{6d} - C_2^{\rm 2T} \big) \Phi^1_{d/2-1}
- \tfrac{d+2}{d-1} \big( \Upsilon^2_{d/2,2} - \half \Upsilon^1_{d/2,1} \big) - \tfrac{1}{3} \Upsilon^1_{d/2-1,1} 
\Big] \, .
\end{split}
\ee
This completes the derivation of the $\beta$-functions for the $\ln(R)$-truncation. 
Note that these $\beta$-functions are consistent in the sense that they incorporate all 
contributions proportional to the interaction monomials spanning the truncation subspace.
As expected from the general discussion in Section \ref{Sect4.1}, the $\beta$-function
for the dimensionful non-local coupling $\ub_k$ is indeed trivial, $\p_t \ub_k = 0$.
Furthermore, the only poles of \eqref{betafct1} occur at the line
\be\label{logRsing} 
\upsilon_{\rm sing} = \half \, (4 \pi)^{-d/2} \, C_0^{\rm 2T} \, \Pt_{d/2}^1, 
\ee
showing explicitly that the singularity at $\lambda = 1/2$ has been resolved,
in full agreement with the arguments given in Section \ref{Sect:4}. 

A second property which can be directly deduced from the $\beta$-functions \eqref{betafct1} is the decomposition
of the coupling constant space in various sectors whose trajectories do not mix under the
RG flow. Since $\beta_g(g=0, \lambda, \upsilon) = 0$ and $\beta_\upsilon(g,\lambda, \upsilon =0)=0$, the $g = 0$- and $\upsilon =0$-planes cannot be crossed by the trajectories. In other words, a RG trajectory starting with
positive $g_{\hat{k}}$ (or $\upsilon_{\hat{k}}$) will never evolve to $g_k < 0$ (or $\upsilon_k < 0$), and vice versa.
%
\subsection{The fixed points of the RG flow}
\label{sect:5.2}
%
In order to gain some first insights on the dynamics arising from \eqref{DGL1},
it is very helpful to analyze the fixed point structure of the $\beta$-functions.
At the fixed points (FP), all $\beta$-functions vanish simultaneously,
\be
\beta_g(g^*, \lambda^*, \upsilon^*) = 0 \, , \quad \beta_\lambda(g^*, \lambda^*, \upsilon^*) = 0 \, , \quad \beta_\upsilon(g^*, \lambda^*, \upsilon^*) = 0 \, ,
\ee
and many properties of the RG flow can be captured by linearizing the RG equation at such points and studying the flow in their vicinity. Introducing the generalized couplings $\vec{u}_k = [g_k , \lambda_k , \upsilon_k]$ and the Jacobi matrix
\be\label{eq:5.15}
{\bf B}=\left(B_{ij}\right)\,,\quad \quad B_{ij} = \left. \frac{\p \beta_{u_i}}{\p u_j} \right|_{\rm FP} \,,
\ee
the linearized flow equation reads
\be\label{linflow}
\p_t u_i \approx \sum_j B_{ij} (u_j - u_j^*) \,. 
\ee
Using the stability coefficients $\theta^i = -\lambda^i$, with $\lambda^i$ the eigenvalue of ${\bf B}$ associated with the right eigenvectors $V^i$, $B_{ij}$ can be diagonalized, so that eq.\ \eqref{linflow} is easily solved. These analytic solutions then provide a good picture of which RG trajectories are dragged into the FP or repelled along an unstable direction. 

Applying this construction to the $\beta$-functions \eqref{betafct1}, we first observe that $\beta_\upsilon = 0$ forces all FP onto the $v_k = 0$ plane. Substituting this condition into the remaining $\beta$-functions, $\beta_g$ and $\beta_\lambda$ simplify considerably and their roots are easily found analytically,
\bea\label{logRFP} \nn
\mbox{GFP}:  && \big\{ g^* = 0 \, , \, \lambda^* = 0 \, , \, \upsilon^* = 0 \, \big\} \, , \\
\mbox{IRFP}:  && \big\{ g^* = 0 \, , \, \lambda^* = \tfrac{\Pt^1_{d/2}}{2 \Phi^2_{d/2}} \, , \, \upsilon^* = 0 \, \big\} \, , \\ \nn
\mbox{NGFP}:  && \big\{ 
g^* = - \, \tfrac{(4 \pi)^{d/2-1} \, d(d-2) \, \Pt^1_{d/2}}{4 \Phi^2_{d/2} \, \big( (d^2-3d-2) \Phi^1_{d/2} + 4 \Upsilon^1_{d/2,1} \big) }  
\, , \, \lambda^* = - \tfrac{(d-2) \, \Pt^1_{d/2}}{4 \phi^2_{d/2}} \, , \, \upsilon^* = 0 \, \big\} \, .
\eea
Here, GFP denotes the Gaussian Fixed Point, while the other FP have providently been labeled 
Infrared Fixed Point (IRFP), and Non-Gaussian Fixed Point (NGFP). 
Let us then discuss the stability properties of these fixed points in turn.

At the GFP, the stability matrix ${\bf B}$ becomes 
\be\label{eq:5.17}
{\bf B}_{\rm GFP} = \left[ 
\begin{array}{ccc}
d-2 & 0 & 0 \\
(4 \pi)^{1-d/2} \big( (d^2-3d-2) \Phi^1_{d/2} + 4 \Upsilon^1_{d/2,1} \big) &
-2 & 0 \\
0 & 0 & -d 
\end{array}
\right]\,,
\ee
with stability coefficients and associated right eigenvectors
\be\label{esGFP}
\begin{array}{ll}
\theta_1  = 2-d \,  , \quad & V^1 = \left[ \, 1 \, , \tfrac{1}{d} (4 \pi)^{1-d/2} \big( (d^2-3d-2) \Phi^1_{d/2} + 4 \Upsilon^1_{d/2,1} \big) \, , \, 0 \, \right]^{\rm T} \, , \\
\theta_2  = 2 \, , & V^2 = \left[ \, 0 \, , \, 1 \, , 0 \, \right]^{\rm T} \, , \\
\theta_3  = d \, , &  V^3 = \left[ \, 0 \, , \, 0 \, , \, 1 \, \right]^{\rm T} \, . 
\end{array}
\ee
This information can be used to solve the RG equation in the vicinity of the GFP, giving
\be
\begin{split}\label{lngfpsol}
g_k = & \alpha_1 \frac{k^{d-2}}{M^{d-2}} \, , \\
\lambda_k = & \alpha_2 \frac{k^{-2}}{M^{-2}} + \alpha_1 \,\frac{(4 \pi)^{1-d/2}}{d}  \big( (d^2-3d-2) \Phi^1_{d/2} + 4 \Upsilon^1_{d/2,1} \big) \,  \frac{k^{d-2}}{M^{d-2}} \, , \\
\upsilon_k = & \alpha_3 \frac{k^{-d}}{M^{-d}} \, .
\end{split}
\ee
Here, $\{\alpha_i\}$ denotes the set of integration constants and $M$ is an arbitrary fixed mass scale. It is then natural
to definite the dimensionful integration constants
\be
G_0 = m_{\rm Pl}^{2-d} \, , \quad \Lambda_0 = \alpha_2 \, m_{\rm Pl}^2 \, , \quad \ub_0 = \alpha_3 \, m_{\rm Pl}^d\,,
\ee 
where we have identified $M$ with the Planck mass $m_{\rm Pl}$ by setting $\alpha_1 = 1$.
By using the relations \eqref{dimlesscc}, we then obtain the scale-dependence of the dimensionful coupling constants in the vicinity of the GFP,
\be
\begin{split}\label{lngfpdsol}
G_k = & \, G_0 \, , \\
\Lambda_k = &  \, \Lambda_0 + 
\frac{(4 \pi)^{1-d/2}}{d} \,  G_0\,  \big( (d^2-3d-2) \Phi^1_{d/2} + 4 \Upsilon^1_{d/2,1} \big) \,  k^{d} \, , \\
\ub_k = & \, \ub_0 \, .
\end{split} 
\ee
These equations show that $G_k, \Lambda_k$ and $\ub_k$ run towards constant, non-zero values as $k \rightarrow 0$. In particular, $\Lambda_k$ does not vanish in the IR unless we make the special choice $\alpha_2 = 0$. Thus the GFP does not determine the behavior of the cosmological constant $\Lambda_0$ in the infrared.

Let us now turn to the IRFP \eqref{logRFP}. Here, the matrix ${\bf B}$ becomes
\be
{\bf B}_{\rm IRFP} = \left[ 
\begin{array}{ccc}
d & 0 & 0 \\
(4 \pi)^{1-d/2} \big( (d^2-3d-2) \Phi^1_{d/2} + 4 \Upsilon^1_{d/2,1} \big) &
2 & \frac{4 (4 \pi)^{d/2}}{(d+1)(d-2) \Phi^2_{d/2}} \\
0 & 0 & -d 
\end{array}
\right] \, , 
\ee 
and the eigensystem of ${\bf B}_{\rm IRFP}$ is
\be
\begin{array}{ll}
\theta_1  = -d \,  , \quad & V^1 = \left[ \, 1 \, , \tfrac{1}{d-2} (4 \pi)^{1-d/2} \big( (d^2-3d-2) \Phi^1_{d/2} + 4 \Upsilon^1_{d/2,1} \big) \, , \, 0 \, \right]^{\rm T} \, , \\
\theta_2  = -2 \, , & V^2 = \left[ \, 0 \, , \, 1 \, , 0 \, \right]^{\rm T} \, , \\
\theta_3  = d \, , &  V^3 = \left[ \, 0 \, , \, - \frac{4 (4\pi)^{d/2}}{\left( d^3+d^2-4d-4 \right) \Phi^2_{d/2}} \, , \, 1 \, \right]^{\rm T} \, . 
\end{array}
\ee
Note that here the stability coefficient $\theta_2$ has now changed sign compared to the eigensystem of the GFP \eqref{esGFP}, 
making the eigendirection $V^2$ IR-attractive (hence the name IRFP). 
 Solving the linearized flow equation results in
\be
\begin{split}
g_k = & \alpha_1 \frac{k^{d}}{M^{d}} \, , \\
\lambda_k = & \alpha_2 \frac{k^{2}}{M^{2}} 
+ \alpha_1 \,\frac{(4 \pi)^{1-d/2}}{d-2} \, \big( (d^2-3d-2) \Phi^1_{d/2} + 4 \Upsilon^1_{d/2,1} \big) \, \frac{k^d}{M^d} 
\\ & \, 
- \alpha_3 \frac{4 (4\pi)^{d/2}}{\left( d^3 + d^2 -4d-4 \right) \Phi^2_{d/2}} \frac{k^{-d}}{M^{-d}} 
+ \frac{\Pt^1_{d/2}}{2 \Phi^2_{d/2}}\, , \\
\upsilon_k = & \alpha_3 \frac{k^{-d}}{M^{-d}} \, ,
\end{split}
\ee
again denoting the integration constants by $\alpha_i$. Using the relations \eqref{dimlesscc}, we then obtain the scale-dependence of the dimensionful coupling constants,
\be\label{FPrunning}
\begin{split}
G_k = & \, G_0 \, \frac{k^2}{m_{\rm Pl}^2}  , \\
\Lambda_k = &  \, \Lambda_0 \frac{k^4}{m_{\rm Pl}^4} + 
\frac{(4 \pi)^{1-d/2}}{d-2} \, \big( (d^2-3d-2) \Phi^1_{d/2} + 4 \Upsilon^1_{d/2,1} \big) \,  G_0 \frac{k^{d+2}}{m_{\rm Pl}^2} \\
& \, - \ub_0 \frac{4 (4\pi)^{d/2}}{\left( d^3 + d^2 -4d-4 \right) \Phi^2_{d/2}} \, k^{2-d} +  
\frac{\Pt^1_{d/2}}{2 \Phi^2_{d/2}} k^2 \, ,  \\
\ub_k = & \, \ub_0 \, .
\end{split} 
\ee
Thus, for $d>2$, the IRFP has an unstable direction in $\ub_k$, driving the RG trajectories with $\ub_0 \not = 0$ away from the FP. For trajectories in the plane $\ub_0 = 0$, however, the IRFP is an infrared attractor for both Newton's constant and the cosmological constant. In this case both $G_k$ and $\Lambda_k$ are \emph{dynamically driven to zero} as $k \rightarrow 0$,
independently of their values at the initial scale $\hat{k}$.

Finally the NGFP is located in the (probably unphysical) region $g < 0$. Determining its stability coefficients, one finds that it is UV attractive for all three couplings $g_k, \lambda_k$ and $\upsilon_k$. In this sense, the properties of this NGFP are very similar to the ones observed in the Einstein-Hilbert truncation, cfg.\ Fig.\ \ref{fig0}, except that its position has now shifted from $g_* > 0, \lambda_* > 0$ to $g_* < 0, \lambda_* < 0$. We will return to the discussion of this particular feature in our concluding remarks in Section \ref{sect:8}. 
\subsection{Numerical solutions of the flow equations}
\label{sect:5.3}
%
After discussing their fixed point structure, we now proceed by integrating the
flow equations \eqref{DGL1} numerically. This requires fixing the space-time dimension
and the cutoff-scheme explicitly, so we will restrict ourselves to $d=4$ and to the optimized
cutoff \eqref{optcutoff} in the following. Given the decomposition of the coupling constant space,
we will furthermore limit our focus on trajectories with $g_k > 0$, as these are the physically most 
relevant ones. For $d=4$ and the optimized cutoff, the FP \eqref{logRFP} are located at
\be
\begin{split}\label{flnfp}
&\mbox{GFP}: \big\{ g^* = 0 \, , \, \lambda^* = 0 \, , \, \upsilon^* = 0 \big\} \, , \qquad \qquad
\mbox{IRFP}: \big\{ g^* = 0 \, , \, \lambda^* =  \tfrac{1}{6} \, , \, \upsilon^* = 0 \big\}\, , \\
&\mbox{NGFP}:  \big\{ g^* = - \tfrac{8 \pi}{9} \, , \, \lambda^* = - \tfrac{1}{6} \, , \, \upsilon^* = 0 \big\}\, .
\end{split}
\ee
We note, in particular, that the NGFP is indeed situated in the region $g<0$ and hence will not play a role in the subsequent discussion.
Moreover, the singular line \eqref{logRsing} is located at
\be\label{sing2}
\upsilon_{\rm sing} = 0.00263 \, .
\ee

\begin{figure}[t]
\epsfxsize=5.8in
\begin{center}
\leavevmode
\epsffile{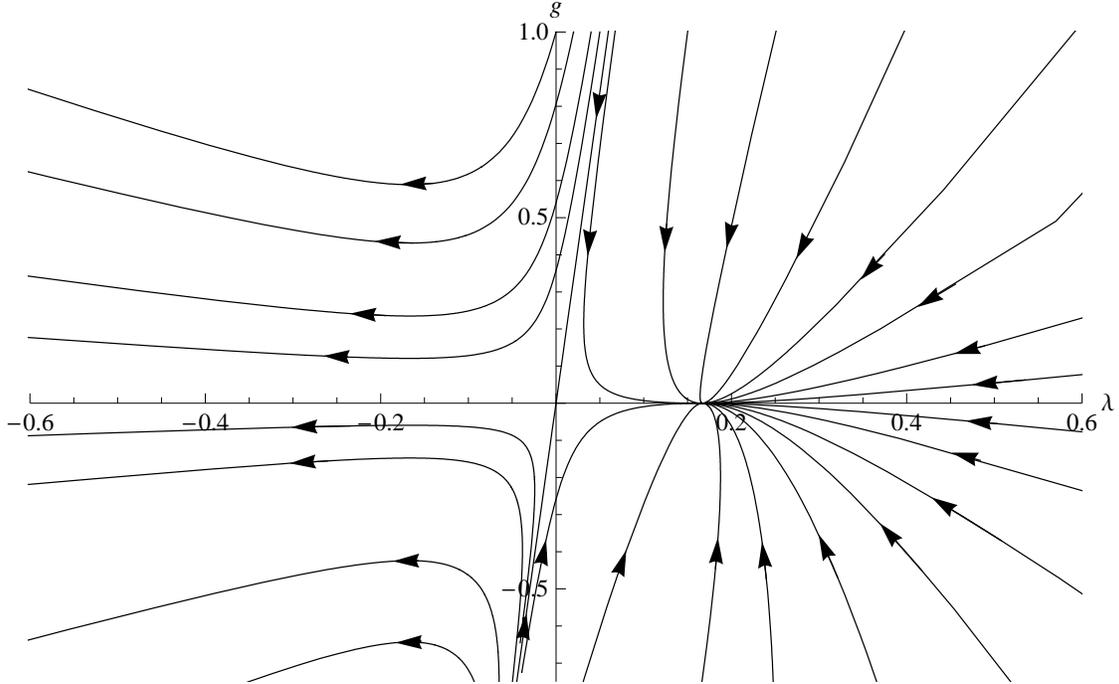}
\end{center}
\parbox[c]{\textwidth}{\caption{\label{eins}{\footnotesize RG flow of the $\ln(R)$-truncation in the fixed plane $\upsilon = 0$. The flow is dominated by the GFP at $g^*=0, \lambda^*=0$ and the IRFP at $g^*=0, \lambda^*= 1/6$. All RG trajectories running to the left of the ``separatrix'' are captured by the IRFP as $k\rightarrow0$.}}} 
\end{figure}

Due to the existence of the IRFP, the most interesting region of the parameter space is the one comprising the RG flow of $g_k, \lambda_k$ on the fixed plane $\upsilon = 0$, which is shown in Figure \ref{eins}.
This diagram clearly illustrates the separation between trajectories with positive and negative values $g_k$. The RG flow in the upper half plane is thereby dominated by the GFP and the IRFP. The RG trajectories in this region fall into three distinct classes, which, in analogy to the Einstein-Hilbert truncation shown in Figure \ref{fig0}, are labelled 
Type I\~a, II\~a, and III\~a, respectively. The class II\~a consists of the single trajectory (``separatrix'') hitting the GFP as $k \rightarrow 0$. The trajectories of Type I\~a run to the left of the separatrix and lead to a negative
cosmological constant $\Lambda_0$ in the infrared. The trajectories of Type III\~a, to the right of the separatrix, are captured by the IRFP as $k \rightarrow 0$, which implies that the corresponding dimensionful couplings $G_k, \Lambda_k$ go to zero as $k \rightarrow 0$. 

\begin{figure}[p]
\begin{center}
$\begin{array}{cc}
\multicolumn{1}{l}{} &
	\multicolumn{1}{l}{} \\ [-0.53cm]
\epsfxsize=2.5in
\epsffile{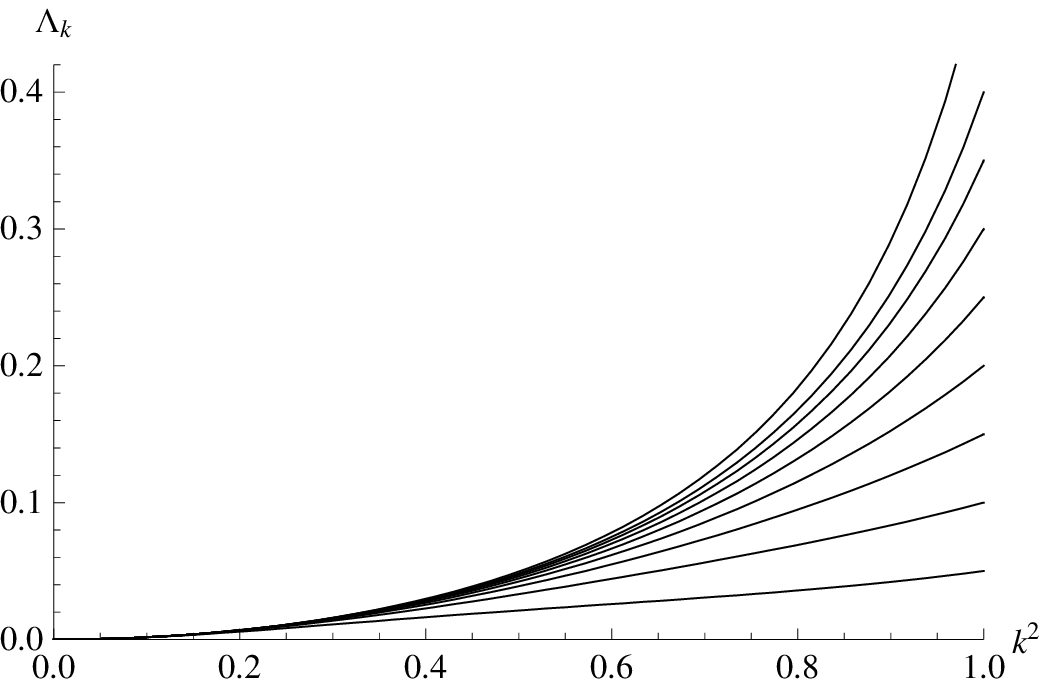} &
	\epsfxsize=2.5in
	\epsffile{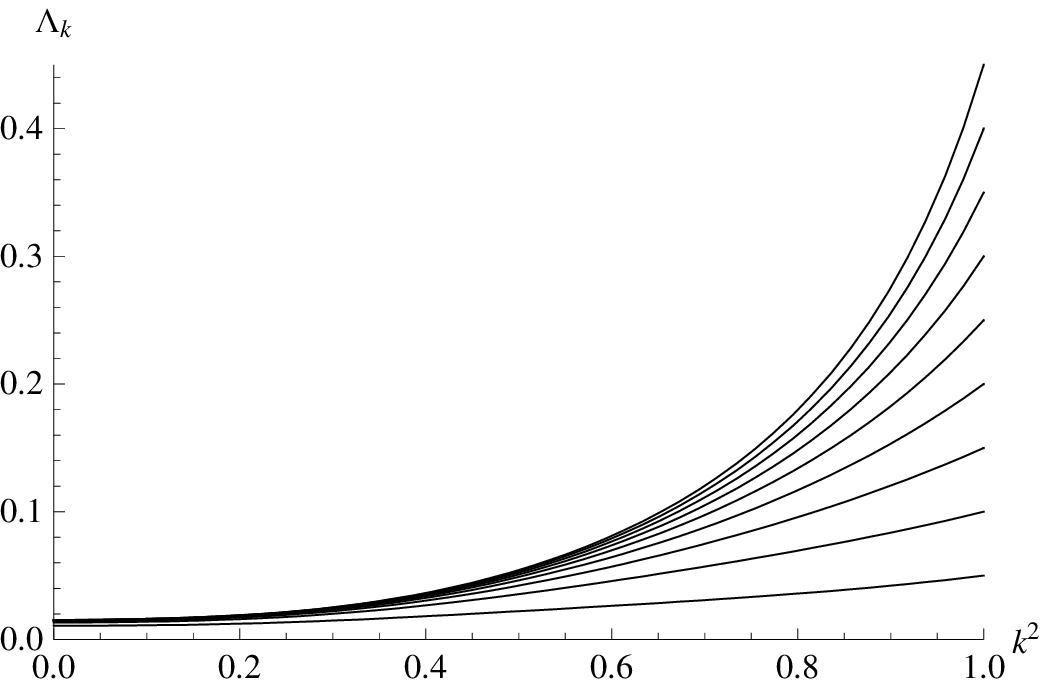} \\ 
\epsfxsize=2.5in
\epsffile{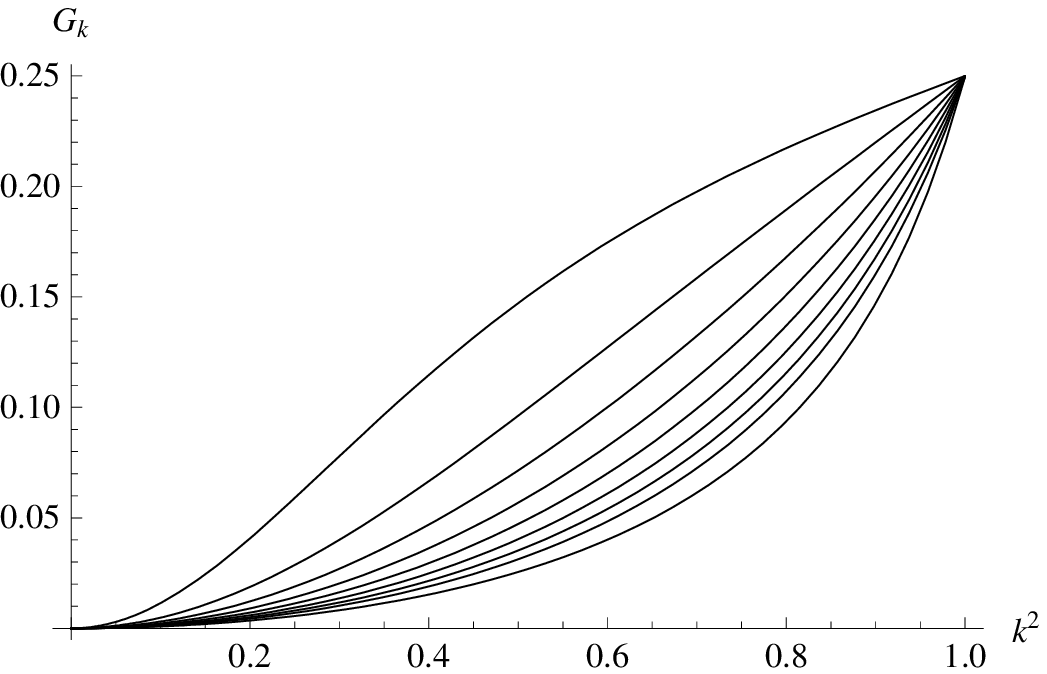} &
	\epsfxsize=2.5in
	\epsffile{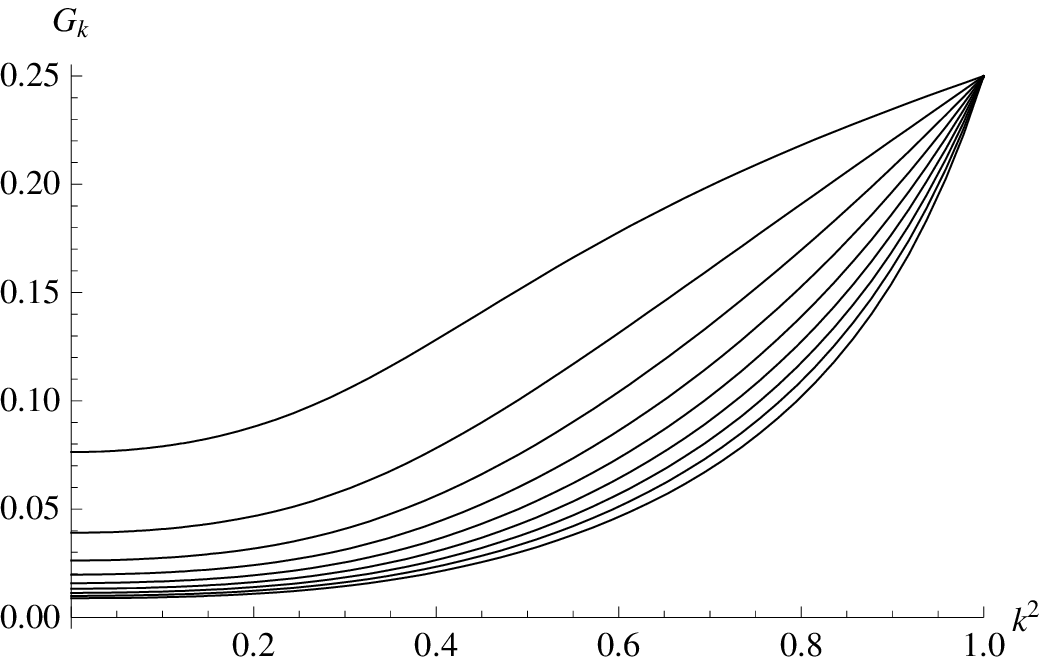} \\ [0.4cm]
\mbox{(a)} & \mbox{(b)} \\
\mbox{}\\
\epsfxsize=2.5in
\epsffile{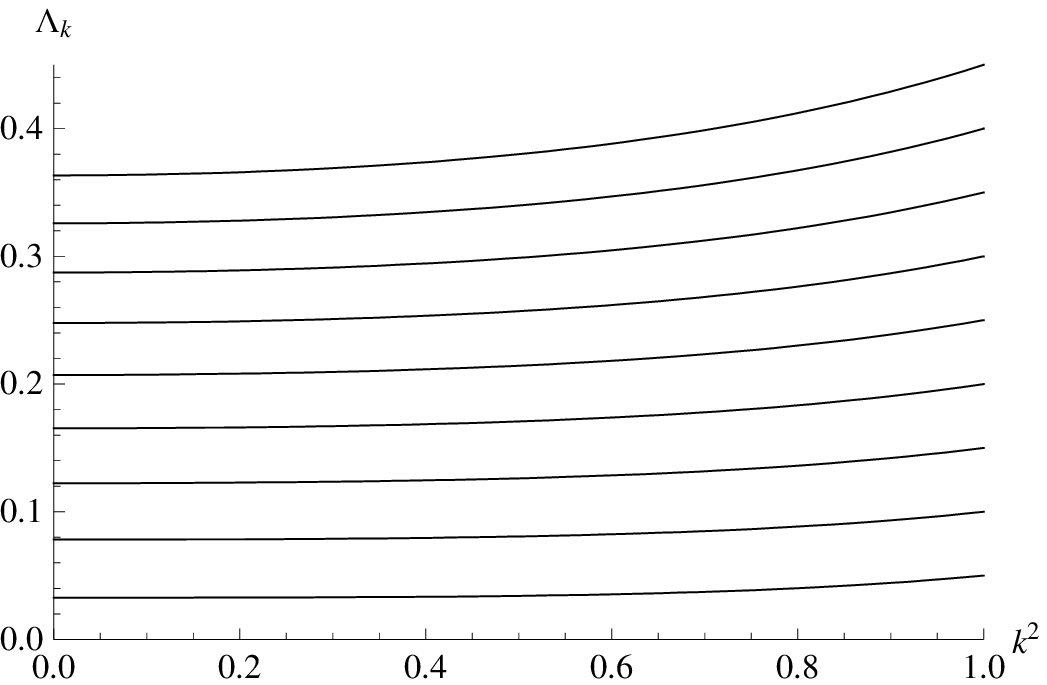} &
	\epsfxsize=2.5in
	\epsffile{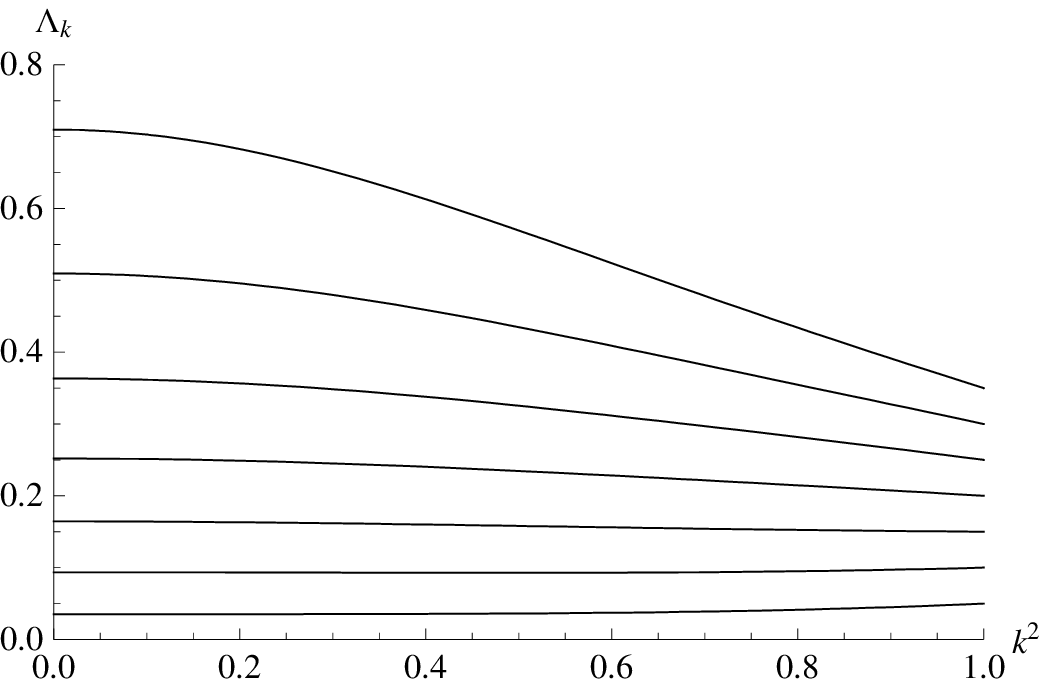} \\ 
\epsfxsize=2.5in
\epsffile{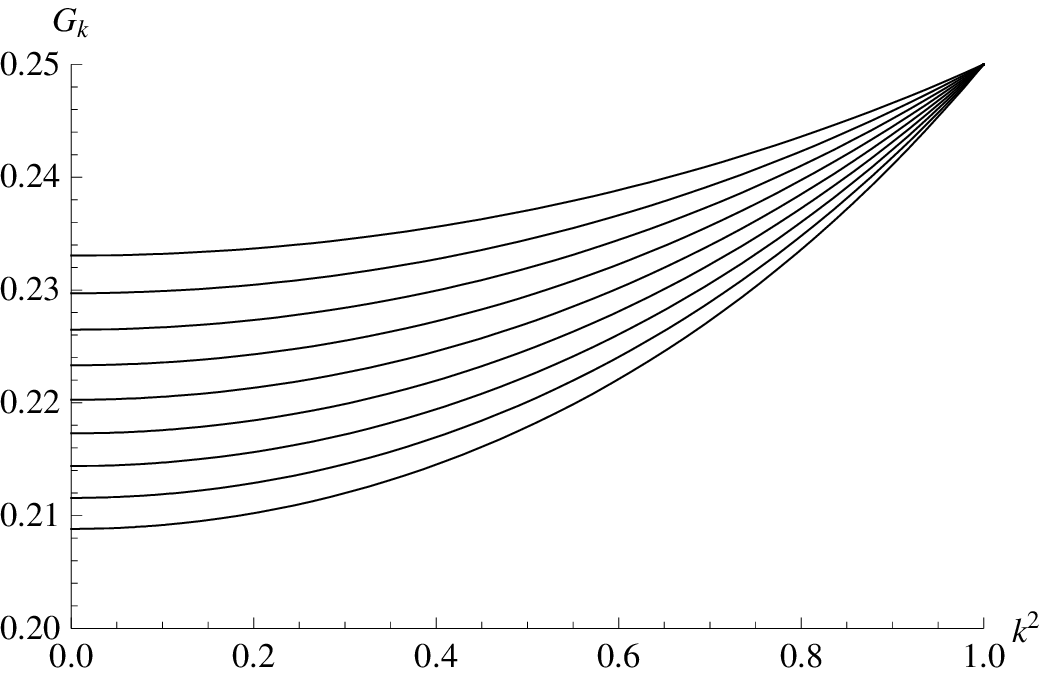} &
	\epsfxsize=2.5in
	\epsffile{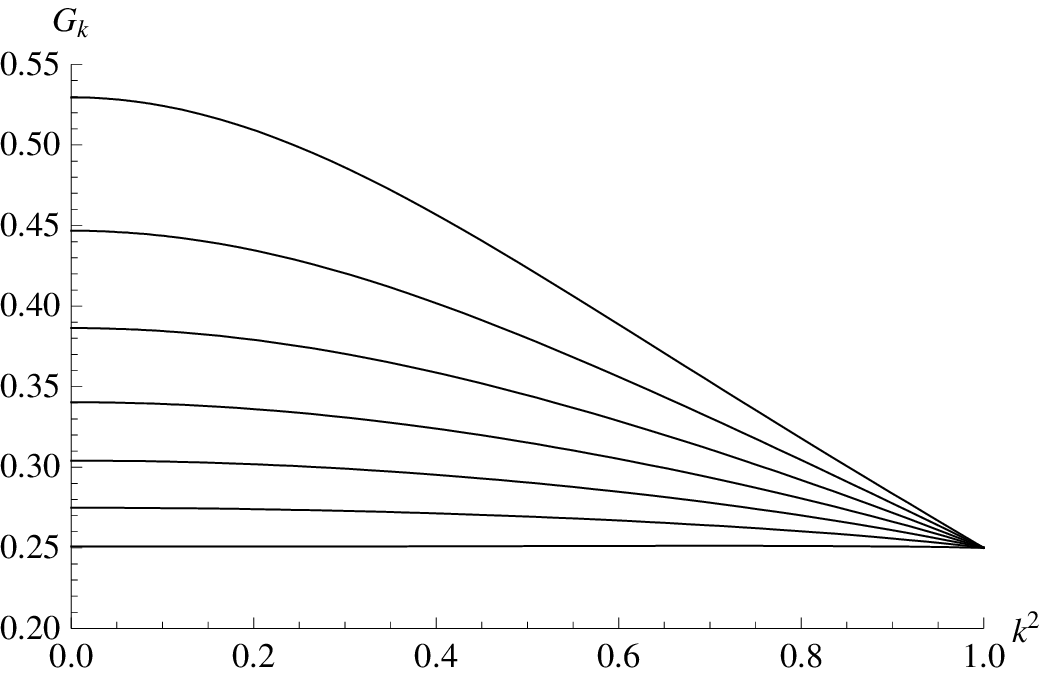} \\ [0.4cm]
\mbox{(c)} & \mbox{(d)} \\
\end{array}$
\parbox[c]{\textwidth}{\caption{\label{fig:dimfuln}{\footnotesize  
Numerical solutions to the $\ln(R)$-truncation flow equations in terms of the dimensionful couplings, for (a) $\ub_{\hat{k}}=0$, (b) $\ub_{\hat{k}}=-5\times10^{-5}$, (c) $\ub_{\hat{k}}=-5\times10^{-2}$ and (d) $\ub_{\hat{k}}=0.01$ imposed at $\hat{k} = 1$. Here, we have taken $G_{\hat{k}}=0.25$, while $\Lambda_{\hat{k}}$ assumes various values. We see that, for  $\ub_{\hat{k}} \lesssim 0$,
 the cosmological constant is quenched in the IR limit, going to zero (a), or small but non-zero (b) values. }}}
\end{center}
\end{figure}

The IR behavior of these classes can also be understood on the level of the linearized solutions \eqref{lngfpdsol} and \eqref{FPrunning}, as follows. The classes I\~a and II\~a correspond to linearized solutions $\alpha_2 <0$ and $\alpha_2 = 0$, respectively. The solutions with $\alpha_2 > 0$ are driven away from the GFP regime and captured by the IRFP \eqref{FPrunning}, giving $G_0 = 0, \Lambda_0 = 0$ independently of any integration constants. The latter feature is
also illustrated in Figure \ref{fig:dimfuln} (a). Due to the absence of an UV FP in the $\ln(R)$-truncation, we do not expect, however, that the classes of RG trajectories discussed here give rise to a well-defined behavior as $k \rightarrow \infty$. This is possibly owed to the fact that our truncation subspace is insufficient for a proper description of quantum gravity in the UV.

Going away from the $\upsilon = 0$-plane, we have to distinguish the two instances $\vb_k < 0$ and $\vb_k > 0$.
To illustrate the behavior in these regions, some examples for trajectories starting with $\Lambda(k = 1) > 0$ are shown in 
Figure \ref{fig:dimfuln}. 
For the trajectories with $\vb_k < 0$ we find that both $G_k, \Lambda_k$ decrease with decreasing values $k$. Comparing the series of diagrams (a), (b), (c), in which the modulus of the non-local coupling is continuously increased, we observe that this ``quenching'' works more efficiently for small non-local coupling $|\vb_k|$, the case of $\vb_k = 0$, corresponding to the IRFP, being the most efficient. 

Finally, for the trajectories with $\vb_k > 0$, we distinguish two different behaviors. Trajectories starting with $\upsilon_{\hat{k}} < \upsilon_{\rm sing}$ run into the singularity and terminate at finite $k_{\rm term} =  (\upsilon_{\hat{k}} / \upsilon_{\rm sing})^{1/4} \hat{k}$, whereas trajectories in the region $\upsilon_{\hat{k}} > \upsilon_{\rm sing}$ are non-singular and can lead to increasing values of $G_k, \Lambda_k$ as $k$ decreases. An example of this latter behavior is shown in Figure \ref{fig:dimfuln} (d).
%
\section{Truncation II: the $R^{-n}$-truncation}
\label{sect:6}
We now turn towards our second class of non-local truncations, the $R^{-n}$-truncations \eqref{Rinvmodel}, for which
\be\label{invRansatz}
f_k(R) = - R + 2 \Lambda_k + 16 \pi G_k \, \mb_k \, R^{-n} \, .
\ee
We begin by deriving the non-perturbative $\beta$-functions of this truncation in Subsection \ref{sect:6.1} and subsequently discuss their properties in Subsections \ref{sect:6.2} and \ref{sect:6.3}.
%
\subsection{Deriving the RG equations}
\label{sect:6.1}
%
Following the calculation of the previous section, we first compute the contribution from the
truncation-dependent traces $\cS^4_k$ and $\cS^5_k$. Substituting the ansatz \eqref{invRansatz}
into \eqref{traces} and evaluating the resulting expressions up to linear order in $R$, we find
\be
\begin{split}
\cS^4_k=& \, \frac{k^d}{(4 \pi)^{d/2}} \, C^{\rm 2T}_0 \, \left\{ \Phi_{d/2}^1 + \half \eta_{\mb} \, \tilde{\Phi}_{d/2}^1 \, \right\} \, \int \! d^dx \sqrt{g} \\ 
&+ \frac{k^{d-2}}{(4 \pi)^{d/2}} \, \Big\{ 
C^{\rm 2T}_0 \, \Big[ \, \frac{d\left((2n-1)+d\right)-4n}{n\,d(d-1)} \, \Big(  \Phi_{d/2}^2  +\half \, \eta_{\mb} \, \tilde{\Phi}_{d/2}^2 \Big) \Big] 
\\ & \qquad 
+ C^{\rm 2T}_2 \, \left[ \Phi_{d/2-1}^1 + \half \, \eta_{\mb} \, \tilde{\Phi}_{d/2-1}^1\right]  \Big\} \, \int \! d^dx  \sqrt{g} R \, , \\
\cS^5_k=&\,\frac{k^d}{(4 \pi)^{d/2}} \left\{ 2 \Upsilon^1_{d/2,1} + \half \eta_{\mb} \tilde{\Upsilon}^1_{d/2,0,1} \right\} \int \! d^dx \sqrt{g} \\
&\,+ \frac{k^{d-2}}{(4 \pi)^{d/2}} \Big\{ 
\tfrac{4n+2+d}{2(n+1)(d-1)} 
\Big[
2 \Upsilon^2_{d/2,2} -  \Upsilon^1_{d/2,0} + \half \eta_{\mb}  ( \tilde{\Upsilon}^2_{d/2,1,1} - \tilde{\Upsilon}^1_{d/2,0,0})  
\Big]\\
& \qquad+ C_2^{\rm S} \left[ 2 \Upsilon^1_{d/2-1,1} + \half \eta_{\mb} \tilde{\Upsilon}^1_{d/2-1,0,1} \right] \Big\} \int \! d^dx  \sqrt{g} R \, .
\end{split}
\ee
Here, $\eta_{\mb}(k)$ is the anomalous dimension \eqref{andim}, and the cutoff-scheme dependent constants 
are defined in \eqref{Phinot}. Conversely, the LHS of \eqref{traces} now reads
\be
\p_t \Gammab_k[g] = \frac{1}{16 \pi G_{\hat{k}}} \, \int \! d^dx \sqrt{g} \Big[ - R \, \p_t Z_{Nk} + 2 \, \p_t(Z_{Nk} \Lambda_k) \Big] 
+ \int \! d^dx \sqrt{g} \frac{1}{R^n} \, \p_t \mb_k \, . 
\ee
Comparing the coefficients multiplying the interaction monomials in the truncation ansatz again leads to three coupled differential equations encoding the scale-dependence of $Z_{Nk}, \Lambda_k$ and $\mb_k$. Rewriting these equations in terms of the dimensionless coupling constants
\be\label{dimlessmu}
g_k  = k^{d-2} \, G_k \, , \quad 
\lambda_k = k^{-2} \Lambda_k \, , 
\quad \mu_k = \, k^{-(d+2n)} \mb_k \,,  
\ee
we then arrive at the RG equations of the $R^{-n}$-truncation,
\be\label{flow2}
\p_t g_k = \beta_g(g, \lambda, \mu) \, , \quad 
\p_t \lambda_k = \beta_\lambda(g, \lambda, \mu) \, , \quad
\p_t \mu_k  = \beta_\mu(g, \lambda, \mu) \,,
\ee
with
\be\label{betafct2}
\begin{split} 
\beta_g(g, \lambda, \mu) = & \, \left(d-2+\eta_N \right) \, g_k \, , \\
\beta_\lambda(g, \lambda, \mu) = & \, - \left( 2 - \eta_N \right) \lambda_k + (4\pi)^{1-d/2} \, g_k 
\left( (d^2-3d-2) \, \Phi^1_{d/2} + 4 \, \Upsilon^1_{d/2,1} \right) \, , \\
\beta_\mu(g, \lambda, \mu) = & \, -\left(d+2n\right) \, \mu_k \,. 
\end{split}
\ee
The anomalous dimension of Newton's constant is given by
\be
\begin{split}
\eta_N=&\,-{\rm H}_{n,d}\,g_k\,,
\end{split}
\ee
where
\be\label{Hnd}
\begin{split}
{\rm H}_{n,d} := & \,\frac{2}{(4 \pi)^{d/2-1}}\Big\{\tfrac{2+d+4n}{(n+1)(d-1)}\left[2 \Upsilon^2_{d/2,2}-\Upsilon^1_{d/2,0}\right]+4\,C_2^{\rm S}\,\Upsilon^1_{d/2-1,1}
 \qquad \\
& \quad \quad -\tfrac{d^2-6-6d\,C_2^{\rm 2T}}{3d}\,\Phi^1_{d/2-1} +2\, \left[ \tfrac{1}{n} \, C_0^{\rm 2T}+\tfrac{2(d-2)}{d(d-1)} \, C_0^{\rm 2T} -\tfrac{d^2-d+1}{d(d-1)} \right] \,\Phi^2_{d/2}\Big\} \,.
\end{split}
\ee
The $d$-dependent constants $C_{2k}^s$ are given in \eqref{B.20}. Observe that ${\rm H}_{n,d}$ is actually a (truncation-dependent) constant, independent of any coupling constants, so that $\eta_N$ is linear in $g_k$. This feature manifestly differs from the $\eta_N$ obtained in the Einstein-Hilbert truncation \cite{MR}, which contained contributions from all orders in $g_k$. Thus, the $R^{-n}$-truncations probably do not capture non-perturbative contributions from an infinite number of graviton loops. Moreover, again confirming the results of Section \ref{Sect:4}, we note that the $\beta$-function for the non-local coupling $\mb_k$ is trivial, so that $\mb_k$ is a $k$-independent constant along a RG trajectory.
We thereby notice that the non-local coupling $\mu_k$ actually decouples from the flow of $g_k, \lambda_k$, in the sense that $\beta_g(g,\lambda,\mu), \beta_g(g,\lambda,\mu)$ are independent of $\mu_k$. Thus, the RG flow for $g_k, \lambda_k$ is not altered by the value of the constant $\mb_k$ and can be studied independently from the flow of $\mu_k$. 

Finally, we observe that the $\beta$-functions have no poles in the coupling space, being well-defined for any value $g,\lambda,\mu$. Similarly to the $\ln(R)$-truncation,
 the coupling constant space decomposes into various regions, whose RG trajectories do not mix under the RG flow. These regions are separated by the fixed planes $g=0$ and $\mu=0$, as well as the fixed plane at $g^* = (d-2)/{\rm H}_{n,d}$. The latter is due to the linearity of $\eta_N$, which implies that $\beta_g(g^*, \lambda, \mu) = 0$ for any value $\lambda, \mu$, so that the $g^*$-plane cannot be crossed by the RG flow either.
%
\subsection{Fixed points of the RG flow}
\label{sect:6.2}
%
In order to get a better idea of the RG flow encoded by \eqref{betafct2}, we follow
the strategy of Subsection \ref{sect:5.2} and analyze the FP of the $\beta$-functions.
The conditions $\beta_\mu(g^*,\lambda^*,\mu^*) = 0$ implies that all FP must be located on the $\mu = 0$-plane.
Solving $\beta_g(g^*,\lambda^*,\mu^*) = 0, \beta_\lambda(g^*,\lambda^*,\mu^*) = 0$ analytically,
we find that there are two roots, yielding a GFP and a NGFP, 
\bea\label{invRFP} \nn
\mbox{GFP}:  && \{ g^* = 0 \, , \, \lambda^* = 0 \, , \, \mu^* = 0 \, \} \, , \\
\mbox{NGFP}: && \{ 
g^* = - \, \tfrac{d-2}{{\rm H}_{n,d}}
\, , \, \lambda^* = - \tfrac{d-2}{{\rm H}_{n,d}}\tfrac{(4\pi)^{1-d/2}
\left[ (d^2-3d-2) \, \Phi^1_{d/2} + 4 \, \Upsilon^1_{d/2,1} \right]}{d} \, , \, \mu^* = 0 \, \} \, .
\eea
Focusing on the NGFP for the moment, we see that its position depends crucially on the sign of ${\rm H}_{n,d}$ which, for fixed $d$, in turn depends on the exponent $n$ of the $R^{-n}$-term in our ansatz. Noting that $g^* < 0$ ($g^* > 0$) for ${\rm H}_{n,d} > 0$, $({\rm H}_{n,d}< 0)$, respectively, we define $n_{\rm crit}$ as the (possibly non-integer) positive real root of the quadratic equation ${\rm H}_{n,d}=0$, assuming it exists. The existence and actual value of the transition point $n_{\rm crit}$ are thereby cutoff-scheme dependent and hence not universal features of our flow equations. Using the optimized cutoff, this particular feature of the $R^{-n}$-truncation will be further investigated in the next subsection.

Let us now proceed by discussing the stability properties of the GFP and NGFP in turn.
Linearizing the RG flow at the GFP, the stability matrix \eqref{eq:5.15} becomes
\be
{\bf \tilde{B}}_{\rm GFP} = \left[ 
\begin{array}{ccc}
d-2 & 0 & 0 \\
(4 \pi)^{1-d/2} \big( (d^2-3d-2) \Phi^1_{d/2} + 4 \Upsilon^1_{d/2,1} \big) &
-2 & 0 \\
0 & 0 & -d-2n 
\end{array}
\right]\,,
\ee
with corresponding stability coefficients and respective right eigenvectors
\be\label{invesGFP}
\begin{array}{ll}
\tilde{\theta}_1  = 2-d \,  , \quad & \tilde{V}^1 = \left[ \, 1 \, , \tfrac{1}{d} (4 \pi)^{1-d/2} \big( (d^2-3d-2) \Phi^1_{d/2} + 4 \Upsilon^1_{d/2,1} \big) \, , \, 0 \, \right]^{\rm T} \, , \\
\tilde{\theta}_2  = 2 \, , & \tilde{V}^2 = \left[ \, 0 \, , \, 1 \, , 0 \, \right]^{\rm T} \, , \\
\tilde{\theta}_3  = 2n+d \, , &  \tilde{V}^3 = \left[ \, 0 \, , \, 0 \, , \, 1 \, \right]^{\rm T} \, . 
\end{array}
\ee
With this information at hand, it is then straightforward to solve the linearized flow equations close to the GFP. The result is here very similar to our findings in the $\ln(R)$-truncation, indicating that $G_k, \Lambda_k$ and $\mb_k$ approach $\alpha_i$-dependent constants as $k \rightarrow 0$. In particular, the IR value of $\Lambda_k$ vanishes
only if the corresponding integration constant $\alpha_2 = 0$.

Turning to the NGFP, we find that the stability matrix ${\bf \tilde{B}}$ becomes
\be
{\bf \tilde{B}}_{\rm NGFP} = \left[ 
\begin{array}{ccc}
2-d & 0 & 0 \\
(4 \pi)^{1-d/2} \big( (d^2-3d-2) \Phi^1_{d/2} + 4 \Upsilon^1_{d/2,1} \big) &
-d & 0 \\
0 & 0 & -d-2n 
\end{array}
\right]\,,
\ee
with eigensystem
\be\label{invesNGFP}
\begin{array}{ll}
\tilde{\theta}_1  = d-2 \,  , \quad & \tilde{V}^1 = \left[ \, 1 \, , \tfrac{1}{d} (4 \pi)^{1-d/2} \big( (d^2-3d-2) \Phi^1_{d/2} + 4 \Upsilon^1_{d/2,1} \big) \, , \, 0 \, \right]^{\rm T} \, , \\
\tilde{\theta}_2  = d \, , & \tilde{V}^2 = \left[ \, 0 \, , \, 1 \, , 0 \, \right]^{\rm T} \, , \\
\tilde{\theta}_3  = 2n+d \, , &  \tilde{V}^3 = \left[ \, 0 \, , \, 0 \, , \, 1 \, \right]^{\rm T} \, . 
\end{array}
\ee
Looking at the stability coefficients $\theta_i$, we note that, for $d > 2$, the NGFP is UV attractive along all eigendirections $V^i$. Solving the linearized flow equation, we then find
\be
\begin{split}\label{invRfpsol}
g_k = & \alpha_1 \frac{k^{2-d}}{M^{2-d}} -\frac{d-2}{H_{n,d}}\, , \\
\lambda_k = & \alpha_2 \frac{k^{-d}}{M^{-d}} 
+ \frac{(4 \pi)^{1-d/2}}{d^2}  \big( (d^2-3d-2) \Phi^1_{d/2} + 4 \Upsilon^1_{d/2,1} \big) \, \left(\alpha_1\,\frac{k^{2-d}}{M^{2-d}}-\frac{d(d-2)}{H_{n,d}}\right) \, ,
\\
\mu_k = & \alpha_4 \frac{k^{-d-2n}}{M^{-d-2n}} \, ,
\end{split}
\ee
where $M$ again denotes a fixed mass scale. This result indicates that the NGFP is indeed a UV attractor for the RG flow, i.e., for $k \rightarrow \infty$, all the dimensionless coupling constants $g_k, \lambda_k, \mu_k$ approach their fixed point values \eqref{invRFP} independently of the integration constants $\alpha_i$. In the case of ${\rm H}_{n,d} < 0 \Leftrightarrow  g^* > 0, \lambda^* > 0$ this FP is the natural generalization of the NGFP shown in Figure \ref{fig0}. 
%
\subsection{Numerical solutions of the flow equations}
\label{sect:6.3}
%
After discussing the general FP structure of the $\beta$-functions \eqref{betafct2},
we now restrict ourselves to $d=4$ and the optimized cutoff \eqref{optcutoff}
and proceed with the numerical investigation of the RG flow. In this case, the FP
\eqref{invRFP} are located at
\be
\begin{split}
&\mbox{GFP}: \{ g^* = 0 \, , \, \lambda^* = 0 \, , \, \mu^* = 0 \} \, , \qquad
\mbox{NGFP}: \{ g^* = \tfrac{2}{{\rm H}_{n,4}} \, , \, \lambda^* =  \tfrac{3}{8 \pi {\rm H}_{n,4}} \, , \, \mu^* = 0 \}\, ,
\end{split}
\ee
\begin{figure}[t]
\begin{center}
$\begin{array}{cc}
\multicolumn{1}{l}{} &
	\multicolumn{1}{l}{} \\ [-0.53cm]
\epsfxsize=3in
\epsffile{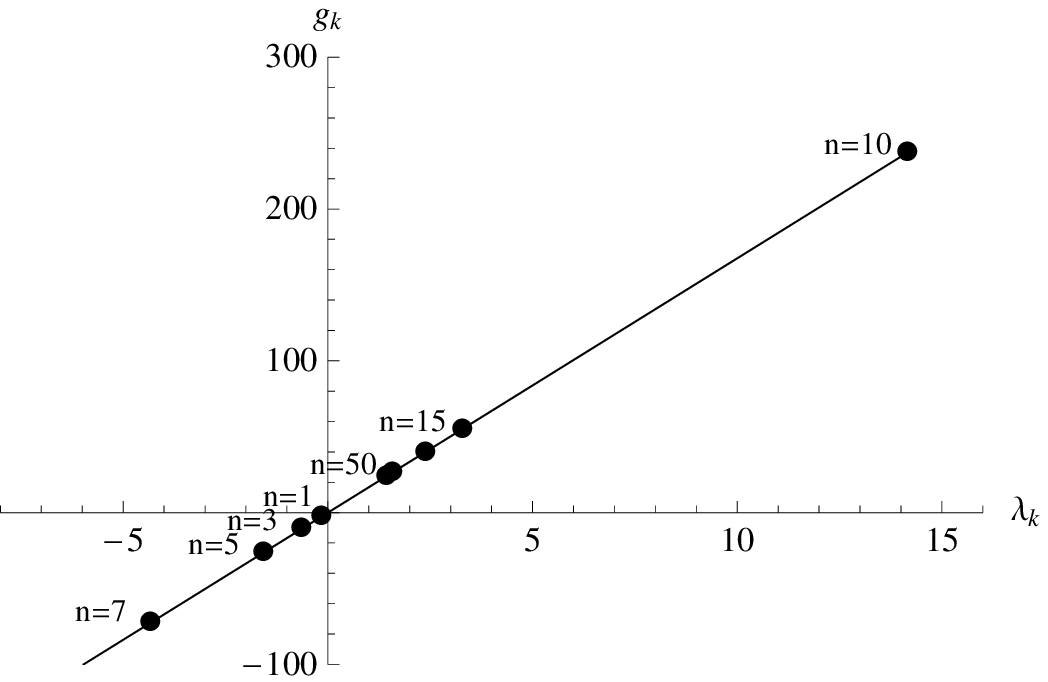} &
	\epsfxsize=3in
	\epsffile{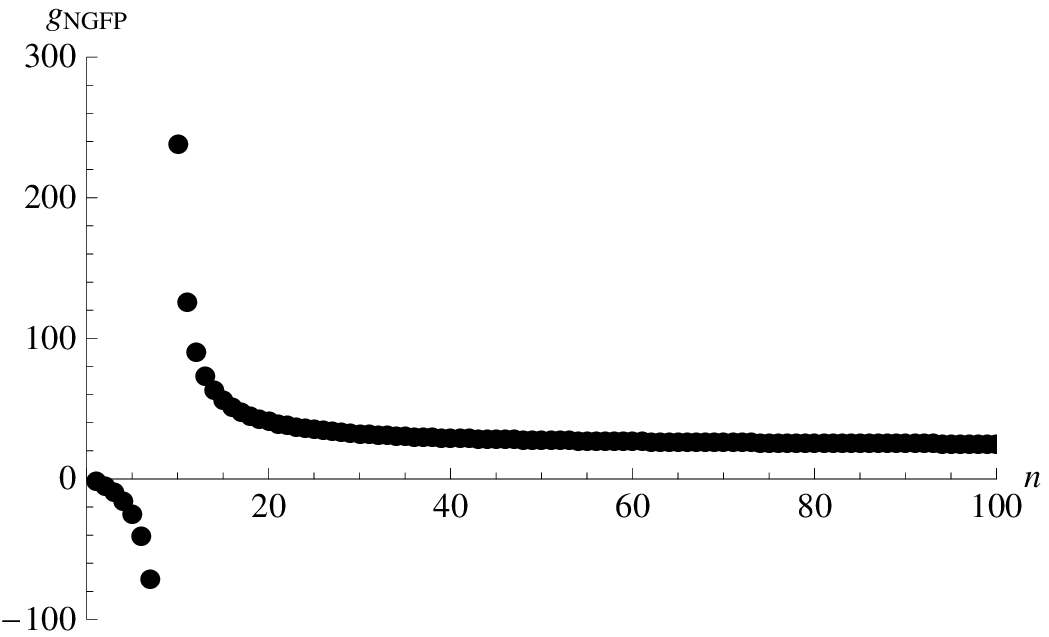}\\ [0.4cm]
\mbox{(a)} & \mbox{(b)}
\end{array}$
\end{center}
\parbox[c]{\textwidth}{\caption{\label{drei}{\footnotesize Coordinates of the NGFP on the $\mu_k=0$-plane for different values of $n$, using $d=4$ and the optimized cutoff. The transition from $g_{\rm NGFP}^* < 0$ to $g_{\rm NGFP}^* > 0$ occurs for $n_{\rm crit} \simeq 9.09$.}}}
\end{figure}
with the function \eqref{Hnd} determining the position of the NGFP, 
\be
{\rm H}_{n,4}=\frac{-7 n^2+57 n+60}{24 \pi n (n+1)}\,.
\ee
We observe that ${\rm H}_{n,4}$ as a function of $n$, indeed has a zero at positive values $n_{\rm crit} \simeq 9.09$.
At this value the NGFP passes from $g_* < 0, \lambda_* < 0$ at $n < n_{\rm crit}$ to $g_* > 0, \lambda_* > 0$ for 
$n > n_{\rm crit}$. This $n$-dependence of the NGFP is illustrated in Figure \ref{drei}. Thus, for $n > n_{\rm crit}$, the $R^{-n}$-truncation indeed gives rise to a NGFP which is a UV attractor for $g_k, \lambda_k, \mu_k$. Based on this observation, we will now discuss the RG flow for truncations with $n<n_{\rm crit}$ and $n > n_{\rm crit}$ in turn.

\begin{figure}[t]
\epsfxsize=0.8\textwidth
\begin{center}
\leavevmode
\epsffile{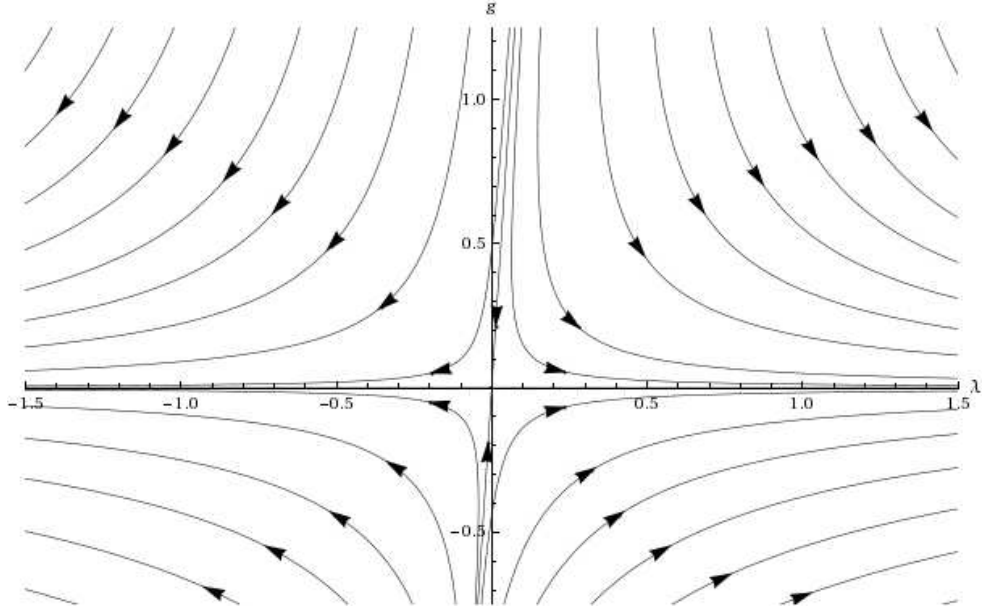}
\end{center}
\parbox[c]{\textwidth}{\caption{\label{zwei}{\footnotesize Typical RG flow of the $R^{-n}$-truncation  for 
$n < n_{\rm crit}$ on the fixed plane $\mu_k = 0$, choosing $n=1$ for explicitness.}}}
\end{figure}

Let us first consider the case $n < n_{\rm crit}$. Choosing $n=1$, the typical RG flow in the 
$\mu = 0$-plane  is exemplified in Figure \ref{zwei}.
This diagram nicely illustrates the separation between trajectories with negative and positive $g_k$. The RG flow on the upper half plane, $g > 0$, is completely dominated by the GFP. In analogy with the previous section, we can again classify the RG trajectories in this region as type I\~a or III\~a, according to whether they lie to the left or to the right of the ``separatrix'', i.e., the RG trajectory of type II\~a hitting the GFP as $k \rightarrow 0$. We note that the trajectories of {\it all three classes} can be continued to the IR, $k \rightarrow 0$, where they give rise to a negative, positive or vanishing value of $\Lambda_k$, respectively. Due to the absence of a UV FP, it is unclear, however, whether these trajectories give rise to a well-defined theory in the UV as $k \rightarrow \infty$.

\begin{figure}[t]
\epsfxsize=0.8\textwidth
\begin{center}
\leavevmode
\epsffile{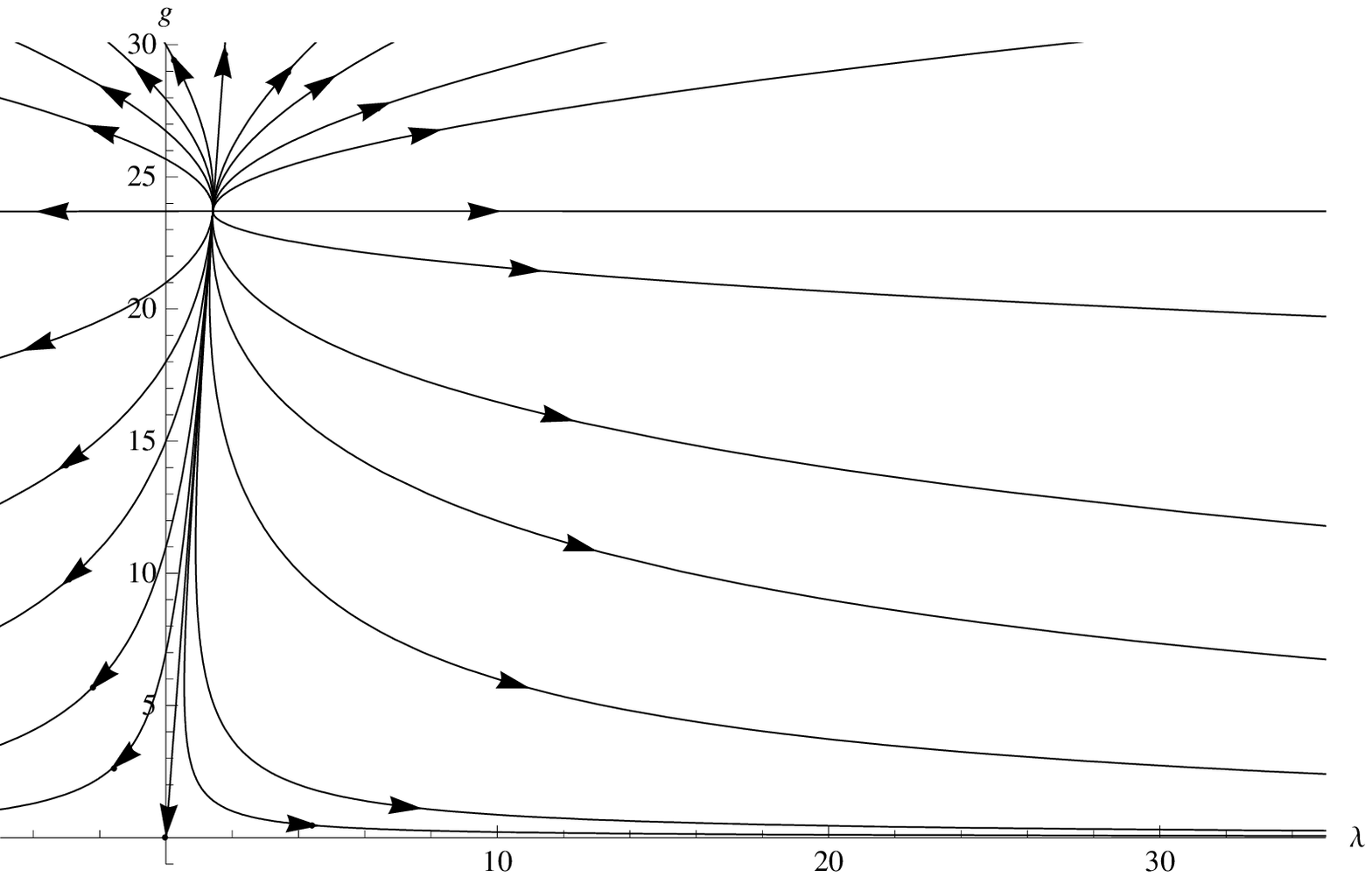}
\end{center}
\parbox[c]{\textwidth}{\caption{\label{vier}{\footnotesize Typical RG flow of the $R^{-n}$-truncations with $n > n_{\rm crit}$, illustrated by the choice $n=100$, in the fixed plane $\mu_k = 0$. The flow is dominated by the interplay between the NGFP at $g^*>0, \lambda^*>0$ and the GFP at the origin. All RG trajectories are well-defined at all scales $0 \le k \le \infty$.}}}
\end{figure}

For $n > n_{\rm crit}$ the RG flow in the upper half plane is governed by the interplay of the NGFP and the GFP. Using the example $n=100$, this is illustrated in Figure \ref{vier}. Here, we see that {\it all} RG trajectories with $g_k > 0$ emanate from the NGFP in the UV. They can then be classified into A-type and B-type trajectories, depending on whether they run below or above the $g^*_{\rm NGFP}$-plane. According to their behavior in the IR, these trajectories can furthermore be divided into type Ia, IIa and IIIa, depending on whether they give rise to a negative, zero (``separatrix'') or positive cosmological constant $\Lambda_0$. The remarkable feature of this phase diagram is that all trajectories with $g_k > 0$ have a well defined IR limit $k = 0$. Taking into account the $\mb_k$-independence of the flow, all RG trajectories on the upper half plane are complete, in the sense that they give rise to a well-defined $\Gammab_k[g]$ {\it on the entire interval} $ 0 \le k \le \infty$.

\begin{figure}[ht]
\begin{center}
$\begin{array}{cc}
\multicolumn{1}{l}{} &
	\multicolumn{1}{l}{} \\ [-0.53cm]
\epsfxsize=3in
\epsffile{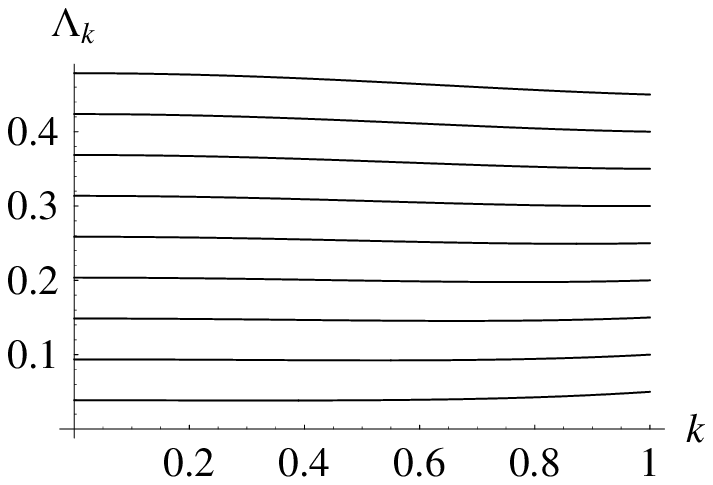} &
	\epsfxsize=3in
	\epsffile{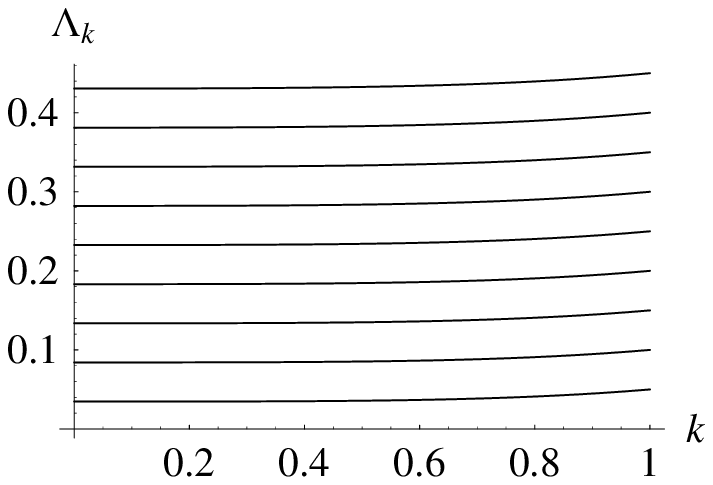} \\ 
\epsfxsize=3in
\epsffile{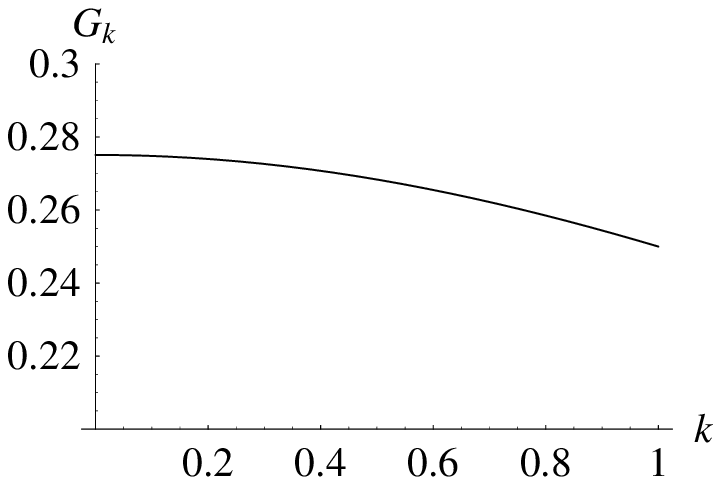} &
	\epsfxsize=3in
	\epsffile{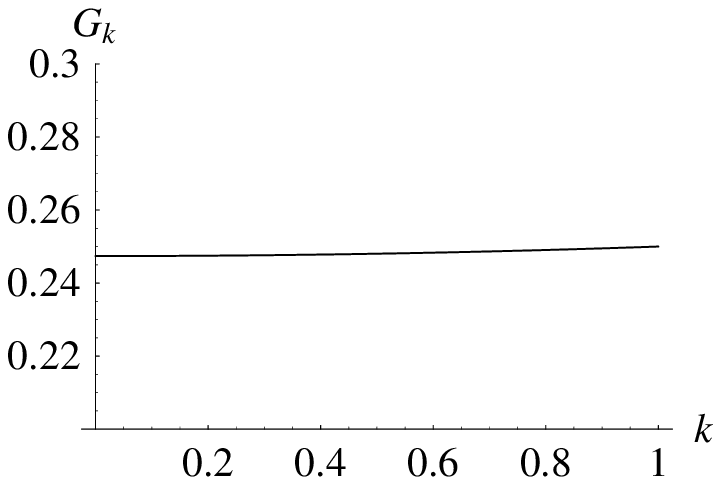} \\ [0.4cm]
\mbox{(a)} & \mbox{(b)}
\end{array}$
\end{center}
\parbox[c]{\textwidth}{\caption{\label{fuenf}{\footnotesize Numerical solutions to the flow equations of the $R^{-n}$-truncations for (a) $n<n_{\rm crit}$ and (b) A-type trajectories with $n > n_{\rm crit}$, in terms of the dimensionful couplings. Imposing the initial conditions at $\hat{k} = 1$ we have set $G_{\hat{k}}=0.25$ with $\Lambda_{\hat{k}}$ taking various values. Due to its decoupling  these results are valid for any value $\mb_{\hat{k}}$.}}}
\end{figure}
The typical RG flow of the dimensionful couplings $G_k, \Lambda_k$ along trajectories of Type III\~a ($n<n_{\rm crit}$) and A-Type IIIa ($n>n_{\rm crit}$) is then shown in the diagrams (a) and (b) of Figure \ref{fuenf}, respectively.
In both cases we find essentially no significant decrease of the cosmological constant as $k \rightarrow 0$. Moreover, due to the sign difference in its $\beta$-function we find that $G_k$ increases (decreases) towards the IR. 
The RG flow of B-type trajectories is qualitatively very similar to the one found in the A-region and therefore not shown explicitly.
%
\section{A differential equation capturing the RG flow of $f_k(R)$}
\label{Sect:7}
%
This section adopts an orthogonal approach to the truncations previously discussed, 
in the sense that we are not restricting ourselves to a specific
form of the function $f_k(R)$, but instead specialize to $d=4$ and a special
choice of cutoff-scheme, the optimized cutoff \eqref{optcutoff}. 
Using the heat-kernel results collected in Appendix \ref{App:B} then allows us to 
evaluate the traces in \eqref{traces} explicitly for arbitrary function $f_k(R)$,
resulting in an autonomous partial differential equation governing the RG flow
of $f(R)$-gravity.
\subsection{Deriving the non-perturbative flow equation}
\label{Sect:7.1}
%
We start with the following observation.\footnote{We are in debt to R.\ Percacci for explaining this point to us.} When the operator traces appearing in \eqref{traces} are evaluated via the heat-kernel expansion, their arguments $W(z)$ enter into the series expansion via the functionals $Q_n[W]$. In $d=4$, the coefficients multiplying the heat-kernel coefficients $\tr[a_{2l}], l \ge 2$, are thereby proportional to $Q_{2-l}[W]$ which are essentially the $l$th derivative of $W(z)$ evaluated at $z=0$ \eqref{Qfct2}. The virtue of working with the optimized cutoff appears when \eqref{optcutoff} is substituted into $W(z)$ and the derivatives are computed explicitly. Then, one finds that starting from a finite value $l_{\rm term}$ all derivatives vanish at $z=0$.\footnote{Strictly speaking, this ``derivative expansion'' truncates for the traces $\cS^1_k$ and $\cS^3_k$ only. The traces $\cS^4_k$ and $\cS^5_k$ give rise to an infinite series of terms which, starting at $l_{\rm term}$ are proportional to $\delta(k^2)$ and its derivatives. In the following we will neglect the contributions of such terms to the flow equation.} Thus, working with the optimized cutoff has the major advantage that the knowledge of a {\it finite} number of heat-kernel coefficients suffices to evaluate the traces \eqref{traces}. We stress that we do not expect this feature to persist once a different cutoff-scheme as, e.g., the sharp cutoff or the exponential cutoff, is used. 

For the purpose of this section, it is useful to redefine $f_k(R)$ by absorbing the wave function renormalization $Z_{Nk}$, working with $\tilde{f}_k(R) := Z_{Nk} f_k(R)$ and dropping the tilde in the following. Furthermore $V$ denotes the volume of $S^4$.

Equipped with the result above, we now proceed with the explicit evaluation of the RHS in \eqref{traces}. In contrast to the truncations discussed previously, we now do not perform a series expansion of the trace arguments with respect to $R$. Evaluating the terms in the first line then gives
\be\label{eq:Suniv}
\begin{split}
\cS^1_k = & \, - \frac{V}{(4 \pi)^2} \left[ 
k^4 \Phi^1_2(- \tfrac{R}{3k^2} ) \left. a_0 \right|_{\rm 0}
+ k^2 \Phi^1_1(-\tfrac{R}{3k^2}) \left. a_2 \right|_{\rm 0}
+ \frac{1}{1 - \tfrac{R}{3 k^2}} \, \left. a_4 \right|_{\rm 0} 
 \right] 
\\ & \, 
+  \left[ 1 + 5 \theta(1-\tfrac{R}{3k^2} ) \right] \, \big[ 1-\tfrac{R}{3k^2} \big]^{-1} \, , \\
\cS^2_k = & \, 10 \, \theta\left(1 - \tfrac{R}{3k^2}\right) \, , \\
\cS^3_k = & \, - \frac{V}{(4 \pi)^2}  \left[ 
k^4 \Phi^1_2(- \tfrac{R}{4k^2} ) \left. a_0 \right|_{\rm 1T}
+ k^2 \Phi^1_1(-\tfrac{R}{4k^2}) \left. a_2 \right|_{\rm 1T}
+ \frac{1}{1 - \tfrac{R}{4 k^2}} \, \left. a_4 \right|_{\rm 1T} 
 \right] \\
 & \, + \left[ 10 \, \theta\left(1 - \tfrac{R}{4k^2} \right) - \theta\left(1 + \tfrac{R}{4k^2} \right) \right] \, \big[1 - \tfrac{R}{4k^2} \big]^{-1} \,.
\end{split}
\ee
Here, the terms proportional to $V$ arise from the evaluation of the complete traces, the heat-kernel coefficients $a_{2k}|_s$ being given in \eqref{B.18} and \eqref{B.19}. Furthermore, the $\Phi^p_n(w)$ denote the threshold functions for the optimized cutoff \eqref{thresopt}. The $V$-independent terms capture the contribution of the finite number of 
$-D^2$-eigenmodes required to complete the traces (cf.\ Subsection \ref{sect3.3}).

The evaluation of the $f_k(R)$-dependent traces $\cS^4_k$ and $\cS^5_k$ is slightly more involved. For the 2T trace, we find
\be\label{eq:S4}
\begin{split}
S^4_k = & \, \frac{V}{2 (4\pi)^2} \, \eta_f \,  
\left[
k^4 \Pt^1_2(w_4) \, a_0|_{\rm 2T} + k^2 \Pt^1_1(w_4) \, a_2|_{\rm 2T} + \frac{1}{1+w_4} \, a_4|_{\rm 2T} - \frac{k^{-2}}{1+w_4} \, a_6|_{\rm 2T}
\right] \\
& \, + \frac{V}{(4\pi)^2} \,  
\left[
k^4 \Phi^1_2(w_4) \, a_0|_{\rm 2T} + k^2 \Phi^1_1(w_4) \, a_2|_{\rm 2T} + \frac{1}{1+w_4} \, a_4|_{\rm 2T}
\right] \\
& \, + \frac{5}{2} \, \frac{\eta_f \, (1+\tfrac{R}{3k^2}) + 2}{1+w_4} \theta(1+\tfrac{R}{3k^2})
+ 5 \frac{ \eta_f \, (1+\tfrac{R}{6k^2}) + 2}{1+w_4} \theta(1+\tfrac{R}{6k^2}) \, ,
\end{split}
\ee
with heat kernel coefficients $a_{2k}|_{\rm 2T}$ listed in \eqref{B.19}. The argument of the threshold functions is given by the dimensionless quantity
\be
w_4 := \frac{f_k}{k^2 \, f_k^\prime} - \frac{R}{3 k^2} \, ,
\ee
and, in analogy to \eqref{andim}, we have defined the ``anomalous dimension'' of $f_k^\prime(R)$ as
\be\label{andimf}
\eta_f(k) := \p_t \ln(f_k^\prime(R)) \, .
\ee

Finally, we evaluate the scalar trace $\cS^5_k$. Setting $d=4$ and comparing its argument $W_5(z)$ with the schematic notation used in \eqref{Qgen} motivates the definition
\be
u_k := 36 f^{\prime \prime}_k \, , \qquad 
v_k := 12 k^{-2} \left( f_k^\prime - 2 R f^{\prime \prime}_k \right) \, , \qquad
w_k := 4 k^{-4} \left( 2 f_k - 2 R f^\prime_k + R^2 f^{\prime \prime}_k \right) \, .
\ee
In terms of these, the derivatives of $W_5(z)|_{z=0}$ become
\be
\begin{split}
W_5(0) = & \, 12 \, \frac{3\p_t f^{\prime \prime}_k + 12 f^{\prime \prime}_k + k^{-2} \left( \p_t f^\prime_k - 2 R \p_t f^{\prime \prime}_k + 2 f^\prime_k - 4 R f^{\prime \prime}_k \right) }{u_k + v_k + w_k} \, , \\
W_5^\prime(0) = & \, -12 \, k^{-4} \, \frac{\p_t f^\prime_k - 2 R \p_t f^{\prime \prime}_k}{u_k + v_k + w_k} \, , \quad \mbox{and} \quad 
W_5^{\prime \prime}(0) =  - 72 k^{-4} \frac{\p_t f_k^{\prime \prime}}{u_k + v_k + w_k} \, .
\end{split}
\ee
As mentioned above, the higher order derivatives $W_5^{\prime\ldots\prime}(0)$ are proportional to $\delta(k^2)$ or derivatives thereof and will not be included in the following. Using the generalized threshold functions \eqref{genthrfct}, we can then explicitly evaluate $\cS_k^5$,
\be\label{eq:S5}
\begin{split}
\cS_k^5 =  \frac{V}{2(4 \pi)^2}  \Big\{ &
36 k^4 \big[ \p_t f^{\prime \prime}_k \tilde{\Upsilon}^1_{2,0,1}(u_k,v_k,w_k) + 4 f^{\prime \prime}_k \Upsilon^1_{2,1}(u_k,v_k,w_k) \big] \\
& +12 k^2 \big[ (\p_t f^\prime_k - 2 R \p_t f^{\prime \prime}_k) \tilde{\Upsilon}^1_{2,0,0}(u_k,v_k,w_k)
+2 (f^\prime_k - 2 R f^{\prime \prime}_k )  \Upsilon^1_{2,0}(u_k,v_k,w_k) \big] \\
& + 36 k^2 \big[
\p_t f^{\prime \prime}_k \, \tilde{\Upsilon}^1_{1,0,1}(u_k,v_k,w_k) + 4 f^{\prime \prime}_k \Upsilon^1_{1,1}(u_k,v_k,w_k)
\big] \, a_2|_0 \\
& + 12 \big[
(\p_t f^\prime_k - 2 R \p_t f^{\prime \prime}_k) \tilde{\Upsilon}^1_{1,0,0}(u_k,v_k,w_k)
+2 (f^\prime_k -2 R f^{\prime \prime}_k) \Upsilon^1_{1,0}(u_k,v_k,w_k)
\big] \, a_2|_0 \\
& \, + W_5(0) \, a_4|_0 - W^\prime_5(0) \, a_6|_0 + W^{\prime \prime}_k(0) \, a_8|_0 
\Big\} \, ,
\end{split}
\ee
with the heat-kernel coefficients $a_{2k}|_0$ from \eqref{B.18}. We thus see that, thanks to the optimized cutoff, the RHS of eq.\ \eqref{traces} can indeed be explicitly computed for general function $f_k(R)$. Substituting the $f_k(R)$-ansatz into the LHS, on the other hand, yields
\be\label{LHS}
\p_t \Gammab_k[g] = 2 \kappa^2 V \, \p_t f_k(R) \, .
\ee

The flow equation governing the scale-dependence of $f_k(R)$ is then obtained by substituting \eqref{eq:Suniv}, \eqref{eq:S4} and \eqref{eq:S5} into the RHS of \eqref{traces} and equating this to \eqref{LHS}. Explicitly substituting the (generalized) threshold functions from Appendix \ref{C.2} then leads to the following partial differential equation for $f_k(R)$,
\be\label{flowdimfull} 
\begin{split}
& \!\!\!2  \kappa^2 V  \, \p_t f_k(R) = \\ 
& \Big[ 1 + 5 \theta\big(1-\tfrac{R}{3 k^2}\big) 
- \tfrac{ V k^4}{2(4\pi)^2} \big( 1 + \tfrac{1}{3} \tfrac{R}{k^2} + \tfrac{29}{1080} \tfrac{R^2}{k^4} \big)\Big]
\Big[1 - \tfrac{R}{3k^2} \Big]^{-1} 
+ 10 \theta\left(1 -\tfrac{R}{3k^2}\right) \\ 
& + \Big[ 10 \theta(1-\tfrac{R}{4k^2}) - \theta(1+\tfrac{R}{4k^2}) 
- \tfrac{Vk^4}{(4\pi)^2}  \big( \tfrac{3}{2} +  \tfrac{R}{4k^2}- \tfrac{67}{1440} \, \tfrac{R^2}{k^4} \big) \Big] 
\Big[ 1 - \tfrac{R}{4k^2}\Big]^{-1} \\ 
& +  \tfrac{Vk^4}{(4 \pi)^2} \Big[ \eta_f \, \big( \tfrac{5}{12} - \tfrac{5}{24}  \tfrac{R}{k^2} - \tfrac{271}{864} \tfrac{R^2}{k^4} + \tfrac{7249}{108864} \tfrac{R^3}{k^6} \big) 
+ \big( \tfrac{5}{2} - \tfrac{5}{6}  \tfrac{R}{k^2} - \tfrac{271}{432}  \tfrac{R^2}{k^4} \big)
\Big] \Big[ 1 + \tfrac{f_k}{k^2 f^\prime_k} - \tfrac{R}{3k^2} \Big]^{-1} \\ 
& + 5  \Big[ 
 \big( \tfrac{1}{2} \eta_f \,(1+\tfrac{R}{3k^2}) + 1\big) \theta(1+\tfrac{R}{3k^2}) 
+ \big( \eta_f \, (1+\tfrac{R}{6k^2}) + 2\big) \theta(1+\tfrac{R}{6k^2}) 
\Big] \,  
\Big[ 1 + \tfrac{f_k}{k^2 f^\prime_k} - \tfrac{R}{3k^2} \Big]^{-1} \\ 
& + \tfrac{1}{8} \tfrac{Vk^8}{(4\pi)^2}
 \Big[
\p_t \fpp_k \big( 9 - \tfrac{91}{60} \tfrac{R^2}{k^4} - \tfrac{29}{90} \tfrac{R^3}{k^6} - \tfrac{181}{10080} \tfrac{R^4}{k^8}\big) 
+  \fpp_k \big( 72 - \tfrac{91}{15} \tfrac{R^2}{k^4} - \tfrac{29}{45} \tfrac{R^3}{k^6} \big) \\ 
& \qquad \qquad \quad + k^{-2} \p_t \fp_k \big(2 + \tfrac{R}{k^2} + \tfrac{29}{180} \tfrac{R^2}{k^4} + \tfrac{37}{4536} \tfrac{R^3}{k^6} \big)
+  k^{-2} \fp_k \big( 12 + 4 \tfrac{R}{k^2} + \tfrac{29}{90} \tfrac{R^2}{k^4} \big)
\Big] \\
& \qquad \qquad \qquad \times \Big[ 2 f_k + 3 k^2 f^\prime_k (1-\tfrac{2R}{3k^2}) + 9 k^4 f^{\prime \prime}_k (1-\tfrac{R}{3k^2})^2 \Big]^{-1} \, .
\end{split}
\ee

In order to discuss its properties, it is useful to rewrite eq.\ \eqref{flowdimfull} in terms of dimensionless quantities.
This is done by first noting that the volume and curvature scalar of the background 4-spheres are related by
$V = 384 \, \pi^2 R^{-2}$. Using this relation to eliminate $V$, eq.\ \eqref{flowdimfull} becomes a partial differential equation in $k,R$. The dimensionful quantities $R$ and $f_k(R)$ are then traded for their dimensionless counterparts,
\be\label{dimless}
\rho := k^{-2} \, R \quad \mbox{and} \quad \cF_k(\rho) := \frac{1}{16 \pi G_{\hat{k}}} \, k^{-4} \, f_k(R/k^2)\,.
\ee
In terms of these,
\be
\p_t f_k = 16 \pi G_{\hat{k}} \, k^4 \, \left( \p_t \cF_k + 4 \cF_k - 2 \rho \cF^{\prime}_k \right) \, ,
\ee
while the anomalous dimension \eqref{andimf} becomes
\be
\eta_f = \frac{1}{\cF^\prime_k} \, \left( \p_t \cF^\prime_k +2 \cF^\prime_k - 2 \rho \cF^{\prime \prime}_k \right) \, ,
\ee
and similar relations hold for the other derivatives of $f_k(R)$. Reexpressing the dimensionful
 quantities in \eqref{flowdimfull} through the dimensionless ones leads to the following autonomous
 partial differential equation governing the RG flow of $\cF_k(\rho)$,
\bea\label{flowdimless} 
&& 384 \pi^2   \left( \p_t \cF_k + 4 \cF_k - 2 \rho \cF^{\prime}_k \right) = \vspace*{20cm} \\ \nn
&& \quad \Big[ 5 \rho^2 \theta\left(1-\tfrac{\rho}{3}\right) 
-  \left( 12 + 4 \, \rho - \tfrac{61}{90} \, \rho^2 \right)\Big]
\Big[1 - \tfrac{\rho}{3} \Big]^{-1}  
+ 10 \, \rho^2 \, \theta\left(1 -\tfrac{\rho}{3}\right) \\ \nn
&& + \Big[ 10 \, \rho^2 \, \theta(1-\tfrac{\rho}{4}) - \rho^2 \, \theta(1+\tfrac{\rho}{4}) 
-   \left( 36 + 6 \, \rho - \tfrac{67}{60} \, \rho^2 \right) \Big] 
\Big[ 1 - \tfrac{\rho}{4}\Big]^{-1} \\ \nn
&& +  \Big[ \eta_f \, \left( 10 - 5  \rho - \tfrac{271}{36}  \rho^2 + \tfrac{7249}{4536}  \rho^3 \right) 
+ \left( 60 - 20  \rho - \tfrac{271}{18}  \rho^2 \right)
\Big] \left[ 1 + \tfrac{\cF_k}{\cF^\prime_k} - \tfrac{\rho}{3} \right]^{-1} \\ \nn
&& + \frac{5\rho^2}{2} \, \Big[ 
 \eta_f \,
\left( (1+\tfrac{\rho}{3}) \theta(1+\tfrac{\rho}{3}) + (2+\tfrac{\rho}{3}) \theta(1+\tfrac{\rho}{6}) \right)
+ 2 \theta(1+\tfrac{\rho}{3}) + 4 \theta(1+\tfrac{\rho}{6}) 
\Big]  
 \left[ 1 + \tfrac{\cF_k}{\cF^\prime_k} - \tfrac{\rho}{3} \right]^{-1} \\ \nn
&& + 
\Big[
\cF^{\prime}_k \, \eta_f \,
\left(6 + 3 \rho + \tfrac{29}{60} \rho^2 + \tfrac{37}{1512} \, \rho^3 \right)
+ \left( \p_t \cF^{\prime \prime}_k  - 2 \rho \cF^{\prime \prime \prime}_k \right) 
\left( 27 - \tfrac{91}{20} \rho^2  - \tfrac{29}{30} \, \rho^3 - \tfrac{181}{3360} \, \rho^4 \right) 
 \\ \nn
&&   + \cF^{\prime \prime}_k \left( 216 - \tfrac{91}{5}  \rho^2 - \tfrac{29}{15}  \rho^3 \right)   
+  \cF^\prime_k \left( 36 + 12  \rho + \tfrac{29}{30}  \rho^2 \right)
\Big] \Big[ 2 \cF_k + 3 \cF^\prime_k (1-\tfrac{2}{3}\rho) + 9 \cF^{\prime \prime}_k (1-\tfrac{\rho}{3})^2 \Big]^{-1} \, . 
\eea

This equation then constitutes the desired autonomous partial differential equation for the RG flow of $f(R)$-gravity. 

A very peculiar feature of \eqref{flowdimless} is the appearance of discontinuities on its RHS, induced by the
$\theta$-function terms. The origin of these contributions can be traced back to the use of the optimized cutoff, for which the finite sums of the type \eqref{eq:3.24} give rise to stepfunctions. The observed discontinuities then probably reflect the price to be paid for being able to truncate the heat-kernel expansion at a finite order. Repeating the construction
above using a different cutoff-scheme will eventually remove these stepfunctions but requires one to carry out the heat-kernel expansion to all orders in $R$. We also note that similar discontinuities have already been
observed when constructing the $\beta$-functions of the $R^2$-truncation using the exponential cutoff-scheme \cite{oliver2}. 
In that case, the discontinuities were not in the parameter $\rho$ but in the space-time dimension $d$. Evaluating the finite sums for the exponential cutoff induced contributions to the flow equation which were proportional to $\delta_{2,d}$ and $\delta_{4,d}$, so that the flow equations could change discontinuously with the space-time dimension. While it is clearly desirable to get a better understanding of such discontinuities, we will nevertheless leave this topic for the time being, and rather elucidate some other properties of the flow equation \eqref{flowdimless}, in the next subsection. 

\subsection{Fixed points of the RG equation}
\label{Sect:7.2}
%
From the dimensionless flow equation \eqref{flowdimless}, it is now straightforward to obtain an ordinary
differential equation describing the fixed functionals $\cF_{*}(\rho)$ of the RG flow of $f(R)$-gravity. By definition, these satisfy $\p_t \cF_k(\rho)=0$ and are consequently solutions of the 
non-linear ordinary differential equation 
\bea\label{fixfunctional} 
&& 768 \pi^2   \left( 2 \cF_{*} -  \rho \cF^{\prime}_{*} \right) = \\ \nn
&&  \quad \left[ 5 \rho^2 \theta\left(1-\tfrac{\rho}{3}\right) - 12 - 4 \, \rho + \tfrac{61}{90} \, \rho^2 \right]
\left[1 - \tfrac{\rho}{3} \right]^{-1} +  10 \rho^2 \theta\left(1-\tfrac{\rho}{3}\right) \\ \nn
&& + \left[  10 \rho^2 \theta\left(1-\tfrac{\rho}{4}\right) -\rho^2 \theta\left(1+\tfrac{\rho}{4}\right) -36 - 6 \, \rho + \tfrac{67}{60} \, \rho^2 \right] 
\left[ 1 - \tfrac{\rho}{4}\right]^{-1} \\ \nn
&& +  \Big[   
 \left( 80 - 30  \rho - \tfrac{271}{9}  \rho^2 + \tfrac{7249}{2268} \rho^3 \right)
  - 2 \rho \tfrac{\cF^{\prime \prime}_{*}}{\cF^\prime_{*}}  
\left( 10 - 5  \rho - \tfrac{271}{36}  \rho^2 + \tfrac{7249}{4536}  \rho^3 \right)
\Big] \left[ 1 + \tfrac{\cF_{*}}{\cF^\prime_{*}} - \tfrac{\rho}{3} \right]^{-1} \\ \nn
&& + 5\rho^2 \Big[\big( 1 - \rho \tfrac{\cF^{\prime \prime}_{*}}{\cF^\prime_{*}} \big)
\big( (1+\tfrac{\rho}{3}) \theta(1+\tfrac{\rho}{3}) + (2+\tfrac{\rho}{3}) \theta(1+\tfrac{\rho}{6}) \big) 
+ \theta(1+\tfrac{\rho}{3}) + 2 \theta(1+\tfrac{\rho}{6}) 
\Big]  
 \left[ 1 + \tfrac{\cF_{*}}{\cF^\prime_{*}} - \tfrac{\rho}{3} \right]^{-1} \\ \nn
&& - 
\Big[
 2 \rho \cF^{\prime \prime \prime}_{*}  
\left( 27 - \tfrac{91}{20} \rho^2  - \tfrac{29}{30} \, \rho^3 - \tfrac{181}{3360} \, \rho^4 \right) 
-  \cF^{\prime \prime}_{*} \left( 216 - 12\rho - \tfrac{121}{5} \, \rho^2 - \tfrac{29}{10} \, \rho^3 - \tfrac{37}{756} \, \rho^4 \right) \\ \nn
&& \qquad \;  
-  \cF^\prime_{*} \left( 48 + 18 \, \rho + \tfrac{29}{15} \, \rho^2 + \tfrac{37}{756} \, \rho^3 \right)
\Big] 
\Big[ 2 \cF_{*} + 3 \cF^\prime_{*} (1-\tfrac{2}{3}\rho) + 9 \cF^{\prime \prime}_{*} (1-\tfrac{\rho}{3})^2 \Big]^{-1} \, .
\eea
Due to its non-linearity, finding exact analytic solutions to this equation is rather involved, and a detailed analysis of 
the RG flows following from the partial differential equations \eqref{flowdimless} and \eqref{fixfunctional} is beyond the scope of the present paper. Nonetheless, let us end our discussion by pointing out some of their properties.

We first investigate the possibility of a GFP arising from the flow equation
 \eqref{fixfunctional}. In this course, we reinstall the dimensionless
 Newton's constant by setting
\be\label{eq:7.16}
\cF_{*}(\rho) = \frac{1}{16 \pi g_*} \tilde{\cF}_{*}(\rho) \, ,
\ee
where $g_*$ indicates that $g_k$ is taken at the fixed point, $\p_t g_k = 0$.
Substituting \eqref{eq:7.16} into \eqref{fixfunctional}, we see that, due to its 
homogeneity properties, the RHS of the flow equation is independent of $g_{*}$.
Schematically, \eqref{fixfunctional} then takes the form
\be\label{eq:7.17}
768 \pi^2 \left( 2 \tilde{\cF}_{*} - \rho \tilde{\cF}_*^\prime \right) = 16 \pi g_* \big[ \ldots \big] \, ,
\ee
with the terms inside the bracket being independent of $g_*$. At the GFP, we have $g_* = 0$,
so that the ``quantum corrections'' on the RHS decouple, and \eqref{eq:7.17} reduces to
\be\label{eq:7.18}
2 \tilde{\cF}_{*}^{\rm GFP} - \rho \tilde{\cF}_*^{\prime {\rm GFP}} = 0 \, .
\ee
This equation has the solution $\tilde{\cF}_{*}^{\rm GFP} = \tilde{\beta}_* \rho^2$, with $\tilde{\beta}_*$ 
an arbitrary integration constant. The corresponding fixed functional is given by
\be\label{eq:7.19}
\Gammab_*^{\rm GFP} = \frac{1}{16 \pi g_*} \int d^4x \sqrt{g} \tilde{\beta}_* R^2 \, .
\ee
We note that, with respect to classical power counting, this is just the marginal operator
of $f(R)$-gravity. Furthermore, for $\tilde{\beta}_*$ finite, the coupling $\beta_* = (16 \pi g_*)^{-1} \tilde{\beta}_*$
gets shifted to infinity as $g_* \rightarrow 0$, which is in agreement
 with the vanishing of the GFP observed in \cite{oliver2}.

\begin{table}[t]
\begin{center}
\begin{tabular}{c|c|c|c|c|c|c|c}
$n$ & $u_0^*$ & $u_1^*$ & $u_2^*$ & $u_3^*$ & $u_4^*$ & $u_5^*$  & $u_6^*$ \\ \hline
$1$ & 0.00523 & -0.0202 &         &         &         &          &         \\
$2$ & 0.00333 & -0.0125 & 0.00149 &         &         &          &         \\
$3$ & 0.00518 & -0.0196 & 0.00070 & -0.0104 &         &          &         \\
$4$ & 0.00505 & -0.0206 & 0.00026 & -0.0120 & -0.0101 &          &         \\
$5$ & 0.00506 & -0.0206 & 0.00023 & -0.0105 & -0.0096 & -0.00455 &         \\
$6$ & 0.00504 & -0.0208 & 0.00012 & -0.0110 & -0.0109 & -0.00473 & 0.00238 \\
\end{tabular}
\end{center}
\parbox[c]{\textwidth}{\caption{\label{t.3}{Location of the NGFP obtained from the fixed functional equation 
\eqref{fixfunctional} by expanding $\cF_k(\rho)$ in a power series in $\rho$ up to order $\rho^n$, including $k$-dependent coupling constants $u_i(k)$, $i = 0,\ldots,n$.}}}
\end{table}

Following \cite{CPR}, the flow equation \eqref{flowdimless} and \eqref{fixfunctional} can also
be used to investigate the UV properties of the RG flow. This has a close relation
to the asymptotic safety conjecture of quantum gravity, as it allows one 
to (perturbatively) test the existence and stability 
properties of the NGFP in truncation spaces incorporating successively higher powers of the scalar curvature.
In this light, we consider the UV limit of \eqref{flowdimless}. For fixed background curvature,
this corresponds to expanding \eqref{flowdimless} for $\rho = R/k^2 \ll 1$. In this limit, all $\theta$-functions have positive arguments and thus contribute to the flow equation.

We then make a polynomial ansatz for $\cF_k(\rho)$, setting
\be\label{eq:7.20}
\cF_k(\rho) = \sum_{i = 0}^n u_i(t) \, \rho^i \; , \quad n \ge 1 \in \Nom \, .
\ee
Here, $u_1(k) = - (16 \pi g_k)^{-1}$, so that a negative $u_1(k)$ actually corresponds to a positive Newton's constant.
Substituting this ansatz into \eqref{flowdimless}, expanding its RHS in powers of $\rho$, and  matching the coefficients up to $\rho^n$ yields the non-perturbative $\beta$-functions for the dimensionless couplings $u_i(t)$,
\be\label{eq:7.21}
\p_t u_i = \beta_{u_i}(u_0, \ldots , u_n) \,, \quad i = 0, \ldots,n \, .
\ee
The resulting $\beta$-functions indeed give rise to a NGFP with $g_* > 0, \lambda_* > 0$, 
whose $n$-dependent position is shown in Table \ref{t.3}.
Following the computation \eqref{eq:5.15} - \eqref{eq:5.17}, we also obtain the stability coefficients of the NGFP. These are summarized in Table \ref{t.4}. In particular, we see that the NGFP attracts only the couplings $u_0, u_1, u_2$ as $k \rightarrow \infty$, while it is UV repulsive for $u_i, i = 3,\ldots,6$. These results are in complete agreement with
the analysis carried out in \cite{CPR} and provides a nice confirmation of their findings.\footnote{The small numerical deviations between the results given in Table \ref{t.3} and \ref{t.4} and \cite{CPR} can be traced back to the contribution of the single mode term in \eqref{traces}.}

\begin{table}[t]
\begin{center}
\begin{tabular}{c|c|c|c|c|l|c|c}
$n$ & $\theta^\prime$ & $\theta^{\prime \prime}$ & $\theta_2$ & $\theta_3$ & $\qquad \; \; \theta_4$ & $\theta_5$ & $\theta_6$ \\ \hline
$1$ & 2.38 & -2.17 &         &          &         &         &         \\
$2$ & 1.26 & -2.44 & 27.0    &          &         &         &         \\
$3$ & 2.67 & -2.26 & 2.07    & -4.42    &         &         &         \\
$4$ & 2.83 & -2.42 & 1.54    & -4.28    &  -5.09   &         &         \\
$5$ & 2.57 & -2.67 & 1.73    & -4.40    &  -3.97 + $ 4.57 \I$ & -3.97 - $4.57 \I$ & \\
$6$ & 2.39 & -2.38 & 1.51    & -4.16    &  -4.67 + $ 6.08 \I$ & -4.67 - $6.08 \I$ & -8.67 \\
\end{tabular}
\end{center}
\parbox[c]{\textwidth}{\caption{\label{t.4}{Stability coefficients for the NGFP with increasing the dimension of the truncation space $n+1$. The first two critical exponents are a complex pair $\theta = \theta^\prime + i \theta^{\prime \prime}$.}}}
\end{table}
    
\section{Summary and discussion}
\label{sect:8}
%
In this paper, we used the functional renormalization group equation (FRGE) for quantum gravity based on the effective average action $\Gamma_k$ to derive a renormalization group (RG) equation for $f(R)$-gravity, where the gravitational Lagrangian is based on a arbitrary function of the scalar curvature. One prime feature of this flow equation is that its operator traces are invariant under constant rescalings of the scale-dependent functionals  $f_k(R) \rightarrow c f_k(R)$. Based on this property, we were able to deduce that non-local interaction terms built from the curvature scalar can be consistently decoupled from the gravitational RG flow. In particular, the coupling constants appearing in the cosmologically motivated $\ln(R)$- and $R^{-n}$-truncations \eqref{logRmodel} and \eqref{Rinvmodel} are constants along a RG trajectory, and thus the corresponding interactions are not created dynamically as a quantum gravity effect.

Combined with the results \cite{frank2}, these observations suggest that all gravitational interactions which do not occur in the heat-kernel expansion can be consistently decoupled from the gravitational RG flow. This does not imply, however, that $\Gamma_k$ is local at all scales $k$. In fact, more complicated non-local interactions built from the curvature scalar and the (inverse) Laplace operator $D^2$, of the type $\int d^dx \sqrt{g} R (D^2)^{-1} R$ or, possibly, $\int d^dx \sqrt{g} R \ln(-D^2) R$, do appear in the late-time expansion of the  heat-kernel \cite{BGVZ} and are therefore most likely dynamically generated. Including such interactions in a truncation ansatz for $\Gamma_k$ is, however, beyond the scope of the formalism employed in this paper and will require its generalization to background metrics which are non-Einstein.  Nevertheless, it would be very interesting to extend the present formalism in such a way to allow for the inclusion of such terms in the truncation ansatz, both from a phenomenological perspective (see, e.g., \cite{Deser:2007jk}), and from a more fundamental point of view, with respect to curing the conformal factor instability \cite{Mazur:1989by,Wetterich:1997bz} or finding a dynamical solution to the cosmological constant problem.

Motivated by recent attempts to explain the origin of dark matter from a modified theory of gravity which includes interactions that become dominant as the curvature scalar becomes small, the main focus of the present paper was 
the study of the non-perturbative RG flows arising from non-local extensions of the Einstein-Hilbert truncation containing  $\int d^dx \sqrt{g} \ln(R)$ ($\ln(R)$-truncation) or $\int d^dx \sqrt{g} R^{-n}$ ($R^{-n}$-truncation) terms in the ansatz for $\Gamma_k$. Including such non-local interactions in the truncation ansatz significantly changes the structure of the non-perturbative $\beta$-functions obtained in the Einstein-Hilbert truncation. Particularly, the IR singularities of the RG trajectories leading to a positive cosmological constant (found in the Einstein-Hilbert truncation and including the ``RG-trajectory realized by nature'' \cite{RW1}) are resolved. Notably, the $R^{-n}$-truncations with $n \geq n_{\rm crit}$ contain RG trajectories which are dragged into the NGFP for $k \rightarrow \infty$, while giving rise to a positive cosmological constant in the IR as $k \rightarrow 0$.

Probably the most surprising result of this paper is the infrared fixed point (IRFP) appearing in the $\ln(R)$-truncation. For RG trajectories sitting on the plane of vanishing non-local coupling, this infrared attractor dynamically drives both Newton's constant and a positive cosmological constant to zero  as $k \rightarrow 0$,  abolishing the need of fine-tuning these couplings at the initial scale. While it is premature to claim that this provides a fullfledged solution to the cosmological constant problem, it is a first explicit example illustrating how $\Lambda_k$ could be dynamically driven to zero through strong quantum gravity RG effects in the infrared. In light of the particular characteristics associated with the inclusion of non-local interaction terms built from the curvature scalar only, it is crucial, however, to show the existence of such an IRFP at the level of more sophisticated truncations in order to establish the physical viability of this attractor.  

At this stage, we feel obliged to comment on the relation between the results obtained by studying the RG flows in this paper and the asymptotic safety conjecture for quantum gravity. Notably, all our truncations give rise to a NGFP which is UV attractive for both Newton's constant and the cosmological constant. However, for some truncations this fixed point is located at negative values of Newton's constant, $g^* < 0$. At first glance, this seems to be problematic for the asymptotic safety scenario. There, the NGFP is expected to arise at a positive value of Newton's constant, so that it provides a well-defined UV limit for the physically relevant RG trajectories with $g > 0$. In order to interpret our findings, 
it is then illustrative to compare the anomalous dimension $\eta_N$ of Newton's constant obtained in the $\ln(R)$ and $R^{-n}$ truncation with the one obtained in the Einstein-Hilbert truncation \cite{MR}. While $\eta_N$ in the Einstein-Hilbert case receives contributions from arbitrary powers of $g$, the $\eta_N$ arising in our non-local truncations are ony linear in $g$. Thus, it is likely that the non-local truncations considered in this paper are insufficient to correctly capture the RG flow of gravity in the UV, even though they are technically extensions of the Einstein-Hilbert truncation. In light of this, the shift of the position of the NGFP could very well reflect a ``truncation artefact'', in the sense that having $g^* >0$ requires a more refined truncation ansatz or the consistent decoupling of the new non-local interactions. Indeed, following the later procedure, the Taylor expansion of our flow equation for $f(R)$-gravity reproduces the results of \cite{CPR}, verifying the existence of the NGFP in truncations including interactions up to order $R^6$. Ultimately, however, one would like to establish the existence of the NGFP (or a possible IRFP) from the partial differential equation governing the RG flow $f_k(R)$ by avoiding the expansion of $f_k(R)$ in a Taylor or Laurent series altogether. In any case, we feel that the exploration of the theory space underlying the FRGE approach to quantum gravity is still in its infancy and one might wonder which features still await discovery.

\subsection*{Acknowledgments}
We thank A.O.\ Barvinsky, D.\ Litim, R.\ Loll, R.\ Percacci, and M.\ Reuter 
for helpful discussions and support. P.F.M.\ is supported by the Netherlands Organization for Scientific Research (NWO) under their VICI program. F.S.\ acknowledges financial support from the ANR grant BLAN06-3-137168.

\begin{appendix}
\section{\label{App:B}Heat-kernel expansion on the $d$-spheres}

%
In order to evaluate the traces on the RHS of the flow equation \eqref{traces}, we make use of
standard heat-kernel techniques. This appendix collects the relevant formulas.
 We thereby follow the discussion and conventions \cite{oliver1}
extending their results by adapting some formulas from \cite{Avramidi:Book}. 
%
\subsection{\label{App:B.1}Heat-kernel coefficients for unconstrained fields}
%
As the key formula for evaluating operator traces built from the covariant Laplacian $-D^2$, we use
the early-time expansion of the heat-kernel
\be\label{etexpansion}
{\rm Tr}\left[ \e^{-\I t (D^2 + Q)}\right] = \left( \frac{\I}{4 \pi t} \right)^{d/2} \! \int \! d^dx \sqrt{g}
\left\{ 
{\rm tr} \, a_0 - \I \, t {\rm tr}\, a_2(x; Q) - t^2 \, {\rm tr} \, a_4(x;Q) + \ldots
\right\}\,.
\ee
Here, $Q$ denotes a matrix potential, which we take to be proportional to the unit matrix in field space, and the $a_{2k}$ are the heat-kernel or Seeley coefficients.\footnote{Note that the Heat-Kernel coefficients defined in \cite{Avramidi:Book} differ from the ones in \eqref{etexpansion} by a factor $(k/2)!$.} Furthermore, ``tr'' denotes a matrix trace in field space running over the tensor indices of the fields on which $-D^2$ acts. The derivation of the partial differential equation governing the RG flow of $f(R)$-gravity in Section \ref{Sect:7} then requires the knowledge of (some of) the $a_{2k}$ up to $k=4$. 

The first two coefficients in the expansion \eqref{etexpansion} are universal in the sense that they do not depend on whether $-D^2$ acts on unconstrained scalar, vector, or symmetric tensor fields:
\be
a_0 = \unit \; , \qquad a_2(x;Q) = P \; , \qquad P := Q + \frac{1}{6} \, R \, \unit \, .
\ee
Here, $\unit$ denotes the unit matrix in the corresponding field space, i.e.,
$\unit_{(0)} := 1$, $[\unit_{(1)}]_{\mu \nu} := g_{\mu \nu}$, and 
$[\unit_{(2)}]_{\mu \nu \rho \sigma} := g_{\mu \rho} g_{\nu \sigma}$. 
Starting from $a_4(x;Q)$, the heat-kernel coefficients include the commutator of two covariant derivatives on field space,
\be
\cR_{\mu \nu} := \left[ D_\mu \, , \, D_\nu \right] = D_\mu D_\nu - D_\nu D_\mu\,.
\ee
The corresponding terms then give different contributions to the $a_{2k}$, depending on whether $-D^2$ acts on scalars, vectors, or tensor fields. For $a_4(Q)$ we have (viz. \cite{oliver1})
\be
\begin{split}
a_4(Q)  = & \, \left[ \tfrac{5d^2 - 7d+6}{360 d (d-1)} R^2 + \tfrac{1}{6} QR + \tfrac{1}{2} Q^2 \right] \unit_{(0)} \, , \\
\left[a_4(Q)\right]_{\mu \nu}  = & \, \left[ \tfrac{5d^3 - 7d^2+6d-60}{360 d^2(d-1)} R^2 + \tfrac{1}{6} QR + \tfrac{1}{2} Q^2 \right] \Big[\unit_{(1)}\Big]_{\mu \nu} \, , \\
\left[a_4(Q)\right]_{\mu \nu \rho \sigma} = & \, \left[ \tfrac{5d^3 - 7d^2+6d-120}{360 d^2(d-1)} R^2 + \tfrac{1}{6} QR + \tfrac{1}{2} Q^2 \right] \Big[ \unit_{(2)} \Big]_{\mu \nu \rho \sigma}  
+ \tfrac{R^2}{3d^2(d-1)^2} \Big[ g_{\mu \nu} g_{\rho \sigma} - g_{\mu \sigma} g_{\nu \rho} \Big] \, , 
\end{split}
\ee
for spin 0,1, and 2, respectively. 
In order to find the contributions from $a_6(x;Q)$, we start from the general formula \cite{Avramidi:Book} and restrict to  spherically symmetric backgrounds. Taking into account the different normalizations this yields
\be
\begin{split}
a_6(Q)  = & \tfrac{1}{3!} \left[ P^3 - \tfrac{d-3}{30 \, d (d-1)} \, P \, R^2 - \tfrac{2(d+2)(d-3)}{945 \, d^2 (d-1)^2} R^3  \right] \, , \\
\left[ a_6(Q) \right]_{\mu \nu}  = & 
\tfrac{1}{3!} \Big[ P^3 - \tfrac{d^2 -3 d+ 30}{30 \, d^2 (d-1)} P R^2 - \tfrac{ 2 d^3- 2 d^2+ 51 d+126}{945 \, d^3 (d-1)^2} R^3  
\Big] \Big[ \unit_{(1)} \Big]_{\mu \nu} \, , \\[1.1ex]
\left[a_6(Q)\right]_{\mu \nu \rho \sigma}  = &
\tfrac{1}{3!}  \Big[ P^3 - \tfrac{d-3}{30 \, d (d-1)} \, P \, R^2 - \tfrac{2(d+2)(d-3)}{945 \, d^2 (d-1)^2} R^3 \Big] \Big[\unit_{(2)} \Big]_{\mu \nu \rho  \sigma} \\ 
& \qquad  - \tfrac{1}{3 d^2 (d-1)^2} \, P R^2 \, \Big[ d g_{\mu \rho} g_{\nu \sigma } - g_{\mu \nu} g_{\rho \sigma} \Big]
\\ & \qquad 
- \tfrac{d+2}{45 d^3 (d-1)^3} \, R^3 \, \Big[ d g_{\mu \rho} g_{\nu \sigma } - g_{\mu \nu} g_{\rho \sigma} \Big]   
\, .
\end{split}
\ee
For $k = 4$ we need the scalar coefficient $a_8(Q)$ only. Using the general formula \cite{Avramidi:Book} this becomes
\be
\begin{split}
a_8(Q) = \frac{1}{4!} \Big[ P^4 - \tfrac{d-3}{15d(d-1)} R^2 P^2 - \tfrac{8\, (d-3) (d+2)}{945 \, d^2 \, (d-1)^2} R^3 P 
+\tfrac{(d-3)(7d^3 -32 d^2 -59d -60)}{18900 \, d^3 \, (d-1)^3} R^4 \Big] \, .
\end{split}
\ee
Consulting the expansion of the traces \eqref{eq:Suniv}, \eqref{eq:S4}, and \eqref{eq:S5}, we then see that the knowledge of these heat-kernel coefficients suffices to evaluate the RHS of \eqref{traces} in Section \ref{Sect:7}.
%
\subsection{Heat-kernel coefficients for fields with differential constraints}
\label{App:B.2}
%
\begin{table}[t]
\begin{center}
\begin{tabular}{|c|c|c|c|c|} \hline
Eigenfunction & Spin $s$ & Eigenvalue $\Lambda_l(d,s)$ & Degeneracy $D_l(d,s)$ & \\ \hline \hline
$T^{lm}_{\mu \nu}(x)$ & 2 & 
$\frac{l(l+d-1)-2}{d(d-1)} R$ & $\frac{(d+1)(d-2)(l+d)(l-1)(2l+d-1)(l+d-3)!}{2(d-1)!(l+1)!}$ & $l=2,3,\ldots$ \bigg. \\ \hline
$T^{lm}_{\mu}(x)$ & 1 & 
$\frac{l(l+d-1)-1}{d(d-1)} R$ & $\frac{l(l+d-1)(2l+d-1)(l+d-3)!}{(d-2)!(l+1)!}$ & $l=1,2,\ldots$ \bigg. \\ \hline
$T^{lm}(x)$ & 0 & 
$\frac{l(l+d-1)}{d(d-1)} R$ & $\frac{(2l+d-1)(l+d-2)}{l!(d-1)!}$ & $l=0,1,\ldots$ \bigg. \\ \hline
\end{tabular}
\end{center}
\parbox[c]{\textwidth}{\caption{\label{t.2}{Eigenvalues and degeneracies of $-D^2$ on the $d$-sphere. Here $T^{lm}(x)$, $T^{lm}_\mu(x)$ and $T^{lm}_{\mu \nu}$ form a complete orthonormal basis for the $-D^2$-eigenfunctions for
scalars, transverse vectors and transverse traceless symmetric tensors. We refer to \cite{oliver1} for more details.}}}
\end{table}

Now, the fields $h_{\mu \nu}^{\rm T}$ and $\xi^{\mu}$  are subject to the constraints given by the TT decomposition and thus, to compute their heat kernel coefficients, we must modify the formulas for spin 1 and 2 fields accordingly. To do so, we first note that the TT decomposition \eqref{TTmetric} and \eqref{TTghost} implies the following relations between the heat-kernel traces over the constrained and unconstrained fields
\be\label{trace2}
\begin{split}
{\rm Tr}_{(1)}\left[ \e^{-\I t (D^2 + \,qR)}\right] = & 
\, {\rm Tr}_{\rm 1T}\left[ \e^{-\I t (D^2 +\, qR)}\right]  + 
{\rm Tr}_{0}\left[ \e^{-\I t (D^2 + \,\tfrac{dq+1}{d}R)}\right] 
-\e^{-\I t \, \tfrac{dq+1}{d}R} \, ,\\
{\rm Tr}_{(2)}\left[ \e^{-\I t (D^2 + \,qR)}\right] = & \,{\rm Tr}_{\rm 2T}\left[ \e^{-\I t (D^2 + \,qR)}\right] 
+ {\rm Tr}_{\rm 1T}\left[ \e^{-\I t (D^2 +\, \left(\tfrac{d+1}{d(d-1)} + q\right)R)}\right] \\
&+ {\rm Tr}_{0}\left[ \e^{-\I t (D^2 +\, \left(\tfrac{2}{d-1} + q\right)R)}\right] 
+ {\rm Tr}_{0}\left[ \e^{-\I t (D^2 + \,qR)}\right]  \\
& - \e^{-\I t \, \left(\tfrac{2}{d-1}+q \right)R} - \left( d+1 \right) \, \e^{-\I t \, \left(\tfrac{1}{d-1}+q \right)R} - \tfrac{d(d+1)}{2} \, \e^{-\I t \, \left(\tfrac{2}{d(d-1)}+q \right)R} \, .
\end{split}
\ee
Here we set $Q = qR\unit$. There are two types of terms appearing on the RHS of these equations, complete traces with respect to the constraint fields and contributions from a discrete set of eigenvalues, which are required to complete the traces.
The relations \eqref{trace2} then allow one to compute the heat-kernel coefficients for the constraint fields from their unconstrained counterparts given in the last subsection. Defining
$a_{2k}|_s := {\rm tr} \left[ \left. a_{2k} \right|_{s} \right]$ with $s = 0, {\rm 1T}, {\rm 2T}$ denoting the spin of the constraint field and performing the trace over vector indices explicitly we find
\be\label{B.1}
\begin{split}
 \left. a_0 \right|_{\rm 1T}  = & \, d-1 \, , \\
 \left. a_2 \right|_{\rm 1T}  = & \, \frac{1}{6d}\, (d+2)(d-3) \, R\, , \\
 \left. a_4 \right|_{\rm 1T}  = & \, \frac{5d^4-12d^3-47d^2-186d+180}{360 \, d^2 \, (d-1)} \, R^2 \, , \\
\end{split}
\ee 
and
\be\label{B.2}
\begin{split}
 \left. a_0 \right|_{\rm 2T}  = & \, \frac{1}{2}(d-2)(d+1) \, , \\
 \left. a_2 \right|_{\rm 2T}  = & \, \frac{(d-5)(d+1)(d+2)}{12 \, (d-1)} \, R\, , \\
 \left. a_4 \right|_{\rm 2T}  = & \, \frac{(d+1)(5d^4 -22d^3-83d^2-392d-228)}{720 \, d \, (d-1)^2} \, R^2 \, , \\
 \left. a_6 \right|_{\rm 2T}  = & \, 
\frac{(d+1)(d+2)(35d^5-287d^4-93d^3-7765d^2+1966d-14016)}{90720 \, d^2 \, (d-1)^3} \, R^3 \, , 
\end{split}
\ee 
for the heat-kernel coefficients of ${\rm Tr}_{\rm 1T}[ \e^{-\I t D^2} ]$ and ${\rm Tr}_{\rm 2T}[ \e^{-\I t D^2} ]$, respectively. Since the traces \eqref{trace2} are evaluated under a Fourier integral (cfg.\ eq.\ \eqref{Ftrans}) it is also
straightforward to deal with the discrete mode terms. Carrying out their (inverse) Fourier transform, these become proportional to $\delta$-functions which allow for an easy evaluation of the subsequent $s$-integral.

With regard to the derivation in Section \ref{Sect:7} it is also convenient to give the numerical value of the heat-kernel coefficients for the special case $d=4$. In the scalar sector, setting $Q=0$, we have
\be\label{B.18}
\begin{split}
\left. a_0 \right|_{\rm 0} = & \, 1 \, , \quad
\left. a_2 \right|_{\rm 0} = \frac{1}{6} \, R \, , \quad
\left. a_4 \right|_{\rm 0} = \frac{29}{2160} \, R^2 \, , \quad
\left. a_6 \right|_{\rm 0} = \frac{37}{54432} \, R^3 \, , \quad
\left. a_8 \right|_{\rm 0} = \frac{149}{6531840} \, R^4 \, ,
\end{split}
\ee
while \eqref{B.1} and \eqref{B.2} yield
\be\label{B.19}
\begin{split}
{\rm tr} \left[ \left. a_0 \right|_{\rm 1T} \right] = & \, 3 \, , \quad
{\rm tr} \left[ \left. a_2 \right|_{\rm 1T} \right] = \frac{1}{4} \, R \, , \quad
{\rm tr} \left[ \left. a_4 \right|_{\rm 1T} \right] = - \frac{67}{1440} \, R^2 \, , \\
{\rm tr} \left[ \left. a_0 \right|_{\rm 2T} \right] = & \, 5 \, , \quad
{\rm tr} \left[ \left. a_2 \right|_{\rm 2T} \right] = -\frac{5}{6} \, R \, , \quad
{\rm tr} \left[ \left. a_4 \right|_{\rm 2T} \right] = - \frac{271}{432} \, R^2 \, , \quad
{\rm tr} \left[ \left. a_6 \right|_{\rm 2T} \right] = - \frac{7249}{54432} \, R^3 \, . \\ 
\end{split}
\ee
Furthermore, when writing down the flow equations in sections \ref{sect:5} and \ref{sect:6}, we used the following
shorthand notation for the $d$-dependent factors of the heat-kernel coefficients, obtained by setting $R=1$ in the formulas above
\be\label{B.20}
\begin{split}
C_{2k}^{\rm S}(d) := a_{2k}|_{\rm 0}(R=1) \, , \quad 
C_{2k}^{\rm 1T}(d) := a_{2k}|_{\rm 1T}(R=1) \, , \quad 
C_{2k}^{\rm 2T}(d) := a_{2k}|_{\rm 2T}(R=1) \, . 
\end{split}
\ee
%
\section{Threshold functions}
\label{App:C}
This appendix collects the details about the standardized dimensionless threshold functions
$\Phi^p_n(w), \Pt^p_n(w)$ and $\Upsilon^p_{n;m}(u,v,w), \tilde{\Upsilon}^p_{n,m,l}(u,v,w)$,
which were used to encapsulate the cutoff-scheme dependence of our flow equations 
in the main part of this paper. We start by giving their general definitions
and some of their properties in Appendix \ref{C.1}, before specializing to the
optimized cutoff \cite{optcutoffl} in Appendix \ref{C.2}.
%
\subsection{General definitions}
\label{C.1}
%
Naturally, the $\beta$-functions derived from the FRGE depend on the 
choice (of the scalar part) of the IR cutoff operators \eqref{eq:3.12},
$R_k = k^2 \, R^{(0)}(-D^2/k^2)$.
At the level of the flow equations, it is convenient to encode this cutoff-scheme dependence
in the standard dimensionless threshold functions
\be\label{thrfct}
\begin{split}
\Phi_n^p(w) := & \, \frac{1}{\Gamma(n)} \int_0^\infty \! dy \, y^{n-1} \, 
\frac{R^{(0)}(y) - y R^{(0) \prime}(y)}{\left( y + R^{(0)}(y) + w \right)^p} \, , \\
\tilde{\Phi}_n^p(w) := & \, \frac{1}{\Gamma(n)} \int_0^\infty \! dy \, y^{n-1} \, 
\frac{R^{(0)}(y)}{\left( y + R^{(0)}(y) + w \right)^p} \, .
\end{split}
\ee
Here, $R^{(0)}(y)$ is a dimensionless profile function determining the shape of the IR cutoff, while the prime denotes the derivative with respect to the argument. We note that both $\Phi^p_n(w)$ and $\Pt^p_n(w)$ satisfy a recursion relation when taking derivatives with respect to their argument,
\be\label{phirec}
\p_w \Phi^p_n(w) = -p \, \Phi^{p+1}_n(w) \, , \qquad \p_w \Pt^p_n(w) = -p \, \Pt^{p+1}_n(w) \, .
\ee

In order to evaluate the last trace in \eqref{traces}, a slight generalization of these threshold functions is required
\be\label{genthrfct}
\begin{split}
\Upsilon^p_{n;m}(u,v,w) := & \, \frac{1}{\Gamma(n)} \int_0^\infty \! dy \, y^{n-1} \,
\frac{\left(y + R^{(0)}(y) \right)^m \, \left( R^{(0)}(y) - y R^{(0) \prime}(y) \right) }{\left( u \left(y + R^{(0)}(y) \right)^2 + v \left(y + R^{(0)}(y) \right) + w \right)^p}
\, , \\
\tilde{\Upsilon}^p_{n,m,l}(u,v,w) := & \, \frac{1}{\Gamma(n)} \int_0^\infty \! dy \, y^{n-1} \,
\frac{\left(y + R^{(0)}(y) \right)^m \, \left(2y + R^{(0)}(y) \right)^l \, R^{(0)}(y) }{\left( u \left(y + R^{(0)}(y) \right)^2 + v \left(y + R^{(0)}(y) \right) + w \right)^p} \, . 
\end{split}
\ee
These are similar to the threshold functions in \cite{oliver2}, but adapted to the form of the traces appearing in \eqref{traces}. The standard threshold functions \eqref{thrfct} are contained in \eqref{genthrfct} as a special case,
\be\label{thrlimit}
\Upsilon^p_{n;0}(0,1,w) = \Phi_n^p(w) \, , \quad \tilde{\Upsilon}^p_{n,0;0}(0,1,w) = \tilde{\Phi}^p_n(w) \, .
\ee
Obviously, the generalized threshold functions are homogeneous of degree $-p$ under a common rescaling of their arguments
\be
\Upsilon^p_{n,m}(\lambda u, \lambda v,\lambda w) = \lambda^{-p} \, \Upsilon^p_{n,m}(u, v, w) \, , \quad 
\tilde{\Upsilon}^p_{n,m,l}(\lambda u, \lambda v,\lambda w) = \lambda^{-p} \, \tilde{\Upsilon}^p_{n,m,l}(u,v,w) \, .
\ee
Furthermore, they satisfy recursion relations similar to \eqref{phirec}
\bea\label{upsrec} \nn
\p_u \, \Upsilon^p_{n,m}(u, v, w) = & \!\! -p \, \Upsilon^{p+1}_{n,m+2}(u,v,w) \, , \; 
& \p_u \, \tilde{\Upsilon}^p_{n,m,l}(u, v, w) =  -p \, \tilde{\Upsilon}^{p+1}_{n,m+2,l}(u,v,w) \, , \\
\p_v \, \Upsilon^p_{n,m}(u, v, w) = & \!\! -p \, \Upsilon^{p+1}_{n,m+1}(u,v,w) \, , \; 
& \p_v \, \tilde{\Upsilon}^p_{n,m,l}(u, v, w) =  -p \, \tilde{\Upsilon}^{p+1}_{n,m+1,l}(u,v,w) \, , \\ \nn
\p_w \, \Upsilon^p_{n,m}(u, v, w) = & \!\! \! \!\! \! -p \, \Upsilon^{p+1}_{n,m}(u,v,w) \, , \;
& \p_w \, \tilde{\Upsilon}^p_{n,m,l}(u, v, w) =  -p \, \tilde{\Upsilon}^{p+1}_{n,m,l}(u,v,w) \, .
\eea
The recursion relations \eqref{phirec} and \eqref{upsrec} are particularly useful when performing the series expansion of 
 $R$-dependent argument as, e.g., in Subsections \ref{sect3.3}, \ref{sect:5.1} and \ref{sect:6.1}.
%
\subsection{Evaluation for the optimized cutoff}
\label{C.2}
%
For numerical evaluations, one needs to fix the form of the cutoff-scheme explicitly. Throughout this paper we thereby employed the so-called optimized cutoff \cite{optcutoffl}
\be\label{optcutoff}
R_k(z) = (k^2 - z) \, \theta(k^2 - z) \, , \quad z = -D^2 \, ,
\ee
with profile function

\be\label{optcut}
R^{(0)}(y) = (1-y) \theta(1-y) \, , \quad y = -D^2/k^2 \, . 
\ee
On the conceptual side, the main virtue of this cutoff is the minimization of the 
(modified) Ward-identities, thereby leading to more reliable and accurate RG flows. On the technical side, 
the optimized cutoff allows to carry out the integrals appearing in the (generalized)
threshold functions analytically. Substituting \eqref{optcut} into \eqref{thrfct},
one obtains
\be\label{thresopt}
\Phi_n^p(w) = \frac{1}{\Gamma(n+1)} \, \frac{1}{(1+w)^p} \, , \quad \tilde{\Phi}_n^p(w) = \frac{1}{\Gamma(n+2)} \, \frac{1}{(1+w)^p} \, .
\ee
Similarly, the generalized threshold functions \eqref{genthrfct} become
\be\label{upsopt}
\begin{split}
\Upsilon^p_{n,m}(u,v,w) =  & \, \frac{1}{\Gamma(n+1)} \, \frac{1}{\left( u + v + w \right)^{p}} \, , \\
\tilde{\Upsilon}^p_{n,m;l}(u,v,w) = & \, \frac{(-1)^n}{\Gamma(n)} \, \frac{ \beta(-1,n,l+1) + \beta(-1,n+1,l+1) }{ (u + v+w)^{p}} \, .
\end{split}
\ee
Here, $\beta(-1,n,l)$ denotes the incomplete beta function. For fixed values $n,l$, these become constants. In particular we have $\beta(-1,2,2) = 5/6$ and $\beta(-1,3,2) = - 7/12$. Observe that the optimized cutoff induces a degeneracy in the generalized threshold functions, in the sense that their RHS is $m$-independ. Making use of the properties of the incomplete beta-function, one can also verify explicitly that \eqref{thresopt} and \eqref{upsopt} agree in the limit \eqref{thrlimit}.
%
\end{appendix}

%
\end{document}